\documentclass[aps,prd,showpacs,nofootinbib,showkeys]{revtex4}
\usepackage{amssymb}
\usepackage{amsmath}
\usepackage[dvipsone]{graphics}
\usepackage{epsfig}
\usepackage{bm}
\usepackage{natbib}
\usepackage{color}

\expandafter\ifx\csname package@font\endcsname\relax\else
\expandafter\expandafter
 \expandafter\usepackage
 \expandafter\expandafter
 \expandafter{\csname package@font\endcsname}%
\fi

\newcommand{\be}{\begin{equation}}
\newcommand{\en}{\end{equation}}
\newcommand{\br}{\begin{eqnarray}}
\newcommand{\er}{\end{eqnarray}}

\renewcommand{\l}{\label}

\renewcommand{\t}{\times}
\renewcommand{\a}{\alpha}
\renewcommand{\b}{\beta}
\newcommand{\g}{\gamma}
\newcommand{\sss}{\scriptscriptstyle}

\newcommand{\p}{\partial}
\renewcommand{\le}{\left}
\renewcommand{\r}{\right}
\def\ct#1{\cite{#1}}
\def\s{Section }

\makeatletter
\@addtoreset{equation}{section}
\makeatother

\def\Quadrat#1#2{{\vcenter{\hrule height #2
  \hbox{\vrule width #2 height #1 \kern#1
    \vrule width #2}
  \hrule height #2}}}
\def\dAl{\mathop{\kern 1pt\hbox{$\Quadrat{8pt}{0.4pt}$} \kern1pt}}

\newcommand{\nn}[1]{\sum\limits_{#1}^\infty}

\newcommand{\nnb}[2]{{\varphi}(#1,{\bm #2})}

\newcommand{\nnbb}[3]{\hat\varphi^{(#1)}(#2, {\bm #3})}

\newcommand{\nnd}[1]{\stackrel{{(#1)}}{\zeta}}
\newcommand{\nnw}{{\bm w}}
\newcommand{\nne}[3]{\stackrel{{(#1)}}{h}_{#2#3}}
\newcommand{\nnt}[3]{\stackrel{{(#1)}}{T}_{#2#3}}
\newcommand{\nnfa}[5]{\hat h^{{(#1)}}_{#2#3}(#4,
 {\bm #5})}

\newcommand{\nnff}[1]{\Psi_{#1}(t, {\bm x})}

\newcommand{\nnxe}{({\bm x}_{{ B}})}

\newcommand{\nnx}{{\bm x}}

\newcommand{\nag}[1]{\hat h_{#1}(u,{\bm w})}

\newcommand{\nxi}{\hat\varphi(u,{\bm w})}

\newcommand{\nnv}[1]{v^{#1}_{{ B}}}
\newcommand{\nnr}[1]{R^{#1}_{{ B}}}
\newcommand{\nnaa}[1]{a^{#1}_{{ B}}}

\begin{document}
\title{Post-Newtonian Reference Frames for Advanced Theory of the Lunar Motion and a New Generation of Lunar Laser Ranging}
\author{Sergei Kopeikin}
\email{kopeikins@missouri.edu}
\affiliation{Department of Physics and Astronomy, University of Missouri-Columbia,  MO 65211, USA}
\author{Yi Xie}
\affiliation{Astronomy Department, Nanjing University, Nanjing, Jiangsu 210093, China}
\affiliation{Department of Physics and Astronomy, University of Missouri-Columbia, MO 65211, USA}
\date{\today}

\begin{abstract}
We construct a set of post-Newtonian reference frames for a comprehensive study of the orbital dynamics and rotational motion of Moon and Earth by means of lunar laser ranging (LLR) with the precision of one millimeter. We work in the framework of a scalar-tensor theory of gravity depending on two parameters, $\beta$ and $\gamma$, of the parameterized post-Newtonian (PPN) formalism and utilize the concepts of the relativistic resolutions on reference frames adopted by the International Astronomical Union (IAU) in 2000. We assume that the solar system is isolated and space-time is asymptotically flat at infinity. The primary reference frame covers the entire space-time, has its origin at the solar-system barycenter (SSB) and spatial axes stretching up to infinity. The SSB frame is not rotating with respect to a set of distant quasars that are assumed to be at rest on the sky forming the International Celestial Reference Frame (ICRF). The secondary reference frame has its origin at the Earth-Moon barycenter (EMB). The EMB frame is locally-inertial with its spatial axes spreading out to the orbits of Venus and Mars, and is not rotating dynamically in the sense that equation of motion of a test particle moving with respect to the EMB frame, does not contain the Coriolis and centripetal forces. Two other local frames -- the geocentric (GRF) and the selenocentric (SRF) frames --  have their origins at the center of mass of Earth and Moon respectively and do not rotate dynamically. Each local frame is subject to the geodetic precession both with respect to other local frames and with respect to the ICRF because of the relative motion. The advantage of dynamically non-rotating local frames is in a more simple mathematical description. The set of the global and three local frames is introduced in order to fully decouple the relative motion of Moon with respect to Earth from the orbital motion of the Earth-Moon barycenter as well as to connect the coordinate description of the lunar motion, an observer on Earth, and a retro-reflector on Moon to directly measurable quantities such as the proper time and the round-trip laser-light distance. We solve the gravity field equations and find out the metric tensor and the scalar field in all frames, which description includes the post-Newtonian definition of the multipole moments of the gravitational field of Earth and Moon. We also derive the post-Newtonian coordinate transformations between the frames and analyze the residual gauge freedom imposed by the scalar-tensor theory on the metric tensor. The residual gauge freedom is used for removal spurious, coordinate-dependent post-Newtonian effects from the equations of motion of Earth and Moon.

\end{abstract}
\maketitle
\newpage
\tableofcontents
\newpage
\section{Introduction}\l{intro}
\subsection{Background}\l{back}
The tremendous progress in technology, which we have witnessed during the last 30 years, has led to enormous improvements of precision in the measuring time and distances within the boundaries of the solar system. Further significant growth of the accuracy of astronomical observations is expected in the course of time. Observational techniques like lunar and satellite laser ranging, radar and Doppler ranging, very long baseline interferometry, high-precision atomic clocks, gyroscopes, etc. have made it possible to start probing not only the static but also kinematic and dynamic effects in motion of celestial bodies to unprecedented level of fundamental interest. Current accuracy requirements make it inevitable to formulate the most critical astronomical data-processing procedures in the framework of Einstein's general theory of relativity. This is because major relativistic effects are several orders of magnitude larger than the technical threshold of practical observations and in order to interpret the results of such observations, one has to build physically-adequate relativistic models. Many current and planned observational projects and specialized space missions can not achieve their goals unless the relativity is taken into account properly. The future projects will require introduction of higher-order relativistic models supplemented with the corresponding parametrization of the relativistic effects, which will affect the observations.

The dynamical modeling for the solar system (major and minor planets), for deep space navigation, and for the dynamics of Earth's satellites and Moon must be consistent with general relativity. Lunar laser ranging (LLR) measurements are particularly important for testing general relativistic predictions and for advanced exploration of other laws of fundamental gravitational physics. Current LLR technologies allow us to arrange the measurement of the distance from a laser on Earth to a corner-cube reflector (CCR) on Moon with a precision approaching 1 millimeter \cite{battat,murthyetal08}. There is a proposal to place a new CCR array on Moon \cite{llr21century}, and possibly to install other devices such as microwave transponders \cite{transponder} for multiple scientific and technical purposes.
Successful human exploration of the Moon strongly demands further significant improvement of the theoretical model of the orbital and rotational dynamics of the Earth-Moon system. This model should inevitably be based on the theory of general relativity, fully incorporate the relevant geophysical processes, lunar libration, tides, and should rely upon the most recent standards and recommendations of the IAU for data analysis \cite{iau2000}.

This paper discusses relativistic reference frames in construction of the high-precise dynamical model of motion of Moon and Earth. The model will take into account all the classical and relativistic effects in the orbital and rotational motion of Moon and Earth at the millimeter level. Although a lot of efforts has been made in this field of dynamic astronomy, there are some controversial issues, which obscure the progress in better understanding of the fundamental principles of the relativistic model of the Earth-Moon system (see, for example, recent discussion \cite{k07,2007PhRvL..98g1102M,2007PhRvL..98v9002M}). It is one of the goals of our investigation and,particularly, this paper to clarify these incomprehensible issues in order to allow the upcoming millimeter LLR to perform one of the most precise fundamental tests of general relativity in the solar system.

From Newton's time, many mathematical astronomers have attempted to create lunar theories capable to yielding predictions as accurate as the best observed positions of Moon. The first attempt was undertaken by Newton himself \ct{1975QB391.N55......}; subsequent contributions, in the framework of his theory of gravitation \ct{newton}, were given by many remarkable astronomers including Euler, Laplace, Delaunay, Newcomb, Brown, Hill, and others \cite{Euler,1754tdll.book.....C,dalam,poisson,lapla,damo,plana,Lubbock,ponte, delau,hill,hansen,airy,tiss,bro,adams,newc,poin,plum,eck,deprit}.
The theories can be grouped into three categories: analytic, numerical-analytic and numerical. The most impressive example of a purely analytic theory is given by Delaunay \cite{delau} whose elaboration took over twenty years. Because of its completely analytic nature, it can be applied to any three-body problem. Examples of semi-analytic theories are those by Laplace \cite{lapla} and Brown-Hill \cite{hill,bro}. For example, the Brown-Hill theory includes 1500 separate terms and was used in the Apollo program of the human exploration of the Moon \ct{gutz}. An entirely numerical theory was devised by Airy \ct{airy}.

After Einstein published his theory of gravitation  \ct{ein}, its effects on the lunar dynamics were worked out by a number of authors: de Sitter \ct{des}, Chazy \ct{chazy}, Eddington \ct{eddi}, Brumberg \ct{brum}, Baierlein \ct{baier}, Lestrade and Bretagnon \ct{lest}, Mashhoon and Theiss \ct{mash}, Soffel et al \ct{sof}. Modern lunar ephemerides fully including the post-Newtonian effects of general relativity are the ELP (Institut de M\'{e}canique C\'{e}leste et de Calcul des \'{E}ph\'{e}m\'{e}rides) \ct{chatc}, LE (Jet Propulsion Laboratory, NASA) \ct{sta}, EPM (Institute of Applied Astronomy, Russian Academy of Sciences) \ct{epm} and PMO (Purple Mountain Observatory, Chinese Academy of Sciences) \ct{pmoe}.

Historical reviews of the Moon-Earth-Sun problem can be found in \ct{cook, gutz}. Introductory general overviews of the main Newtonian and post-Newtonian features of the lunar motion are given by Roy \ct{roy}, Soffel \ct{sof89}, and Brumberg \ct{brum}.

\subsection{Lunar Laser Ranging}\l{lulr}

Lunar laser ranging (LLR) is a technique based on a set
of laser stations on Earth and corner retro-reflectors (CCR) located on a visible (near) side of Moon \ct{bender1973,alley83} making the natural reference frame to a mutual study of geophysical and selenophysical processes.
Indeed, LLR technique is currently the most
effective way to study the interior of Moon and dynamics
of the Moon-Earth system. The most important contributions
from LLR include: detection of a molten lunar core \ct{williams2007}
and measurement of its influence on Moon's orientation \ct{moonfluidcore} and tidal dissipation \ct{moondissipation,mooncore}; detection of lunar free libration
along with the forced terms from Venus \ct{moonlibration} and the internal excitation mechanisms \ct{libration}; an accurate test of the strong principle of equivalence for
massive bodies  \ct{sep1976,sep1998} also known
as the Nordtvedt effect \cite[Section 8.1]{willbook};
and setting of a stringent limit on time variability of the universal
gravitational constant and (non)existence of long-range
fields besides the metric tensor \ct{2001LNP...562..317N}.
LLR analysis has also given access to more subtle tests of
relativity \ct{1991ApJ...382L.101M,1996PhRvD..54.5927M,test2004,2006PhRvD..74d5001B,2007PhRvL..99x1103B,2008PhRvD..78b4033S},
measurements of Moon's tidal acceleration \ct{1978Sci...166..977C,1994AnShO..15..129X,2002A&A...387..700C} and geodetic
precession of the lunar orbit \ct{1987PhRvL..58.1062B,1989AdSpR...9S..75D}, and has provided orders-of-magnitude improvements
in the accuracy of the lunar ephemeris \ct{1996IAUS..172...37N,2007HiA....14..472K,pmoe,2008AdSpR..42.1378K,2008AIPC..977..254S} and its three-dimensional
rotation \ct{1999A&A...343..624C,2003LPI....34.1161W}. On the geodesy front, LLR contributes
to the determination of Earth orientation parameters,
such as nutation, precession (including relativistic geodetic
precession), polar motion, UT1, and to the
long-term variation of these effects \ct{2008ASSL..349..457M,2008JGeod..82..133M}. LLR also contributes to
the realization of both the terrestrial and selenodesic reference
frames \ct{1996AnShO......169H,1999JGeod..73..125H}. The Satellite Laser Ranging (SLR) realization of a dynamically-defined
inertial reference frame \ct{1990IAUS..141..173S} in contrast to the kinematically-realized frame of VLBI \cite[Section 6]{2000afce.conf.....W}, offers new possibilities for mutual
cross-checking and confirmation \ct{2008JGeod..82..133M} especially after the International Laser Ranging Service (ILRS) was established in September 1998 to support programs in
geodetic, geophysical, and lunar research activities and to provide the International Earth Rotation Service (IERS)
with products important to the maintenance of an accurate International Terrestrial Reference Frame (ITRF) \cite{ilrs02}.

Over the years, LLR has benefited from a number of
improvements both in observing technology and data modeling,
which led to the current accuracy of post-fit residuals
of $\simeq$2 cm (see, for example, \ct{2002SPIE.4546..154M} and \cite[Section 11]{ilrs03}) Recently,
sub-centimeter precision in determining range distances
between a laser on Earth and a retro-reflector on Moon
has been achieved \ct{battat,murthyetal08}. As precision of LLR
measurements was gradually improving over years from a
few meters to few centimeters, enormous progress in understanding
evolutionary history of the Earth–Moon orbit and
the internal structure of both planets has been achieved.
With the precision approaching 1 millimeter and better, accumulation
of more accurate LLR data will lead to new, fascinating
discoveries in fundamental gravitational theory, geophysics, and physics of lunar interior \ct{2007IJMPD..16.2127M} whose unique interpretation will intimately rely upon our ability to develop a systematic theoretical approach to analyze the sub-centimeter LLR data \ct{2008AdSpR..42.1378K}.

\subsection{EIH Equations of Motion in N-body Problem}\l{eqmo}

Nowadays, the theory of the lunar motion
should incorporate not only the numerous Newtonian perturbations but has to deal with much
more subtle relativistic phenomena being currently incorporated to the ephemeris codes \ct{chatc,sta,epm,pmoe}. Theoretical approach, used for construction of the ephemerides, accepts that the post-Newtonian description of the planetary motions can be achieved with the Einstein-Infeld-Hoffmann (EIH) equations of motion of point-like masses \ct{eih}, which have been  independently derived by Petrova \ct{petrova} and Fock \cite[Section 6]{fockbook} for massive fluid balls as well as by Lorentz and Droste \ct{lord1,lord2,lord3} under assumptions that the bodies are spherical, homogeneous and consist of incompressible fluid. These relativistic equations are valid in the barycentric frame of the solar system with time coordinate $t$ and spatial coordinates $x^i\equiv{\bm x}$.

Due to the covariant nature of general theory of relativity the barycentric coordinates are not unique and are defined up to the space-time transformation \ct{vab,brum,sof89}
\br\l{gt1}
t&\mapsto& t-\frac{1}{c^2}\sum_{\sss{B}}\nu_{\sss{B}}\frac{G{\rm M}_{\sss{B}}}{R_{\sss{B}}}\le({\bm R}_{\sss{B}}\cdot{\bm v}_{\sss{B}}\r)\;,
\\\l{gt2}
{\bm x}&\mapsto&{\bm x}-\frac{1}{c^2}\sum_{\sss{B}}\lambda_{\sss{B}}\frac{G{\rm M}_{\sss{B}}}{R_{\sss{B}}}{\bm R}_{\sss{B}}\;,
\er
where summation goes over all the massive bodies of the solar system ($B=1,2,...,N$); $G$ is the universal gravitational constant; $c$ is the fundamental speed in the Minkowskian space-time; a dot between any spatial vectors, ${\bm a}\cdot{\bm b}$, denotes an Euclidean dot product of two vectors ${\bm a}$ and ${\bm b}$;  ${\rm M}_{\sss {B}}$ is mass of a body B; ${\bm x}_{\sss{B}}={\bm x}_{\sss{B}}(t)$ and ${\bm v}_{\sss{B}}={\bm v}_{\sss{B}}(t)$ are coordinates and velocity of the center of mass of the body B; ${\bm R}_{\sss{B}}={\bm x}-{\bm x}_{\sss{B}}$ is a relative distance from  a field point ${\bm x}$ to the body B;
$\nu_{\sss B}$ and $\lambda_{\sss{B}}$ are constant, but otherwise free parameters being responsible for a particular choice of the barycentric coordinates. We emphasize that these parameters can be chosen arbitrary for each body B of the solar system. Standard textbooks \cite{vab,brum,sof89} (see also \cite[section 4.2]{willbook}) assume that the coordinate parameters are equal for all bodies, that is $\nu_1=\nu_2=...\nu_N=\nu$ and $\lambda_1=\lambda_2=...\lambda_N=\lambda$. These simplifies the choice of coordinates and their transformations, and allows one to identify the coordinates used by different authors. For instance, $\nu=\lambda=0$ corresponds to harmonic or isotropic coordinates \ct{fockbook}, $\lambda=0$ and $\nu=1/2$ realizes the standard coordinates used in the book of Landau and Lifshitz \ct{lali} and in PPN formalism \ct{willbook}. The case of $\nu=0, \lambda=2$ corresponds to the Gullstrand-Painlev\'e coordinates \ct{painleve,gullstrand}. We prefer to have more freedom in transforming EIH equations of motion and do not equate the coordinate parameters for different massive bodies. Physically, it means that the space-time around each body is covered locally by its own coordinate grid, which matches smoothly with the other coordinate charts of the massive bodies in the buffer domain, where the different coordinate charts overlap.

If the bodies in N-body problem are numbered by indices B, C, D, etc., and the coordinate freedom is described by equations (\ref{gt1}), (\ref{gt2}), EIH equations have the following form (compare with \cite[equation 88]{vab})
\br\label{eih1}
{\rm M}_{\sss{B}}a^i_{\sss{B}}&=&
F^i_{\sss N} + \frac1{c^2} F^i_{\sss{EIH}}\;,
\er
where the Newtonian force
\br\label{eih2}
F^i_{\sss N} &=&-
\sum_{\sss{C\neq B}}\frac{G{\rm M}_{\sss{B}}{\rm M}_{\sss C}R^i_{\sss{BC}}}
{R^3_{\sss{BC}}}\;,
\er
the post-Newtonian perturbation
\br\l{eih3}
F^i_{\sss{EIH}}&=&-
\sum_{\sss{C\neq B}}\frac{G{\rm{M}}_{\sss{B}}{\rm{M}}_{\sss C}R^i_{\sss{BC}}}
{R^3_{\sss{BC}}}\Biggl\{(1+\lambda_{\sss{C}}) v^2_{\sss{B}}-(4+2\lambda_{\sss{C}})(\bm v_{\sss{B}}
\cdot\bm v_{\sss C})+(2+\lambda_{\sss{C}})v^2_{\sss C}
-\frac32\left(\frac{\bm R_{\sss{BC}}\cdot\bm v_{\sss C}}{R_{\sss{BC}}}\right)^2
\\\nonumber&&
-3\lambda_{\sss{C}}\left[\frac{\bm R_{\sss{BC}}\cdot({\bm v}_{\sss{B}}-{\bm v}_{\sss C})}{R_{\sss{BC}}}\right]^2
-(5-2\lambda_{\sss{B}})\frac{G{\rm M}_{\sss{B}}}{R_{\sss{BC}}}
-(4-2\lambda_{\sss{C}})\frac{G{\rm M}_{\sss C}}{R_{\sss{BC}}}
\\\nonumber&&
-\sum_{\sss{D\neq B,C}}G{\rm M}_{\sss D}\left[\frac{1}{R_{\sss{CD}}}
+\frac{4-2\lambda_{\sss{D}}}{R_{\sss{BD}}}
-\le(\frac{1+2\lambda_{\sss{C}}}{2R^3_{\sss{CD}}}-\frac{\lambda_{\sss{C}}}{R^3_{\sss{BD}}}+
\frac{3\lambda_{\sss{D}}}{R_{\sss{BD}}R^2_{\sss{BC}}}-\frac{3\lambda_{\sss{D}}}{R_{\sss{CD}}R^2_{\sss{BC}}}\r)(\bm R_{\sss{BC}}\cdot\bm R_{\sss{CD}})\right]
\Biggr\}\\\nonumber&&
-\sum_{\sss{C\neq B}}\Biggl\{
\frac{G{\rm M}_{\sss{B}}{\rm M}_{\sss C}(v^i_{\sss C}-v^i_{\sss{B}})}{R^3_{\sss{BC}}}
\biggl[(4-2\lambda_{\sss{C}})(\bm v_{\sss{B}}\cdot\bm R_{\sss{BC}}) -(3-2\lambda_{\sss{C}})
(\bm v_{\sss C}\cdot\bm R_{\sss{BC}})\biggr]
\\\nonumber&&
+\frac{G{\rm M}_{\sss{B}}{\rm M}_{\sss C}}{R_{\sss{BC}}}
\sum_{\sss{D\neq B,C}}G{\rm M}_{\sss D}R^i_{\sss{CD}}\le(\frac{7-2\lambda_{\sss{C}}}{2R^3_{\sss{CD}}}+\frac{\lambda_{\sss{C}}}{R^3_{\sss{BD}}}+
\frac{\lambda_{\sss{D}}}{R_{\sss{CD}}R^2_{\sss{BC}}}-\frac{\lambda_{\sss{D}}}{R_{\sss{BD}}R^2_{\sss{BC}}}\r)\Biggr\}\;,
\er
and ${\bm v}_{\sss B}={\bm v}_{\sss B}(t)$ is velocity of the body B, ${\bm a}_{\sss{B}}=\dot{\bm v}_{\sss{B}}(t)$ is its acceleration,  ${\bm R}_{\sss{BC}}={\bm x}_{\sss{B}}-{\bm x}_{\sss{C}}$, ${\bm R}_{\sss{CD}}={\bm x}_{\sss{C}}-{\bm x}_{\sss{D}}$ are relative distances between the coordinates of the bodies.

EIH equations (\ref{eih1})--(\ref{eih3}) differ from the equations of the PPN formalism \cite[equation 3]{1969ApJ...158...81E} employed in particular at JPL for actual calculation of the ephemerides of the major planets by the fact that the right side of equation (\ref{eih3}) has been resolved into radius-vectors and velocities of the massive bodies and does not contain second derivatives (accelerations). This elimination of the high-order time derivatives from a perturbed force is a standard practice in celestial mechanics for calculation of the perturbed motion.

Barycentric coordinates ${\bm x}_{\sss{B}}$ and velocities ${\bm v}_{\sss{B}}$ of the center of mass of body $B$ are adequate theoretical quantities for description of the world-line of the body with respect to the center of mass of the solar system. However, the barycentric coordinates are global coordinates covering the entire solar system. Therefore, they have little help for efficient physical decoupling of the post-Newtonian effects existing in the orbital and rotational motions of a planet and for the description of motion of planetary satellites around the planet. The problem stems from the covariant nature of EIH equations, which originates from the fundamental structure of space-time manifold and the gauge freedom of the general relativity theory.

This freedom is already seen in the post-Newtonian EIH equations (\ref{eih3}) as it explicitly depends on the choice of spatial coordinates through parameters $\lambda_{\sss C}, \lambda_{\sss D}$. At the same time the EIH force does not depend on parameters $\nu_{\sss C}$, which means that transformation (\ref{gt1}) of the barycentric coordinate time does not affect the post-Newtonian equations of motion of the solar system bodies. Each term, depending explicitly on $\lambda_{\sss C}$ and $\lambda_{\sss D}$ in equation (\ref{eih3}), has no direct physical meaning as it can be eliminated after making a specific choice of these parameters. In many works on experimental gravity and applied relativity researches fix parameters $\lambda_{\sss C}=\lambda_{\sss D}=0$, which corresponds to working in harmonic coordinates. Harmonic coordinates simplify EIH equations to large extent but one has to keep in mind that they have no physical privilege anyway, and that a separate term or a limited number of terms from EIH equations of motion can not be measured \ct{brum}. This is because the coordinate description of motion of the bodies does not exist independently of observable quantities and must be connected to them via equations of light propagation.

EIH equations of motion can be recast to another form proposed by Brumberg \ct{brum}. It is based on a simple property of decomposition of a vector of relative distance between any two bodies in an algebraic sum of two vectors connecting the two bodies with any other body of the N-body system. Let us take as an example a 4-body problem. Radius-vectors connecting each pair of the bodies are: ${\bm R}_{12}={\bm x}_1-{\bm x}_2$, ${\bm R}_{13}={\bm x}_1-{\bm x}_3$, ${\bm R}_{14}={\bm x}_1-{\bm x}_4$, ${\bm R}_{23}={\bm x}_2-{\bm x}_3$, ${\bm R}_{24}={\bm x}_2-{\bm x}_4$, ${\bm R}_{34}={\bm x}_3-{\bm x}_4$. However, only three of the six vectors are algebraically independent. Indeed, if one takes the first three vectors as independent the others can be expressed in terms of them: ${\bm R}_{23}={\bm R}_{13}-{\bm R}_{12}$, ${\bm R}_{24}={\bm R}_{14}-{\bm R}_{12}$, ${\bm R}_{34}={\bm R}_{14}-{\bm R}_{13}$. Analogous reasoning is valid for any number of the bodies in the N-body problem. This property allows us to reshuffle terms in the original EIH equation and to recast it to the following form \ct{brum}
\br\label{eh1}
a^i_{\sss{B}}&=&-
\sum_{\sss{C\neq B}}\frac{G{\rm M}_{\sss C}R^i_{\sss{BC}}}{R^3_{\sss{BC}}}+\frac{G}{c^2}\sum_{\sss{C\neq B}}{\rm M}_{\sss C}\le({\cal A}_{\sss BC}R^i_{\sss{BC}}+{\cal B}_{\sss BC}V^i_{\sss{BC}}\r)\;,
\er
where $V^i_{\sss{BC}}=\dot R^i_{\sss{BC}}=v^i_{\sss{B}}-v^i_{\sss{C}}$ is the relative velocity between the bodies, the dot over function denotes a time derivative, and the coefficients of the post-Newtonian acceleration are
\br\l{eh2}
{\cal A}_{\sss BC}&=&-
\frac{1}{R^3_{\sss{BC}}}\Biggl\{(1+\lambda_{\sss{C}}) v^2_{\sss{B}}-(4+2\lambda_{\sss{C}})(\bm v_{\sss{B}}
\cdot\bm v_{\sss C})+(2+\lambda_{\sss{C}})v^2_{\sss C}
\\\nonumber&&-\frac32\left(\frac{\bm R_{\sss{BC}}\cdot\bm v_{\sss C}}{R_{\sss{BC}}}\right)^2
-3\lambda_{\sss{C}}\left(\frac{\bm R_{\sss{BC}}\cdot{\bm V}_{\sss{BC}}}{R_{\sss{BC}}}\right)^2
-(5-2\lambda_{\sss{B}})\frac{G{\rm M}_{\sss{B}}}{R_{\sss{BC}}}
-(4-2\lambda_{\sss{C}})\frac{G{\rm M}_{\sss C}}{R_{\sss{BC}}}\Biggr\}
\\\nonumber&&
+\sum_{\sss{D\neq B,C}}\frac{G{\rm M}_{\sss D}}{R^3_{\sss{BC}}}\left[\frac{1-2\lambda_{\sss D}}{R_{\sss{CD}}}
+\frac{4-\lambda_{\sss{D}}}{R_{\sss{BD}}}-\le(\frac{1+2\lambda_{\sss{C}}}{2R^3_{\sss{CD}}}-\frac{\lambda_{\sss{C}}}{R^3_{\sss{BD}}}+
\frac{3\lambda_{\sss{D}}}{R_{\sss{BD}}R^2_{\sss{BC}}}-\frac{3\lambda_{\sss{D}}}{R_{\sss{CD}}R^2_{\sss{BC}}}\r)(\bm R_{\sss{BC}}\cdot\bm R_{\sss{BD}})\right]
\\\nonumber&&
+\sum_{\sss{D\neq B,C}}G{\rm M}_{\sss D}\left[\frac{4}{R_{\sss{BC}}R^3_{\sss{CD}}}+\frac{\lambda_{\sss C}}{R_{\sss{BC}}R^3_{\sss{BD}}}-\frac{\lambda_{\sss C}}{R_{\sss{CD}}R^3_{\sss{BD}}}-\frac{7-2\lambda_{\sss D}}{2R_{\sss{BD}}R^3_{\sss{CD}}}    \right]\;,
\\\nonumber\\\nonumber\l{eh3}
{\cal B}_{\sss BC}&=&
\frac{1}{R^3_{\sss{BC}}}
\biggl[(4-2\lambda_{\sss{C}})(\bm v_{\sss{B}}\cdot\bm R_{\sss{BC}}) -(3-2\lambda_{\sss{C}})
(\bm v_{\sss C}\cdot\bm R_{\sss{BC}})\biggr]\;.
\er
Equations (\ref{eh1})--(\ref{eh3}) have been derived by Brumberg \cite[pages 176-177]{brum}.

\subsection{Gravitoelectric and Gravitomagnetic Forces}\l{xik}

Brumberg's form of EIH equations of motion can be further modified to separate the, so-called, gravitoelectric and gravitomagnetic forces in N-body problem \ct{1988IJTP...27.1395N}. Straightforward re-arrangement of the terms depending on velocities reveal that equations (\ref{eh1})--(\ref{eh3}) can be represented in the form being similar to the Lorentz force in electrodynamics
\br\label{eh4}
a^i_{\sss{B}}&=&\sum_{\sss{C\neq B}}\left[E^i_{\sss{BC}}+\frac{4-2\lambda_{\sss C}}{c}\left({\bm v}_{\sss B}\times{\bm H}_{\sss{BC}}\right)^i-\frac{3-2\lambda_{\sss C}}{c}\left({\bm v}_{\sss C}\times{\bm H}_{\sss{BC}}\right)^i\right]
\er
where $E^i_{\sss{BC}}$ is called the gravitoelectric force, and the terms associated with the cross products $\left({\bm v}_{\sss B}\times{\bm H}_{\sss{BC}}\right)^i$  and $\left({\bm v}_{\sss C}\times{\bm H}_{\sss{BC}}\right)^i$ are referred to as the gravitomagnetic force \ct{1988IJTP...27.1395N}.

The gravitoelectric force is given by
\br\l{eh5}
E^i_{\sss{BC}}&=&\left(-\frac{G{\rm M}_{\sss C}}{R^3_{\sss{BC}}}+\frac{G{\rm M}_{\sss C}}{c^2}{\cal E}_{\sss{BC}}\right)R^i_{\sss{BC}}\;,
\er
where the first term is the Newtonian force of gravity and the post-Newtonian correction
\br\l{eh6}
{\cal E}_{\sss BC}&=&-
\frac{1}{R^3_{\sss{BC}}}\Biggl\{3(-1+\lambda_{\sss{C}}) v^2_{\sss{B}}+3(1-2\lambda_{\sss{C}})(\bm v_{\sss{B}}
\cdot\bm v_{\sss C})-(1-3\lambda_{\sss{C}})v^2_{\sss C}
\\\nonumber&&-\frac32\left(\frac{\bm R_{\sss{BC}}\cdot\bm v_{\sss C}}{R_{\sss{BC}}}\right)^2
-3\lambda_{\sss{C}}\left(\frac{\bm R_{\sss{BC}}\cdot{\bm V}_{\sss{BC}}}{R_{\sss{BC}}}\right)^2
-(5-2\lambda_{\sss{B}})\frac{G{\rm M}_{\sss{B}}}{R_{\sss{BC}}}
-(4-2\lambda_{\sss{C}})\frac{G{\rm M}_{\sss C}}{R_{\sss{BC}}}\Biggr\}
\\\nonumber&&
+\sum_{\sss{D\neq B,C}}\frac{G{\rm M}_{\sss D}}{R^3_{\sss{BC}}}\left[\frac{1-2\lambda_{\sss D}}{R_{\sss{CD}}}
+\frac{4-\lambda_{\sss{D}}}{R_{\sss{BD}}}-\le(\frac{1+2\lambda_{\sss{C}}}{2R^3_{\sss{CD}}}-\frac{\lambda_{\sss{C}}}{R^3_{\sss{BD}}}+
\frac{3\lambda_{\sss{D}}}{R_{\sss{BD}}R^2_{\sss{BC}}}-\frac{3\lambda_{\sss{D}}}{R_{\sss{CD}}R^2_{\sss{BC}}}\r)(\bm R_{\sss{BC}}\cdot\bm R_{\sss{BD}})\right]
\\\nonumber&&
+\sum_{\sss{D\neq B,C}}G{\rm M}_{\sss D}\left[\frac{4}{R_{\sss{BC}}R^3_{\sss{CD}}}+\frac{\lambda_{\sss C}}{R_{\sss{BC}}R^3_{\sss{BD}}}-\frac{\lambda_{\sss C}}{R_{\sss{CD}}R^3_{\sss{BD}}}-\frac{7-2\lambda_{\sss D}}{2R_{\sss{BD}}R^3_{\sss{CD}}}    \right]\;,
\er
The gravitomagnetic force is given by equation
\br\l{eh7}
H^i_{\sss BC}&=&-\frac1{c}\left({\bm V}_{\sss{BC}}\times{\bm E}_{\sss{BC}}\right)^i=
\frac{G{\rm M}_{\sss C}}{c}\frac{\left({\bm V}_{\sss{BC}}\times{\bm R}_{\sss{BC}}\right)^i}{R^3_{\sss{BC}}}\;,
\er
where the dot means a time derivative. As one can see, the gravitomagentic force is proportional to the Newtonian force multiplied by the factor of $v/c$, where $v$ is the relative velocity between two gravitating bodies. Equation (\ref{eh7}) can be also obtained by making use of a linearized Lorentz transformation from the static to a moving frame of the body \ct{1988IJTP...27.1395N,2007GReGr..39.1583K}. Similar arguments work in electrodynamics for physical explanation of the origin of magnetic field of a uniformly moving charge \cite[\s 24]{lali}.

Recently, there was a lot of discussions about whether LLR can measure the gravitomagnetic field $H^i_{\sss BC}$ \ct{2007PhRvL..98g1102M,2007PhRvL..98v9002M,test2004,k07,2008PhRvD..78b4033S}. The answer to this question is subtle and requires more profound theoretical consideration involving the process of propagation of the laser pulses in a curved space-time of the Earth-Moon system. We are hoping to discuss this topic in an other publication.  Nevertheless, what is evident already now is that equation (\ref{eh4}) demonstrates a strong correlation of the gravitomagnetic force of each body with the choice of coordinates. For this reason, by changing the coordinate parameter $\lambda_{\sss C}$ one can eliminate either the term $\left({\bm v}_{\sss B}\times{\bm H}_{\sss{BC}}\right)^i$  or $\left({\bm v}_{\sss C}\times{\bm H}_{\sss{BC}}\right)^i$ from EIH equations of motion (\ref{eh4}). It shows that the strength of the factual gravitomagnetic force is coordinate-dependent, and, hence, a great care should be taken in order to properly interpret the LLR "measurement" of such gravitomagnetic terms in consistency with the covariant nature of the general theory of relativity and the theory of astronomical measurements in curved space-time outlined in papers \ct{1981rcse.conf..283B,1962rdgr.book..441S}, in the textbooks by Synge \cite{syngebook}, by Infeld and Plebansky \ct{inpl}, and by Brumberg \cite{vab}.

\subsection{The Principle of Equivalence in the Earth-Moon System}\l{peq}

Let us discuss in this section the case of the Earth-Moon system moving in the gravitational field of Sun neglecting other planets of the solar system. This is a three-body problem, where two bodies - Earth (index E) and Moon (index M) - form a bounded binary system perturbed by the tidal gravitational field of a third body - Sun (index S). Brumberg \ct{vab,brm} extended the Hill-Brown theory of motion of Moon to the post-Newtonian approximation by making use of an Euclidean translation of the barycentric coordinates of Moon to the geocenter (see also Baierlein \ct{1967PhRv..162.1275B})
\be\label{trc}
{\bm r}={\bm x}_{\sss{M}}-{\bm x}_{\sss{E}}\;,
\en
and introducing a vector of the Newtonian center of mass of the Earth-Moon system, ${\bm R}$, in such a way that the distance Earth-Sun -- ${\bm r}_{\sss{ES}}$, and that Moon-Sun -- ${\bm r}_{\sss{MS}}$, are given by the Newtonian-like equations
\be\label{mse}
{\bm r}_{\sss{ES}}={\bm R}-\frac{{\rm M}_{\sss{M}}{\bm r}}{{\rm M}_{\sss{E}}+{\rm M}_{\sss{M}}}\;,\qquad\qquad {\bm r}_{\sss{MS}}={\bm R}+\frac{{\rm M}_{\sss{E}}{\bm r}}{{\rm M}_{\sss{E}}+{\rm M}_{\sss{M}}}\;.
\en
In these new variables EIH equations (\ref{eh4}) for the geocentric motion of Moon and the center-of mass of the Earth-Moon system, assumes the form
\br\l{meih1}
\ddot{\bm r}&=&-\frac{G\le({\rm M}_{\sss{E}}+{\rm M}_{\sss{M}}\r)}{r^3}{\bm r}+G{\rm M}_{\sss{S}}\le(\frac{{\bm r}_{\sss{ES}}}{r^3_{\sss{ES}}}-\frac{{\bm r}_{\sss{MS}}}{{ r}^3_{\sss{MS}}}\r)+\frac{G{\rm M}_{\sss{S}}}{c^2}\le(A{\bm R}+B{\bm V}+C{\bm r}+D{\bm v}\r)\;,
\\\l{meih2}
\ddot{\bm R}&=&-\frac{G\le({\rm M}_{\sss{S}}+{\rm M}_{\sss{E}}+{\rm M}_{\sss{M}}\r)}{{\rm M}_{\sss{E}}+{\rm M}_{\sss{M}}}\le(\frac{G{\rm M}_{\sss{E}}}{r^3_{\sss{ES}}}{\bm r}_{\sss{ES}}+\frac{G{\rm M}_{\sss{M}}}{r^3_{\sss{MS}}}{\bm r}_{\sss{MS}}\r)+\frac{G{\rm M}_{\sss{S}}}{c^2}\le(A'{\bm R}+B'{\bm V}+C'{\bm r}+D'{\bm v}\r)\;,
\er
where functions $A$, $A'$, $B$, $B'$, $C$, $C'$, $D$, $D'$ depend on relative coordinates ${\bm R}$, ${\bm r}$ of the bodies and their velocities ${\bm V}=\dot{\bm R}$, ${\bm v}=\dot{\bm r}$. Exact analytic form of these functions is notoriously sophisticated and can be found, for example, in the book of Brumberg \cite[Section 5]{brum}. Let us neglect post-Newtonian corrections to the gravitational field of the planets, Earth and Moon, and leave only the Schwarzschild gravitational field of Sun. Then, the main terms in these functions read
\br\l{meih3}
A&=&8(\lambda_{\sss S}-2)\frac{G{\rm M}_{\sss{S}}}{R^6}\le({\bm R}\cdot{\bm r}\r)+3(1+\lambda_{\sss S})\frac{{\bm V}^2}{R^5}\le({\bm R}\cdot{\bm r}\r)-2(1+\lambda_{\sss S})\frac{{\bm V}\cdot{\bm v}}{R^3}\\\nonumber&&+3\lambda_{\sss S}\frac{{\bm R}\cdot{\bm V}}{R^5}\le[2{\bm R}\cdot{\bm v}+2{\bm r}\cdot{\bm V}-\frac{5}{R^2}({\bm R}\cdot{\bm V})({\bm R}\cdot{\bm r})\r]\;,
\\\label{meih4}
B&=&-2\frac{\lambda_{\sss S}-2}{R^3}\le[{\bm R}\cdot{\bm v}+{\bm r}\cdot{\bm V}-\frac{3}{R^2}({\bm R}\cdot{\bm V})({\bm R}\cdot{\bm r})\r]\;,
\\\label{meih5}
C&=&-2(\lambda_{\sss S}-2)\frac{G{\rm M}_{\sss{S}}}{R^4}-(1+\lambda_{\sss S})\frac{{\bm V}^2}{R^3}+\frac{3\lambda_{\sss S}}{R^5}({\bm R}\cdot{\bm V})^2\;,
\\\l{meih6}
D&=&-2(\lambda_{\sss S}-2)\frac{{\bm R}\cdot{\bm V}}{R^3}\;,
\\\l{meih7}
A'&=&C\;,
\\\l{meih8}
B'&=&D\;,
\\\l{meih9}
C'&=&0
\\\l{meih10}
D'&=&0\;,
\er
where parameter $\lambda_{\sss S}$ describes the gauge freedom in choosing coordinates of the Schwarzschild's problem for Sun \ct{vab,brum}.

The reader should notice that the equations (\ref{meih1})-(\ref{meih10}) are still EIH equations of motion in the solar barycentric coordinates expressed in terms of the relative distances between the bodies. Newtonian part of equation (\ref{meih1}) of the orbital motion of Moon around Earth couples with vector ${\bm R}$  of the Earth-Moon center of mass only through the tidal terms. This can be seen by expanding the second term in right hand side of equation (\ref{meih1}) in powers of $r/R$:
\be\l{meih11}
G{\rm M}_{\sss{S}}\le(\frac{{\bm r}_{\sss{ES}}}{r^3_{\sss{ES}}}-\frac{{\bm r}_{\sss{MS}}}{{ r}^3_{\sss{MS}}}\r)=-\frac{G{\rm M}_{\sss{S}}}{R^3}\le[{\bm r}-3{\bm N}({\bm r}\cdot{\bm N})\r]+...\;,
\en
where dots denote small terms of the higher order of magnitude. Comparing with the Newtonian term one can confirm that the Newtonian tidal perturbation (\ref{meih11}) is smaller than the Newtonian term by a factor of $\simeq ({\rm M}_{\sss{S}}/{\rm M}_{\sss{E}})(r/R)^3$.

More important is to note that the Newtonian tidal perturbation (\ref{meih1}) of the lunar orbit is a coupling of the second (quadrupole) derivative of the Newtonian gravitational potential of Sun $U_{\sss{S}}=M_{\sss{S}}/R$ with vector ${\bm r}$ of the lunar orbit
\be\l{meih12}
G{\rm M}_{\sss{S}}\le(\frac{r^i_{\sss{ES}}}{r^3_{\sss{ES}}}-\frac{r^i_{\sss{MS}}}{{ r}^3_{\sss{MS}}}\r)=\frac{\partial^2U_{\sss{S}}}{\partial x^i\partial x^j}r^j+...\;,
\en
where here and everywhere else the repeated (dummy) indices mean the Einstein summation rule, for example, $A_\a B^\a\equiv A_0B^0+A_1B^1+A_2B^2+A_3B^3$, $A_iB^i\equiv A_1B^1+A_2B^2+A_3B^3$, and so on.
Equation (\ref{meih12}) elucidates that the first derivatives of the solar potential does not perturb the lunar orbit in the Newtonian approximation. The first derivatives of the potential are associated with the affine connection (the Christoffel symbols) of the space-time manifold in a metric theory of gravity \cite{lali,mtw,lali}. Hence, their disappearance from the Newtonian equations of the relative motion of Moon around Earth is a consequence of the principle of equivalence. This principle states that the Christoffel symbols of the background gravitational field can be eliminated on the world line of a particle falling freely in this field. The Earth-Moon system can be considered in a first approximation as such a particle, composed of Earth and Moon and located at the Earth-Moon barycenter, which falls in the field of Sun in accordance with equation (\ref{meih2}).

Structure of the post-Newtonian force in equation (\ref{meih1}) seems to violate the principle of equivalence because it contains terms, which are explicitly proportional to the Christoffel symbols, which are the first derivatives of the solar gravitational potential $U_{\sss{S}}$, coupled with velocities of Sun and Moon. However, the principle of equivalence is exact, and must be valid not only in the Newtonian theory but in any approximation beyond it. The contradiction can be resolved if one investigates the residual gauge freedom of the post-Newtonian terms in equations of motion (\ref{meih1})-(\ref{meih10}) more carefully.

\subsection{The Residual Gauge Freedom}\l{rgfr}

The primary gauge freedom of EIH equations of motion is associated with the transformations (\ref{gt1})-(\ref{gt2}) of the barycentric coordinates of the solar system, which are parameterized by parameters $\nu$ and $\lambda$. We have noticed that the post-Newtonian perturbations in the lunar equations of motion are made up of the Christoffel symbols, which admit a certain freedom of coordinate transformations. This freedom remains even after fixing the coordinate parameters $\nu$ and $\lambda$ in equations {\ref{meih1}), (\ref{meih2}).  It is associated with the fact that the Earth-Moon system moves in tidal gravitational field of Sun and other planets, which makes the local background space-time for this system not asymptotically-flat. The residual freedom remains in making transformations of the local coordinates attached to the Earth-Moon system. It induces the gauge transformation of the metric tensor and the Christoffel symbols and changes the structure of the post-Newtonian terms in EIH equations of motion of the Earth-Moon system. The residual gauge freedom is explicitly revealed in the linear dependence of the post-Newtonian force in equation (\ref{meih1}) on the orbital velocity ${\bm V}$ of the Earth-Moon system with respect to Sun. This dependence seems to point out to violation of the principle of relativity according to which an observer can not determine one's velocity of motion with respect to an external coordinate frame by making use of local measurements that are not sensitive to the curvature of space-time (the second derivatives of the solar gravitational potential). LLR is a local measurement technique, which does not observe Sun directly, and, hence, should not be able to determine velocity of the Earth-Moon system with respect to it as it appears in equations (\ref{meih1}) because those velocity-dependent terms are not gauge-invariant and have no absolute physical meaning.

Thus, we face the problem of investigation of the residual gauge freedom of the lunar equations of motion, which goes beyond the choice of the barycentric coordinates by fixing a specific value of the gauge parameter $\lambda$ in equations (\ref{meih3})-(\ref{meih8}). This freedom is actually associated with the choice of the local coordinates of the Earth-Moon barycentric frame as well as the geocentric and selenocentric reference frames. Proper choice of the local coordinates removes all non-physical degrees of freedom from the metric tensor and eliminates spurious (non-measurable) terms from the post-Newtonian forces in the relative equations of motion of Moon. If one ignores the residual gauge freedom and operates, for example, with the Newtonian definitions (\ref{trc})-(\ref{mse}) of the relative coordinates between the bodies, the gauge-dependent terms will infiltrate the equations of motion causing possible misinterpretation of LLR observations. This problem is well-known in
cosmology where the theory of cosmological perturbations is designed essentially in terms of the gauge-independent variables so that observations of various cosmological effects are not corrupted by the spurious, coordinate-dependent signals \ct{1992PhR...215..203M}. Similarly to cosmology, the residual gauge degrees of freedom existing in the relativistic three-body problem, can lead to misinterpretation of various aspects of gravitational physics of the Earth-Moon system \ct{2008arXiv0809.3392K,k07}, thus, degrading the value of extremely accurate LLR measurements for testing fundamental physics of space-time and deeper exploration of the lunar interior \ct{2008AdSpR..42.1378K}.

The residual gauge freedom of the three body problem (Sun-Earth-test particle) was studied by Brumberg and Kopeikin \ct{bk-nc}, Klioner and Voinov \ct{klv}, and Damour, Soffel and Xu \ct{dsx4}. They found that the post-Newtonian equations of motion of a test body (artificial satellite) can be significantly simplified by making use of a four-dimensional space-time transformation from the solar barycentric coordinates $x^\alpha=(ct, {\bm x})$, to the geocentric coordinates $X^\alpha=(cT, {\bm X})$
\br\l{gct1}
T&=&t+\frac1{c^2}A(t,{\bm r}_{\sss E})+\frac1{c^4}B(t,{\bm r}_{\sss E})+O\le(\frac1{c^5}\r)\;,
\\\l{gct2}
X^i&=&x^i-x^i_{\sss E}(t)+\frac1{c^2}C^i(t,{\bm r}_{\sss E})+O\le(\frac1{c^4}\r)\;,
\er
where the gauge functions $A(t,{\bm x})$, $B(t,{\bm x})$, $C^i(t,{\bm x})$ are polynomials of the geocentric distance ${\bm r}_{\sss E}={\bm x}-{\bm x}_{\sss E}(t)$ of the field point ${\bm x}$ from Earth's geocenter, which barycentric coordinates are ${\bm x}_{\sss E}(t)$. Coefficients of these polynomials are functions of the barycentric time $t$ that are determined by solving a system of ordinary differential equations, which follow from the gravity field equations and the tensor law of transformation of the metric tensor from one coordinate chart to another \ct{1988CeMec..44...87K}.

Contrary to the test particle, the Moon is a massive body, which makes the exploration of the residual gauge freedom of the lunar motion more involved. This requires introduction of one global (SSB) frame and three local reference frames associated with the Earth-Moon barycenter, the geocenter, and the center of mass of Moon (selenocenter). It should be clearly understood that any coordinate system can be used for processing and interpretation of LLR data since any viable theory of gravity obeys the Einstein principle of relativity, according to which there is no preferred frame of reference \ct{lali,fockbook,mtw}. For this reason,
we do not admit a privileged coordinate frame in rendering analysis of the
LLR data irrespectively of its accuracy. It means that our approach is insensitive to PPN parameters $\alpha_1$, $\alpha_2$, $\alpha_3$, etc., which describe the preferred frame and preferred location effects in gravitational physics. Accepting the Einstein principle of relativity leads to discarding any theory of gravity based on a privileged frame (aether) \ct{2006dclw.conf..163E} or admitting a violation of the Lorentz invariance \ct{2008APS..DMP.I3001K}. The class of scalar-tensor theories of gravity, which have two PPN parameters - $\beta$ and $\gamma$ \cite{willbook,1992CQGra...9.2093D}, is in agreement with the principle of relativity and it will be used in this paper.

The principle of relativity also assumes that a randomly chosen, separate term in the post-Newtonian equations of motion of massive bodies and/or light can not be physically interpreted as straightforward as in the Newtonian physics. The reason is that the post-Newtonian transformations (\ref{gt1})-(\ref{gt2}) and (\ref{gct1})-(\ref{gct2}) of the barycentric and local coordinates, change the form of the equations of motion so that they are not form-invariant. Therefore, only those post-Newtonian effects, which do not depend on the frame transformations can have direct physical interpretation. For example, the gauge parameters $\nu$ and $\lambda$ entering transformations (\ref{gt1})-(\ref{gt2}) and EIH equations (\ref{meih1})-(\ref{meih10}) can not be determined from LLR data irrespectively of their accuracy because these parameters define the barycentric coordinates and can be fixed arbitrary by observer without any relation to observations. This point of view has been argued by some researchers who believe that separate terms in the barycentric EIH equations of motion of Moon do have direct physical meaning, at least those of them, which are associated with gravitomagnetism \ct{1988IJTP...27.1395N,2008PhRvD..78b4033S,2001LNP...562..317N}. These gravitomagnetic terms can be easily identified in quations (\ref{meih1})--(\ref{meih2}) as being proportional to the velocity of motion of Moon, ${\bm v}$ and that of the Earth-Moon barycenter ${\bm V}$. Such orbital velocity-dependent terms in equations of motion of gravitating bodies are associated with the {\it extrinsic} gravitomagnetic field as opposed to the {\it intrinsic} gravitomagnetism caused by rotational currents of matter \ct{2006IJMPD..15..305K}. It is remarkable that all the orbital velocity-dependent terms can be eliminated from the orbital equations of motion (\ref{meih1})--(\ref{meih2}) by choosing the Gullstrand-Painlev\'e (GP) coordinates \ct{painleve,gullstrand} with $\lambda_{\sss S}=2$, which makes the equation coefficients $B=D=B'=0$. It means that the {\it extrinsic} gravitomagnetic force, which is directly caused by the orbital motions of Earth and Moon, can not be measured by LLR technique \ct{k07},  -- only the tidal {\it extrinsic} gravitomagnetic field of Sun can be measured \cite{2008arXiv0809.3219C}. Papers \ct{2008arXiv0809.3219C,2008arXiv0809.4014I} discuss whether the {\it intrinsic} gravitomagnetism can be measured with LLR or not.

\subsection{Towards a New Lunar Ephemeris}\l{reh}

Existing computer-based theories of the lunar ephemeris \ct{chatc,sta,epm,pmoe} consist of three major blocks:
\begin{enumerate}
\item[(1)] the barycentric EIH equations (\ref{eih1})-(\ref{eih3}) of orbital motion of Moon, Earth, Sun, and other planets of the solar system with the gauge parameters $\nu=1/2$,  $\lambda=0$ - the standard PPN coordinates;
\item[(2)] the Newtonian rotational equations of motion of Moon and Earth;
\item[(3)] the barycentric post-Newtonian equations of motion for light rays propagating from laser to CCR on Moon and back in standard coordinates with the gauge parameters $\nu=1$,   $\lambda=0$.
\end{enumerate}
This approach is straightforward but it does not control gauge-dependent terms in EIH equations of motion associated with the choice of the gauge-fixing parameters $\nu$ and $\lambda$. Particular disadvantage of the barycentric approach in application to the lunar ephemerides is that it mixes up the post-Newtonian effects associated with the orbital motion of the Earth-Moon barycenter around Sun with those, which are attributed exclusively to the relative motion of Moon around Earth. This difficulty is also accredited to the gauge freedom of the equations of motion in three-body problem and was pointed out in papers \ct{bk-nc,dsx4,2000A&A...363..335T}. Unambiguous decoupling of the orbital motion of the Earth-Moon barycenter from the relative motion of Moon around Earth with apparent identification of the gauge-dependent degrees of freedom in the metric tensor and equations of motion is highly desirable in order to make the theory more sensible and to clean up the LLR data processing software from the fictitious coordinate-dependent perturbations, which do not carry out any physically-relevant information and may accumulate errors in numerical ephemerides of Moon.

This goal can be rationally achieved if the post-Newtonian theory of the lunar motion is consistently extended to account for mathematical properties offered by the scalar-tensor theory of gravity and the differential structure of the space-time manifold. Altogether it leads us to the idea that besides the global barycentric coordinates of the solar system one has to introduce three other local reference frames. The origin of these frames should be fixed at the Earth-Moon system barycenter, Earth's center of mass (geocenter), and Moon's center of mass (selenocenter). We distinguish the Earth-Moon barycenter from the geocenter because Moon is not a test particle, thus, making the Earth-Moon barycenter displaced from the geocenter along the line connecting Earth and Moon and located approximately 1710 km below the surface of Earth. Mathematical construction of each frame is reduced to finding a metric tensor by means of solution of the gravity field equations with an appropriate boundary condition \ct{fock1,fockbook}. The gauge freedom of the three-body problem is explored by means of matching the set of the metric tensors defined in each reference frame in the overlapping domains of their applicability associated with the specific choice of boundary conditions imposed in each frame on the metric tensor. This matching procedure is an integral part of the equations defining the local differential structure of the manifold \ct{eisen,dfn}, which proceeds from a requirement that the overlapping space-time domains covered by the local reference frames, are diffeomorphic.

The primary objective of the multi-frame post-Newtonian theory of the lunar ephemeris is the development of a new set of analytic equations to revamp the LLR data processing software in order to suppress the spurious gauge-dependent solutions, which may overwhelm the existing barycentric code at the millimeter accuracy of LLR measurements, thus, plunging errors in the interpretation of selenophysics, geophysics and fundamental gravitational physics. Careful mathematical construction of the local frames with the post-Newtonian accuracy will allow us to pin down and correctly interpret all physical effects having classical (lunar interior, Earth geophysics, tides, asteroids, etc.) and relativistic nature. The gauge freedom in the three-body problem (Earth-Moon-Sun) should be carefully examined by making use of a scalar-tensor theory of gravity and the principles of the analytic theory of relativistic reference frames in the solar system \ct{1988CeMec..44...87K,bk89,dsx1} that was adopted by the XXIV-th General
Assembly of the International Astronomical Union \ct{iau2000,2007AIPC..886..268K} as a standard for data processing of high-precision astronomical observations.

The advanced post-Newtonian dynamics of the Sun-Earth-Moon system must include the following structural elements:
\begin{enumerate}
\item construction of a set of astronomical reference frames decoupling orbital dynamics of the Earth-Moon system from the rotational motion of Earth and Moon with the full account of the post-Newtonian corrections and elimination of the gauge modes;
\item relativistic definition of the integral parameters like mass, the center of mass, the multipole moments of the gravitating bodies;
\item derivation of the relativistic equations of motion of the center-of-mass of the Earth-Moon system with respect to the barycentric reference frame of the solar system;
\item derivation of the relativistic equations of motion of Earth and Moon with respect to the reference frame of the Earth-Moon system;
\item derivation of the relativistic equations of motion of CCR on the Moon (or a lunar orbiter that is deployed with CCR) with respect to the selenocentric reference frame;
\item derivation of the relativistic equations of motion of a laser with respect to the geocentric reference frame.
\end{enumerate}
These equations must be incorporated to LLR data processing software operating with observable quantities, which are proper times of the round trip of the laser pulses between the laser on Earth and CCR on Moon. The computational advantage of the new approach to the lunar ephemeris is that it separates clearly physical effects from the choice of coordinates. This allows us to get robust measurement of true physical parameters of the LLR model and give them direct physical interpretation. The new approach is particularly useful for comparing different models of the lunar interior and for making the fundamental test of general theory of relativity.

There is a practical consideration when we do LLR computer model improvements - a change in the LLR code must have some advantages either for computation, or accuracy, or a more complete and detailed model including adding solution parameters.
Spacecraft missions use the output of the orbit integrator and that imposes another practical matter. The output ephemeris must be consistent with the conventions used in the spacecraft orbit determination program \ct{odp}. That means that the new LLR code must be compatible with the solar system barycentric frame, scale and time.

One should also make a distinction between analytical models, which play an important role in understanding of fundamental gravitational physics, and models for numerical computation/prediction of astronomical events and phenomena. For the numerical computations it basically does not matter if there are gauge-dependent terms that cancel so long as the computations are internally consistent. For understanding what is going on analytically and how gravitational physics actually works, it does matter what terms cancel and what does not. The analytic LLR model, which we are going to work out, pursues mostly the goals of the fundamental physics. It will refine our understanding of the test of general relativity in the Earth-Moon system and the physics of the lunar interior that are the primary concerns of the scientific exploration.

\subsection{Main Objectives of The Present Paper}

This paper deals with the precise analytic construction of the relativistic reference frames in the Earth-Moon system moving in the field of Sun and other planets of the solar system. We shall also identify the post-Newtonian gauge modes and eliminate them from the solutions of the gravity field equations. Although our final goal is to develop a practical LLR code having accuracy of one millimeter, the overall development will be as close to the covariant spirit of modern physical theories as possible.

First of all, we discuss the scalar-tensor theory of gravity in \s\ref{stt}. We formulate the field equations for the metric tensor and the scalar field and describe the model of matter used in our analytic calculations. Powerful mathematical approach developed for calculation of motion of compact astrophysical objects, like neutron stars and/or black holes, employs the model of matter in the form of the "multipole moments", which are the integrals over the volume of the bodies from the unspecified "effective" tensor of energy-momentum \ct{2002LRR.....5....3B}. The matter in this approach is "skeletonized" to push calculations as much forward as possible to the non-linear regime of the gravity field equations. Similar matter "skeleton" is used in a covariant derivation of equations of motion proposed by Dixon \ct{dixon1}. These approaches are useless for development of the LLR model because one has to know the internal motion of matter inside Earth and Moon in order to describe the motion of the laser station and CCR with respect to Earth and to Moon respectively. Hence, we use the tensor of energy-momentum specified by a continuous distribution of matter's density, current, and stress.

Theoretical principles of the post-Newtonian celestial mechanics of N-body system are formulated in \s\ref{tp}. We explain the need of separation of the problem of motion in the internal and external counterparts and the post-Newtonian approximation scheme. Current mathematical knowledge of the post-Newtonian approximations is rather outstanding \ct{2005PhRvD..72d4024B} and we rely upon it to secure the consistency of our derivation.

Post-Newtonian reference frames are constructed in \s\ref{prf}. They include the solar system barycentric (SSB) frame, the Earth-Moon barycentric (EMB) frame, the geocentric (GRF) frame, and the selenocentric (SRF) frame. Each of these frames is associated with the world line of the center of mass of the corresponding gravitating system or a gravitating body. The hierarchical structure of the reference frames corresponds to the hierarchy of masses in the problem under consideration.
Each frame has its own region of mathematical applicability, which is reflected in a specific mathematical structure of solutions of the field equations describing behavior of the metric tensor and the scalar field in the corresponding coordinate charts.

The post-Newtonian coordinate transformations between the frames are derived in \s\ref{pntb}. The derivation is based on a simple fact that the coordinate charts of the frames overlap in a ceratin region of the space-time manifold. Hence, the metric tensor and the scalar field expressed in different coordinates, must admit a smooth tensor transformation to each other. This transformation of the physical fields establishes a system of ordinary differential and algebraic equations for the functions entering the coordinate transformation. The overall procedure is called the method of matched asymptotic expansions, which was used in general relativity for the first time by D'Eath \ct{das1,das2} and applied in the theory of astronomical reference frames in our work \ct{1988CeMec..44...87K}.

Gauge-independent derivation of the post-Newtonian equations of motion of Moon and Earth in various reference frames as well as a systematic post-Newtonian algorithm of LLR data processing with the precision of 1 millimeter will be given elsewhere.

\section{The Scalar-Tensor Theory of Gravity}\l{stt}

Post-Newtonian celestial mechanics describes orbital and rotational motions of extended bodies on a curved space-time manifold described by the metric tensor obtained as a solution of the field equations of a metric-based theory of gravitation in the slow-motion and weak-gravitational field approximation. Class of viable metric theories of gravity ranges from the canonical general theory of relativity \cite{mtw,lali} to a scalar-vector-tensor theory of gravity recently proposed by Bekenstein \cite{2007astro.ph..1848B} for description the motion of galaxies at cosmological scale. It is inconceivable to review all these theories in the present paper and we refer the reader to \ct{willLRR} for further details.  We shall build the theory of lunar motion and LLR in the framework of a scalar-tensor theory of gravity introduced by Jordan \cite{1949Natur.164..637J,1959ZPhys.157..112J} and Fiertz \cite{1956AcHPh.29..128F}, and re-discovered independently by Brans and Dicke \cite{1961PhRv..124..925B,1962PhRv..125.2163D,1962PhRv..126.1875D}. This theory extends the Lagrangian of general theory of relativity by introducing a long range scalar field minimally coupled with gravity field causing a deviation of metric gravity from pure geometry. The presence of the scalar field highlights the geometric role of the metric tensor and makes physical content of the gravitational theory more rich. Equations of the scalar-tensor theory of gravity have been used in NASA Jet Propulsion Laboratory (JPL) and other international space centers for construction of the barycentric ephemerides of the solar system bodies \cite{chatc,sta,epm,pmoe}. We adopt the scalar-tensor theory of gravity for developing the advanced post-Newtonian dynamics of the Earth-Moon system.

\subsection{The Field Equations}\l{fieq}
Gravitational field in the scalar-tensor theory of gravity is described by the metric tensor $g_{\a\b}$ and a long-range scalar field $\phi$ loosely coupled with gravity by means
of a function $\theta(\phi)$.
The field equations in the
scalar-tensor theory are derived from the action \cite{willbook}
\be
     S=\frac{c^3}{16\pi}\int\biggl(\phi R                  \label{10.1}
     - \theta(\phi)\frac{\phi^{,\alpha}\phi_{,\alpha}}
     {\phi} - \frac{16\pi}{ c^4} {\cal L}(g_{\mu\nu},\,\Psi)\biggr)\sqrt{-g}\; d^4 x\; ,
\en where the first, second and third terms in the right side of equation (\ref{10.1}) are
the Lagrangian densities of gravitational field, scalar field and matter
respectively, $g={\rm det}[g_{\alpha\beta}]<0$ is the determinant of the metric tensor
$g_{\alpha\beta}$, $R$ is the Ricci scalar, $\Psi$ indicates dependence of the matter Lagrangian ${\cal L}$ on the
matter fields, and $\theta(\phi)$ is
the coupling function, which is kept unspecified for the purpose of further parametrization of the deviation from general relativity. This makes the theory, we are working with, to be
sufficiently universal.

For the sake
of simplicity we postulate that the
self-coupling potential of the scalar field is identically zero so that the scalar field does not interact
with itself.
This is because this paper deals with a weak gravitational field and one does not expect that this potential can lead to
measurable relativistic effects within the boundaries of
the solar system \cite{willLRR}. However, the self-coupling property of the scalar field leads to its non-linearity, which can be important in strong gravitational fields of neutron stars and black holes, and its
inclusion to the theory may lead to interesting physical consequences \cite{1992CQGra...9.2093D,1993PhRvL..70.2220D}.

Field equations for the metric tensor are obtained by variation of
action (\ref{10.1}) with respect to $g_{\alpha\beta}$. It yields \cite{willbook}
 \br\label{10.2}
   R_{\mu\nu}&=&
   \frac{8\pi}{\phi c^2}\left(T_{\mu\nu}-\frac12 g_{\mu\nu}T\right)+\theta(\phi)\frac{\phi_{,\mu}\phi_{,\nu}}{\phi^2}
   + \frac{1}{\phi}\left(\phi_{;\mu\nu}+\frac12 g_{\mu\nu}{\dAl}_g\phi\right)\;,
\er where
\be\label{covd}
{\dAl}_g\equiv g^{\mu\nu}\frac{\partial^2}{\partial x^\mu\partial x^\nu}-g^{\mu\nu}\Gamma^\alpha_{\mu\nu}
\frac{\partial}{\partial x^\alpha}\;
\en
is
the Laplace-Beltrami operator \cite{mtw,eisen}, and $T_{\mu\nu}$
is the tensor of energy-momentum (TEM) of matter comprising the N-body (solar) system. The variational principle defines it
by equation \cite{lali} \be
  \frac{c^2}2\sqrt{-g}\;T_{\mu\nu}\equiv                         \label{10.3}
  \frac{\partial(\sqrt{-g}{\cal L})}{\partial g^{\mu\nu}}-
  \frac{\partial}{\partial x^{\alpha}}
  \frac{\partial(\sqrt{-g}{\cal L})}{\partial g^{\mu\nu}{}{}_{,\alpha}} \;.
\en
Equation for the scalar field is obtained by
variation of action (\ref{10.1}) with respect  to $\phi$. After making use of the contracted form of equation (\ref{10.2}) it yields \cite{willbook}\be
  {\dAl}_g\phi=\frac1{3+2\theta(\phi)}
  \left(\frac{8\pi}{c^2}\; T-\phi_{,\alpha}\;\phi^{,\alpha}\;    \label{10.4}
  \frac{d\theta}{d\phi}\right)\;.
\en

In what follows, we shall also utilize another version of the
Einstein equations (\ref{10.2}) which is obtained after conformal
transformation of the metric tensor \cite{1992CQGra...9.2093D}\be
   \tilde g_{\mu\nu}=\frac{\phi}{\phi_0} g_{\mu\nu}\qquad,\qquad\tilde g^{\mu\nu}=\frac{\phi_0}{\phi} g^{\mu\nu}\;.
                            \label{13.18}
\en
Here $\phi_0$ denotes the background value of the scalar field that may be a gradually-changing function of time due to the cosmological expansion \cite{willbook}.
It is worth noting that the determinant $\tilde g$ of the conformal metric tensor relates to the determinant
$g$ of the metric $g_{\mu\nu}$ as $\tilde g=(\phi/\phi_0)^4 g$.
The conformal transformation of the metric tensor leads to
the conformal transformation of the
Christoffel symbols and the Ricci tensor. Denoting the conformal Ricci tensor by $\tilde R_{\mu\nu}$, one can
reduce the field equations (\ref{10.2}) to more simple form \cite{1992CQGra...9.2093D}\be
   \tilde R_{\mu\nu}=\frac{8\pi}{\phi c^2}
   \Bigl(T_{\mu\nu}-\frac12 g_{\mu\nu}T\Bigr)+               \label{13.19}
   \frac{2\theta(\phi)+3}{2\phi^2}\,\phi_{,\mu}\,
   \phi_{,\nu}\,.
\en

The metric tensor $g_{\mu\nu}$ is called the physical (Jordan-Fierz)
metric \cite{1992CQGra...9.2093D} because it is used for making real measurements of time intervals, angles,
and space distances. The conformal metric $\tilde g_{\mu\nu}$ is
called the Einstein metric \cite{1992CQGra...9.2093D}. Technically,
it is more convenient for doing mathematical calculations than the Jordan-Fierz
metric. Indeed, if the last (quadratic with respect to the scalar
field) term in equation (\ref{13.19}) is omitted, it becomes
similar to the Einstein equations of general relativity.
In this paper, we prefer to construct the parameterized post-Newtonian
theory of the lunar motion directly in terms of the physical Jordan-Fierz metric.
The conformal metric will be used in discussing propagation of light and the lunar laser ranging somewhere else.

\subsection{The Energy-Momentum Tensor}\l{temo}

Gravitational field and matter, which is a source of this field, are tightly connected via the Bianchi identity of the field equations for the metric tensor \ct{lali,mtw}. The Bianchi identity makes four of ten components of the metric tensor fully independent so that they can be chosen arbitrary. This freedom is usually fixed by picking up a specific gauge condition, which imposes four restrictions on four components of the metric tensor but no restriction on the scalar field. The gauge condition is associated with a specific class of coordinates, which are used for solving the field equations. The Bianchi identity also imposes four limitations on the tensor of energy-momentum of matter, which are microscopic equations of motion of the matter \cite{lali,mtw}. Thus, in order to find the gravitational and scalar fields, and determine motion of the gravitating bodies in N-body (solar) system one has to make several steps: \begin{itemize}
\item[(1)] to specify a model of matter composing of the N-body system,
\item[(2)] to specify the gauge condition imposed on the metric
tensor $g_{\alpha\beta}$,
\item[(3)] to simplify (reduce) the field equations by making use of the gauge freedom,
\item[(4)] to solve the reduced field equations,
\item[(5)] to derive equations of motion of the bodies from the conditions of compatibility of the reduced field equations with the gauge conditions.
\end{itemize}

We assume that the solar system is isolated, which means that we neglect any
influence of our galaxy on the solar system and ignore cosmological effects. This makes the space-time asymptotically-flat so that the barycenter of the solar
system can be set at rest. We assume that matter
of the solar system is described by tensor of energy-momentum of matter with equation of state which is kept arbitrary. There were numerous discussions in early times of development of relativistic celestial mechanics about the role of the energy-momentum tensor of matter in derivation of equations of motion of gravitating bodies. There are two basic models of matter - the field singularity and a continuous distribution of matter. The model of bodies as field singularities was advocated by Einstein and his collaborators \ct{eih,inpl}. The model of bodies consisting of a continuous distribution of matter was preferred  by Lorentz and Droste \ct{lord1,lord2,lord3}, Fock \ct{fockbook}, Chandrasekhar with collaborators \ct{1969ApJ...158...55C,1970ApJ...160..153C}, and others. Damour \ct{1983grr..proc...58D} and Sch\"afer \ct{1985AnPhy.161...81S} succeeded in derivation relativistic equations of motion of self-gravitating bodies, which were modeled by distributions (delta-functions), up to 2.5 post-Newtonian approximation. However, the same equations were derived by Kopeikin \ct{k85} and Grishchuk and Kopeikin \ct{gk83,gk86} for self-gravitating bodies consisting of perfect fluid (see comparison of two approaches in \cite{d89}). It is pretty clear now that any model of matter is appropriate for analysis of the problem of motion of self-gravitating and extended bodies, if mathematical analysis is performed in consistency with physical limitations on the bodies imposed by the field equations. Our goal is to construct a post-Newtonian theory of motion of Earth and Moon with respect to each other and with respect to the other bodies of the solar system. This relativistic analysis should match with the classic models of matter adopted in dynamical astronomy and geophysics. For this reason, we shall model the solar system bodies as consisting of the continuous distribution of matter.

Following Fock \cite{fockbook,fock1} and Papapetrou \cite{pap1,pap2} we define the
energy-momentum tensor as \be
c^2 T^{\a\b}=\rho\left(c^2+\Pi\right)u^\a
u^\b+\pi^{\a\b}\;,    \label{11.1} \en
where $\rho$ and
$\Pi$ are the density and the specific internal energy of matter in the matter's
co-moving frame,
$u^\alpha=dx^\alpha/cd\tau$ is the dimensionless 4-velocity of the matter with $\tau$
being the proper time along the world line of matter's volume element, and
$\pi^{\alpha\beta}$ is a symmetric stress tensor
being orthogonal to the 4-velocity of matter
\be\label{pz1}
u^\alpha\pi_{\alpha\beta}=0\;.
\en
Equation (\ref{pz1}) means that the stress tensor has only spatial components in the frame co-moving with matter.
If one neglects contribution of the off-diagonal components of the stress tensor, it is reduced to a stress tensor of a perfect fluid
\be\label{perf}
\pi^{\alpha\beta}=\left(g^{\alpha\beta}+u^\alpha u^\beta\right)p\;,\en
 where $p$ is an isotropic pressure. Perfect-fluid approximation is used, for example, in PPN formalism \ct{willbook} but it is not sufficient in the Newtonian
theory of motion of the solar system bodies because the tidal and dissipative forces affect their orbital and rotational motions (see, for example,  \cite{1978ppi..book.....Z,1996ARep...40..681M,1993CeMDA..57..295B,1988JGR....93.6216C,1963AmJPh..31...70D}). It is not difficult to incorporate the general model of the stress tensor to the post-Newtonian approximations (see, for example, \ct{dsx1,kovl}). Therefore, we discard the model of the perfect-fluid and incorporate the anisotropic stresses to the post-Newtonian theory of motion of the solar system bodies.

We have noted that due to the Bianchi identity the energy-momentum tensor is conserved, that is obeys to the microscopic equation of motion
\be\label{h2} T^{\a\b}{}{_{;\b}}=0\;,\en
where the semicolon denotes the covariant derivative and repeated indices mean the Einstein summation rule. The conservation of the energy-momentum tensor leads to the equation of continuity \cite{mtw}
\be\label{pz2}
\left(\rho u^\alpha\right)_{;\alpha}=\frac{1}{\sqrt{-g}}\left(\rho\sqrt{-g} u^\alpha\right)_{,\alpha} =0\;,
\en
and to the second law of thermodynamics that is expressed as
a differential relationship between the specific internal energy
and the stress tensor \cite{mtw}
\be \label{11.2}
\rho u^{\alpha}\Pi_{,\alpha}+
      \pi^{\alpha\beta}
      u_{\alpha;\beta}=0\;.
\en
These equations set certain limitations on the structure of the tensor of energy-momentum. They will be employed later for solving the field equations and for derivation of the equations of motion of the bodies.

\section{Theoretical Principles of the Post-Newtonian Celestial Mechanics}\l{tp}
\subsection{External and Internal Problems of Motion}\l{glc}
The post-Newtonian theory of motion of extended celestial bodies described in this paper is based on the scalar-tensor theory of gravity and is a natural extension of PPN formalism for massive point-like particles as described by Nordtvedt and Will \cite{1972ApJ...177..757W,1972ApJ...177..775N,willbook}. PPN formalism contains 10 parameters characterizing different type of deviations from general relativity. It also assumes the existence of a privileged PPN coordinate frame violating the principle of relativity for the metric tensor. PPN privileged frame is associated with the isotropy of the cosmic microwave background radiation. Solar system is moving with respect to this frame.

The present paper does not deal with the privileged-frame effects as the scalar-tensor theory of gravity is Lorentz-invariant. For this reason, we can assume the solar system frame being at rest with the origin located at the solar system barycenter. Heliocentric frame does not coincide with the SSB frame as Sun moves around the SSB at the distances not exceeding two solar radii \ct{hardorp}.  The SSB frame is global with the gravitational field described by the metric tensor, which approaches the Minkowskian metric $\eta_{\a\b}$ at infinity. It means that the global coordinates represent the inertial coordinates of the Minkowskian space-time at infinity. Harmonic coordinates are particularly useful as they simplify the Einstein equations and reduce them to the hyperbolic system of equations \ct{andec}. For this reason, harmonic coordinates were advocated by Fock \ct{fockbook} who believed in their physical privilege. This point of view was confronted by Infeld \ct{inpl} and is currently considered as outdated \ct{mtw,lali}.
Adequate physical description of the global SSB frame is the primary goal of the {\it external problem} of relativistic celestial mechanics \ct{fockbook,d89}. However, the global SSB frame is not sufficient for solving the problem of motion of extended bodies at the post-Newtonian approximation for two reasons.

First, the motion of matter is naturally split in two components -- the orbital motion of the center of mass of each body and the internal motion of matter with respect to the body's center of mass. The SSB frame is fully adequate for describing the orbital dynamics. However, description of the internal motion of matter demands the introduction of a local frame attached to each gravitating body. If a group of the bodies form a gravitationally bounded sub-system, like Earth and Moon, or the sub-system of satellites of major planets, it is natural to introduce the local frame associated with the center of mass of the sub-system. This will allow us to separate the dynamics of the relative motion of the bodies inside the sub-system from the orbital motion of the center of mass of the sub-system itself with respect to the SSB frame. Adequate physical description of the internal motions at the post-Newtonian level of accuracy constitutes the main goal of the {\it internal problem} of relativistic celestial mechanics \ct{fockbook,d89}.

Second, the post-Newtonian celestial mechanics is tightly connected to the geometric properties of the space-time manifold being characterized by the metric tensor, the affine connection (the Christoffel symbol), the curvature tensor and topology. Thus, relativistic description of motion of the celestial bodies is to reflect the diffeomorphic properties of the manifold's geometric structure associated with the set of overlapping coordinate charts and corresponding transformations between them \cite{dfn,arno}. The metric tensor in a local frame of each body must match with the tidal gravitational field of external bodies, hence, it diverges as distance from the body goes to infinity. Therefore, the local coordinates cover only a limited domain (world tube) in space-time around the body under consideration, and the process of their construction must be reconciled with the principle of equivalence \cite{k88,th}.

Newtonian mechanics of N-body system describes translational motion of the bodies in a single global coordinate frame, $x^i$, with the origin placed at the center of mass of all bodies. Local coordinates, $w^i$, are used for description of rotational motion of the bodies, and they are constructed by a simple spatial translation of the global coordinates to the center of mass of each body under consideration. Time in the Newtonian theory is absolute, and, hence, does not change when one transforms the global to local coordinates. Newtonian space is also absolute, which makes the difference between the global and local coordinates physically insignificant.

The theory changes dramatically as one switches from the Newtonian concepts to a consistent relativistic theory of gravity. One still needs a global coordinate frame to describe translational motion of the bodies with respect to one another and the local frames for description of the internal processes inside the bodies. However, there is no longer the absolute time and the absolute space, which are replaced with a Riemannian space-time manifold and a rather complicated set of relativistic differential equations for geometric (gravitational) variables and other fields. Construction of the post-Newtonian global and local frames is now a matter of boundary conditions imposed on the field equations \cite{fockbook}. The principle of relativity should be satisfied when the law of transformation from the global to local coordinates associated with each body (or a sub-system of the bodies) is derived. Not only should it be consistent with the Lorentz transformation but must account for the gauge freedom of the relativistic theory of gravity as well. Time and spatial coordinates are transformed simultaneously making up a class of four-dimensional coordinate transformations \ct{iau2000}.

\subsection{Post-Newtonian Approximations}\l{pnap}
\subsubsection{Small Parameters}
Field equations (\ref{10.2}) and (\ref{10.4}) of the scalar-tensor theory of gravity represent a system of eleventh non-linear differential equations in partial
derivatives. The challenge is to find their solution for the case of N-body system represented by Sun and planets which are not considered as test bodies. Exact solution of this problem is not known and may not exist. Hence, one has to resort to approximation methods. Two basic methods are known: the post-Minkowskian and the post-Newtonian approximations \cite{d89}. Post-Newtonian approximations assume that matter moves slowly and its gravitational field is weak everywhere -- the conditions, which are satisfied within the solar system. For this reason, we use the post-Newtonian approximations in this paper.

post-Newtonian approximations are based on assumption that expansion of the metric tensor in the near zone of a source of gravity can be done in inverse powers of the fundamental speed $c$ that is equal to the speed of light in vacuum. This expansion may be not analytic in higher post-Newtonian approximations in a certain class of coordinates \ct{kake,bld1986}. Exact formulation of a set of basic axioms required for doing the post-Newtonian expansion was given by Rendall \ct{rend}. Practically, it requires to have several small parameters characterizing the source of gravity. They are:
$\epsilon_i\sim v_i/c$, $\epsilon_e\sim v_e/c$, and $\eta_i\sim U_i/c^2$, $\eta_e\sim U_e/c^2$, where $v_i$ is a characteristic velocity of motion of matter inside a body, $v_e$ is a characteristic velocity of the relative motion of the bodies with respect to each other, $U_i$ is the internal gravitational potential of each body, and $U_e$ is the external gravitational potential between the bodies. If one denotes a characteristic radius of a body as $L$ and a characteristic distance between the bodies as $R$, the internal and external gravitational potentials will be $U_i\simeq GM/L$ and $U_e\simeq GM/R$, where $M$ is a characteristic mass of the body.
Due to the virial theorem of the Newtonian gravity \ct{lali} the small parameters are not independent. Specifically, one has
$\epsilon_i^2\sim\eta_i$ and $\epsilon_e^2\sim\eta_e$. Hence,
parameters $\epsilon_i$ and $\epsilon_e$ are sufficient in doing post-Newtonian approximations. Because within the solar system these parameters do not significantly differ from each other, we shall not distinguish between them when doing the post-Newtonian iterations. In what follows, we shall use notation $\epsilon\equiv 1/c$ to mark the presence of the fundamental speed $c$ in the post-Newtonian terms.

Besides the small relativistic parameters $\epsilon$ and $\eta$, post-Newtonian approximations utilize one more small parameter. This parameter is $\delta=L/R$, and it characterizes dependence of the gravitational field outside the bodies on their internal structure and shape. Parameter $\delta$ has no direct relationship to relativity unless the bodies are not compact astrophysical objects like neutron stars or black holes. This is the case of strong gravitational field when the size $L$ of the body approaches its gravitational radius, $L\simeq r_g=(2GM/c^2)$. In this situation $\delta\simeq\eta_e\simeq\epsilon_e^2$ making post-Newtonian approximations more laborious.

It is well-known that in the Newtonian mechanics gravitational field of a spherically-symmetric body is the same as the field of a single point-like particle having the same mass as the body \ct{chandr87}. This is what Damour calls {\it the effacing principle} \cite{1983grr..proc...58D,d89}. It suggests that for spherically-symmetric bodies parameter $\delta=L/R$ does not play any role in the Newtonian approximation. Our study \cite{kovl,2006gr.qc....12017K} reveals that the effacing principle is violated in the first post-Newtonian approximation of the scalar-tensor theory of gravity so that terms of the order of $(\beta-1)\epsilon^2\delta^2$ appear in the translational equations of motion of spherically-symmetric bodies.

Notice that in general relativity, where the PPN parameter $\beta=1$, the effacing principle in equations of motion is violated only by terms of the order of $\epsilon^2\delta^4$ \ct{kovl} as all terms of the order $\epsilon^2\delta^2$ can be eliminated after making an appropriate choice of the center of mass of the bodies \cite{kovl}. For compact relativistic stars $\epsilon^2\delta^4\simeq\epsilon^{10}$, which makes the first post-Newtonian approximation for this objects much smaller than 2.5 post-Newtonian approximation $(\sim\epsilon^5)$, where the radiation-reaction force due to emission of gravitational waves appears for the first time \ct{1983grr..proc...58D,k85,1985AnPhy.161...81S}. This remark corrects Damour's consideration on the compatibility of different orders of post-Newtonian approximations (see \cite[pages 163 and 169]{d89}) and fully justifies our result of the post-Newtonian calculation of the gravitational radiation-reaction force by the Fock-Chandrasekhar method \ct{k85,gk83,gk86}.

If the bodies are not spherically-symmetric, parameter $\delta$ appears in the Newtonian and post-Newtonian approximations as a result of expansion of gravitational field in multipoles. The size of the multipole of multipolarity ${\rm n}$ depends on the parameter of non-sphericity of the body, $J_{\rm n}$,  related to the elastic properties of matter, which are characterized for a self-gravitating body by  Love's numbers $\kappa_{\rm n}$. Generally, they are different for each multipole \ct{1978ppi..book.....Z,loven1,getino}. The present paper will account for all gravitational multipoles of the solar system bodies without making finite truncation of the multipolar series.

\subsubsection{The Post-Newtonian Series}

One assumes that the scalar field can be expanded in a power
series around its background value $\phi_0$, that is
\be                                                         \label{aa}
\phi= \phi_0(1+\zeta)\;,
\en
where $\zeta$ is a dimensionless
perturbation of the scalar field around its background value. In principle, the
background value $\phi_0$ of the scalar field can depend on time
due to cosmological expansion of the universe that may be interpreted as a secular change in the universal gravitational constant $G$ (see below).
According to theoretical expectations \cite{1993PhRvL..70.2217D} and experimental data \cite{willbook,willLRR} the post-Newtonian perturbation $\zeta$ of the scalar field must have a very
small magnitude, so that we can expand all quantities depending on
the scalar field in the Taylor series using the absolute value of
$\zeta$ as a small parameter. In particular, the post-Newtonian decomposition of the
coupling function $\theta(\phi)$ can be written as
\be
\theta(\phi)=\omega+\omega'\;\zeta+O(\zeta^2)\;, \label{10.5}
\en
where $\omega\equiv\theta(\phi_0)$, $\omega'\equiv
\left(d\theta/d\zeta\right)_{\phi=\phi_0}$, and we impose the boundary condition such that
$\zeta$ approaches zero as the distance from the solar system grows to infinity.

We look for solution of the field equations (\ref{10.4}), (\ref{10.2})
in the form of a Taylor expansion of the metric tensor and the
scalar field with respect to parameter $\epsilon\equiv 1/c$
 such that
\br                                                 \label{exp}
g_{\alpha\beta}&=&\eta_{\alpha\beta}+\epsilon\nne{1}{\alpha}{\beta}+
\epsilon^2\nne{2}{\alpha}{\beta}+\epsilon^3
\nne{3}{\alpha}{\beta}+\epsilon^4
\nne{4}{\alpha}{\beta}+O(\epsilon^5)\;.
\er
The generic post-Newtonian expansion of the metric tensor is not analytic \cite{kake,bld1986,d89}. However, the non-analytic terms emerge only in higher post-Newtonian approximations and do not affect results of the present paper since we restrict ourselves by the first post-Newtonian approximation. Notice also that the linear, with respect to $\epsilon$, terms in the metric tensor expansion (\ref{exp}) can be eliminated by coordinate adjustments \cite{th}. These terms correspond to a non-orthogonality of the local coordinate frame and/or a residual rotation of spatial axes \cite{th}. Reference frames with such properties are not used in astronomy and geophysics. Therefore, we assume that all coordinates used in this paper are non-rotating and orthogonal, so that the linear term in expansion (\ref{exp}) is absent.

Various components of the metric tensor and the scalar field have in the first post-Newtonian approximation the following form
\br
g_{00}&=&-1+\epsilon^2\nne{2}{0}{0}+\epsilon^4\nne{4}{0}{0}+O(\epsilon^5)
\;,       \label{11.4}
\\
g_{0i}&=&\epsilon^3\nne{3}{0}{i}+O(\epsilon^5) \;,    \label{11.5}
\\
g_{ij}&=&\delta_{ij}+ \epsilon^2\nne{2}{i}{j}                     \label{11.6}
    +O(\epsilon^4)  \;,
\\
\zeta&=&\epsilon^2 \nnd{2}+O(\epsilon^4) \;, \label{11.7} \er
where
$\nne{n}{\alpha}{\beta}$ and $\nnd{n}$ denote terms of the order
$\epsilon^n\;\;(n=2,3,4)$.
In what follows, we shall use notations:
\be                                                             \label{not}
h_{00}\equiv\nne{2}{0}{0}\;,\quad
l_{00}\equiv\nne{4}{0}{0}\;,\quad
h_{0i}\equiv\nne{3}{0}{i}\;,\quad
h_{ij}\equiv\nne{2}{i}{j}\;,\quad h\equiv\nne{2}{k}{k}\;,\en
and
\be\label{not1}
\varphi\equiv(\omega+2)\nnd{2}\;.
\en

Post-Newtonian expansion of the metric tensor and the scalar field
introduces a corresponding expansion of the energy-momentum tensor
\br
   T_{00}&=&\nnt{0}{0}{0}+\epsilon^2\nnt{2}{0}{0}+O(\epsilon^4) \;,              \label{11.8}
\\
   T_{0i}&=&\epsilon\nnt{1}{0}{i}+O(\epsilon^3) \;,              \label{11.9}
\\
   T_{ij}&=&\epsilon^2\nnt{2}{i}{j}+O(\epsilon^4) \;,              \label{11.10}
\er where $\nnt{n}{\alpha}{\beta}\;\;(n=0,1,2,3...)$ denote terms of the order
$\epsilon^n$. In the first post-Newtonian approximation the components of the energy-momentum tensor were derived by Fock \ct{fockbook}
\br
\nnt{0}{0}{0}&=&\rho^*\;, \label{11.21}
\\
 \nnt{1}{0}{i}&=&-\;\rho^*v^i\;,                       \label{11.23}
\\
 \nnt{2}{i}{j}&=&\rho^*v^iv^j+\pi^{ij}\;,          \label{11.24}
\\
 \nnt{2}{0}{0}&=&\rho^*\left(\frac{v^2}{2}+\Pi-                 \label{11.22}
                  h_{00}-
                  \frac{h}{2}\right)\;,
\er
where $v^i=c u^i/u^0$ is the 3-dimensional velocity of matter.

Fock also introduced the invariant
density of matter \cite{fockbook}
\be
   \rho^*\equiv\sqrt{-g}u^0\rho=\rho+\frac{1}{2}\epsilon^2\rho\left(v^2+h\right)+O(\epsilon^4)\;,                               \label{11.19}
\en
which is a useful mathematical tool in relativistic hydrodynamics \ct{mtw,willbook}. The reason is that the invariant density, unlike density $\rho$, obeys the exact equation of continuity (\ref{pz2}) that can be recast to a Newtonian-like form \cite{fockbook} \br
    c\rho^*_{\,,0}+                      \label{11.20}
    (\rho^*v^i)_{,i}&=&0\;,
\er
where $f_{\;,0}\equiv(1/c)\p f/\p t$.
Equation (\ref{11.20}) is valid in any post-Newtonian approximation and it makes calculation of time derivative of a volume integral of any function $f(t,{\bm x})$ simple
\be\l{dmq}
\frac{d}{dt}\int\limits_{V_A}\rho^*(t,{\bm x})f(t,{\bm x})d^3x=\int\limits_{V_A}\rho^*(t,{\bm x})\frac{df(t,{\bm x})}{dt}d^3x\;,
\en
where the total time derivative
\be\l{qiw}
\frac{d}{dt}=\frac{\p}{\p t}+v^i\frac{\p}{\p x^i}\;,
\en
and one assumes in equation (\ref{dmq}) that the body A moves, and its shape and internal structure depend on time.

\subsection{The Post-Newtonian Field Equations}\l{pnfi}

The post-Newtonian field equations can be derived after substitution the post-Newtonian series of the previous section to the covariant equations (\ref{10.2}) and (\ref{10.4}), and arranging terms in the order of smallness with respect to parameter $\epsilon\equiv 1/c$. However, these post-Newtonian equations will preserve the gauge freedom of the original covariant field equations, which will make their solution depending on four arbitrary functions. It is a common practice to eliminate this arbitrariness by imposing a gauge condition. This is equivalent to a choice of a class of specific coordinates. It should be understood that at this stage of the post-Newtonian iteration procedure, the gauge condition have not rigidly fixed the coordinates as yet, so that a large freedom of coordinate transformations remains. This class of transformations is associated with the residual gauge freedom, which plays an essential role in relativistic celestial mechanics of N-body system.

In general-relativistic celestial mechanics the harmonic gauge condition
\be\l{gytv}
   \left(\sqrt{-g}\;g^{\mu\nu}\right)_{,\nu}=0\;,
\en
is used the most often \ct{fockbook,dsx1,brum,bk89,iau2000}.
The most convenient gauge condition in the scalar-tensor theory of gravity were proposed by Nutku \cite{1969ApJ...158..991N,1969ApJ...155..999N} as a generalization of the harmonic gauge
\be
   \left(\frac{\phi}{\phi_0}\;\sqrt{-g}\;g^{\mu\nu}\right)_{,\nu}=0\;.                   \label{11.3}
\en
Post-Newtonian expansion of gauge condition (\ref{11.3}) yields
\br
   c\left(\frac{2\varphi}{\omega+2}+h_{00}+h\right)_{,0}&=&2h_{0k,k}\;, \label{11.11}
\\
    \left(\frac{2\varphi}{\omega+2}-h_{00}+h\right)_{,i}&=&  \label{11.12}
    2h_{ik,k}\,.
\er
It is worth noting that in the first post-Newtonian approximation, equations
(\ref{11.11}), (\ref{11.12}) do not restrict the metric tensor component
$\nne{4}{0}{0}\equiv l_{00}$, which is directly obtained from the field equation without further limitations.

The post-Newtonian field equations for the scalar field and the metric tensor
  are obtained from equations
(\ref{10.4}) and (\ref{10.2}) after making use of the
post-Newtonian expansions, given by equations(\ref{11.4})--(\ref{11.10}), and the gauge conditions (\ref{11.11}), (\ref{11.12}).
The scalar-tensor theory of gravity with variable coupling
function $\theta(\phi)$ has two constant parameters, $\omega$ and $\omega'$, characterizing deviation from general
relativity. They are related to the standard PPN parameters $\gamma$ and $\beta$ as follows \cite{willbook}
\br \gamma=\gamma(\omega)&=&\frac{\omega+1}{\omega+2}\;,                               \label{11.27}
 \\
 \beta=\beta(\omega)&=&1+ \frac{\omega'}{(2\omega+3)(2\omega+4)^2}\;.        \label{11.28}
 \er
General relativity is obtained as a limiting case of the scalar-tensor
theory when parameters $\gamma=\beta=1$. Notice that in order to get this limit convergent, parameter $\omega'$ must grow slower than $\omega^3$ as $\omega$ approaches infinity. Currently, there are no experimental data restricting the functional behavior of $\omega'\sim\omega^3\beta(\omega)$. This makes parameter $\beta$ to be a primary target for experimental study in the near-future gravitational experiments \ct{2008ASSL..349.....D,2008arXiv0802.0582A} including advanced LLR \ct{test2004}.

The scalar field perturbation (\ref{not1}) is expressed in terms of $\gamma$ as
\be
\nnd{2}=(1-\gamma)\varphi\;. \label{10.6aa}
\en
The background scalar field $\phi_0$ and the
parameter of coupling $\omega$ determine the observed numerical
value of the universal gravitational constant
\be G =\frac
{2\omega+4}{2\omega+3 }\,\phi_0^{-1}\;. \label{10.6}
\en
Had the background value $\phi_0$ of the scalar
field driven by cosmological evolution, the measured value of the
universal gravitational constant would depend on time and one could hope
to detect it experimentally. The best upper limit on time
variability of $G$ is imposed by lunar laser ranging (LLR) as $|\dot{G}/G|<(4\pm 9)\times 10^{-13}$ yr$^{-1}$
\cite{2004PhRvL..93z1101W}.

After making use of the definition of the tensor of energy-momentum, equations (\ref{11.21})--(\ref{11.22}),
and that of the PPN parameters, equations (\ref{11.27})--(\ref{10.6}), one obtains
the final form of the post-Newtonian field equations:
\br
\label{11.29}
&&\dAl\varphi=-4\pi G\rho^*\;,
\\\nonumber
&&\dAl\left\{ h_{00}+\epsilon^2\left[l_{00}+\frac{h_{00}^2}{2}+2(\beta-1)\varphi^2 \right]\right\}=
\\&&\qquad
-8\pi G\rho^*\biggl\{1+\epsilon^2\left[(\gamma+\frac12)\,v^2+\Pi+
       \gamma\,\frac{\pi^{kk}}{\rho^*}-\frac{h}{6}-(2\beta-\gamma-1)\varphi\right]\biggr\}            +\epsilon^2 h_{<ij>}h_{00,ij}\label{11.33}
\,,
\\
 &&\dAl h_{0i}= 8\pi G(1+\gamma)                    \label{11.31}
   \rho^*v^i\;,
\\                                  \label{11.32}
 &&  \dAl h_{ij}=-8\pi G\gamma\rho^*\delta_{ij}  \;,
\er
where
$\dAl\equiv\eta^{\mu\nu}\partial_\mu\partial_\nu $ is the D'Alembert (wave) operator of the Minkowskian space-time,
and $H_{<ij>}\equiv H_{ij}-\delta_{ij}H/3$ is the symmetric trace-free (STF) part of the spatial components of
the metric tensor (the STF tensors are thoroughly discussed in \ct{thor,bld1986}).
Equations (\ref{11.29})--(\ref{11.32}) are valid in the class of coordinates defined by the gauge condition (\ref{11.3}). We shall study the residual gauge freedom of this coordinates in full
details in next sections of the paper.

\subsection{Conformal Harmonic Coordinates}\label{gfree}

By making use of the conformal metric tensor one can recast
equation(\ref{11.3}) to the same form as the
harmonic gauge condition (\ref{gytv}) in general relativity \cite{pap1,fockbook}
\be
   (\sqrt{-\tilde g}\,\tilde g^{\mu\nu})_{,\nu}=0\,.       \label{13.20}
\en
Equation (\ref{11.3}) or (\ref{13.20}) can be re-written as follows
\be                                                 \label{gau}
g^{\mu\nu}\Gamma^\alpha_{\mu\nu}=\left(\ln\frac{\phi}{\phi_0}\right)^{,\alpha}\;,
\en
so that the Laplace-Beltrami operator (\ref{covd}) assumes the form
\be                                             \label{co}
{\dAl}_g\equiv g^{\mu\nu}\left(\frac{\partial^2}{\partial x^\mu\partial x^\nu}-\frac{1}{\phi}\frac{\partial\phi}
{\partial x^\mu}\frac{\partial}{\partial x^\nu}\right)\;.
\en
Dependence of this operator on the scalar field is a property of the adopted gauge condition.

Any function $F(x^\alpha)$ satisfying the homogeneous Laplace-Beltrami equation, $\dAl_gF(x^\alpha)=0$,
is called harmonic. Notice that the Nutku gauge condition (\ref{11.3}) assumes that ${\dAl}_g x^\alpha=-(\ln\phi)^{,\alpha}\not=0$, so any coordinate $x^\alpha$ which obeys the gauge condition (\ref{gau}) is not a harmonic function on a space-time manifold endowed with the Jordan-Fiertz metric. Nonetheless, such non-harmonic coordinates are more convenient in the scalar-tensor theory of gravity because they allow us to eliminate more non-physical terms from the field equations than the harmonic gauge does. We shall call the class of
coordinates being singled out by the Nutku conditions (\ref{11.3}) as conformal harmonic coordinates. This is because these coordinates are harmonic functions in the conformal Einstein metric.

The conformal harmonic coordinates have many properties
similar to the harmonic coordinates in general relativity. The choice of the conformal harmonic coordinates for
constructing the theory of the lunar motion is justified
by the following three factors: (1) the conformal harmonic coordinates approach harmonic coordinates in general relativity when
the scalar field $\phi\rightarrow \phi_0$ so that $\beta=\gamma=1$,
(2) the conformal harmonic coordinates are natural for scalar-tensor parametrization of equations used in
resolutions of the IAU 2000 General Assembly \cite{iau2000} on relativistic
reference frames, (3) the gauge condition (\ref{11.3}) significantly
simplifies the post-Newtonian field equations, thus, facilitating their solution.
Harmonic coordinates were used by Klioner and Soffel \cite{2000PhRvD..62b4019K} for construction of post-Newtonian reference frames in PPN formalism and the difficulties associated with this choice have been analyzed in our paper \cite[appendix A]{kovl}.

Gauge condition (\ref{gau}) does not fix coordinates uniquely. Let us change
coordinates
\be\label{11.12a}
x^\alpha\mapsto w^\alpha=w^\alpha\left(x^\alpha\right)\;,
\en
but keep the gauge condition (\ref{gau}) the same. Simple calculation shows that in such case the new coordinates $w^\alpha$ must satisfy a homogeneous
wave equation
\be                                                         \label{11.b}
g^{\mu\nu}(x^{\beta})\frac{\partial^2 w^\alpha}{\partial x^\mu\partial x^\nu}=0\;,
\en
which have an infinite set of non-trivial solutions defining the entire set of the local coordinates $w^\a$ on the space-time manifold of the solar system.
Equation (\ref{11.b}) describes the residual gauge freedom existing in the class of the conformal harmonic coordinates restricted by the gauge condition (\ref{gau}). This equation of the residual gauge freedom in the scalar-tensor theory of gravity is the same as in the case of the harmonic coordinates in general relativity.

\subsection{Microscopic Post-Newtonian Equations of Motion}\label{pnem}
The macroscopic post-Newtonian equations of motion of matter consist of: \begin{enumerate}
\item the equation of
continuity, \item the thermodynamic equation of internal energy relating the elastic energy $\Pi$ and the stress tensor
$\pi_{\alpha\beta}$, \item the Navier-Stokes equation, which converts to the Euler equation in case of a perfect fluid.\end{enumerate}

The equation of continuity in arbitrary conformal harmonic coordinates $w^\a=(w^0,w^i)=(cu, {\bm w})$ has the most simple form for the invariant
density $\rho^*$ and reads
\be\label{kp1}
\frac{\partial\rho^*}{\partial u}+\frac{\partial\left(\rho^*\nu^i\right)}{\partial w^i}=0\;.
\en
This equation is exact and takes into account all post-Newtonian corrections, as follows from the
definition of the invariant density $\rho^*$ and equation (\ref{11.20}).

The thermodynamic equation relating the internal elastic energy $\Pi$ and the stress tensor $\pi_{\alpha\beta}$ is required in the first post-Newtonian approximation only in a linear order where the stress tensor is completely characterized by its
spatial components $\pi_{ij}$. Hence, one has from equation (\ref{11.2}) the following differential equation
\be\label{kp2}
\frac{d\Pi}{du}+\frac{\pi_{ij}}{\rho^*}\frac{\partial\nu^i}{\partial w^j}=O(\epsilon^2)\;,
\en
where the operator of convective time derivative is $d/du\equiv\partial/\partial u+\nu^i\partial/\partial w^i$.

The macroscopic equation of motion of a volume element of matter follows from the spatial part of the law of conservation of the energy-momentum tensor, $T^{i\nu}{}_{;\nu}=0$. This yields the Navier-Stokes equation because the stress tensor accounts for anisotropic stresses. In case of a perfect fluid, the stresses are reduced to isotropic pressure and the Navier-Stokes equation becomes the Euler equation which is employed in the PPN formalism \cite{willbook}, but not in this paper.
The post-Newtonian Navier-Stokes equation is
\br\label{kp3}&&
\rho^*\frac{d}{du}\left\{\nu^i+\epsilon^2\left[\left(\frac12\nu^2+\Pi+\frac12 h_{00}+\frac{h}{3}\right)\nu^i+
 h_{0i}\right]\right\}+\epsilon^2\,\frac{\partial\left(\pi_{ij}\nu^j\right)}{\partial u}=\\\nonumber\\
\nonumber&&\frac12\rho^*\frac{\partial h_{00}}{\partial w^i}-\frac{\partial\pi_{ij}}{\partial w^j}+\epsilon^2
\left\{\rho^*\left[\frac12\frac{\partial l_{00}}{\partial w^i}+\frac14\left(\nu^2+2\Pi+ h_{00}\right)
\frac{\partial h_{00}}{\partial w^i}+\frac16\,\nu^2\,\frac{\partial h}{\partial w^i}+\nu^k\,
\frac{\partial h_{0k}}{\partial w^i}\right]\right.\\\nonumber\\\nonumber&&\left.
+\frac16\,\pi_{kk}\,\frac{\partial h}{\partial w^i}+\frac12
\pi_{ik}\left(\frac{\partial  h_{00}}{\partial w^k}-\frac13\frac{\partial  h}{\partial w^k}\right)
+\frac12\left( h_{00}-\frac13 h\right)\frac{\partial\pi_{ik}}{\partial w^k}
\right\}+O(\epsilon^4)\;,
\er
where gravitational potentials $h_{00}$, $l_{00}$, $h_{0i}$, $h_{ij}$, and $h=h_{ii}$ are the metric tensor components
defined by equation (\ref{not}). Notice that the scalar field enters the macroscopic equation of motion of matter only implicitly through the metric tensor components.

\section{Post-Newtonian Reference Frames}\l{prf}
\subsection{Coordinates and Observables}
Physically adequate relativistic description of the lunar motion is not conceivable without a self-consistent theory of relativistic reference frames in the Earth-Moon system as it orbits the barycenter of the solar system. The solar system has a hierarchic structure associated with the diversity of masses of the solar system bodies and the presence of planetary satellite systems, which form a set of gravitationally bounded sub-systems of the solar system consisting of a planet and its satellites. The most massive body of the solar system is Sun. Therefore, in accordance with the heliocentric point of view the planets orbit Sun because their masses are significantly smaller than that of Sun. Nevertheless, major planets of the solar system, like Jupiter and Saturn, pull Sun by their gravitational fields strong enough making it to revolve at some distance (less than two solar radii \ct{hardorp}) around a common solar-system barycenter (SSB). A global, solar-system barycentric frame is required to describe the orbital motion of Sun and planets around the SSB. On the other hand, rotational motion of Sun and planets is more natural to describe in their local frames associated with each of the bodies. Many planets have their natural satellites which orbit the planet. A planet with its satellites form a sub-system of the solar system which can be considered as essentially isolated from the rest of the solar system. This is because the principle of equivalence applied to the sub-system, reduces gravitational attraction of the external bodies to the tidal force that is much smaller than the gravity of the planet. If the ratio of masses of a planet and its satellites is not negligibly small, it is convenient to introduce a local coordinate frame associated with the barycenter of the sub-system consisting of the planet and its satellites. This is because motion of the barycenter of the sub-system around the SSB approximates the Keplerian ellipse fairly well, while the planet and its satellites oscillates on their orbits around the barycenter of the sub-system.  The hierarchic structure of the coordinate frames in the solar system leads to a natural decomposition of orbital motion of each body in the solar system in a trigonometric series of fundamental harmonics like in case of the Fourier expansion.

From this point of view the theory of the lunar motion should introduce three local coordinate frames attached correspondingly to Earth's center of mass (geocenter), to Moon's center of mass (selenocenter), and to the barycenter of the Earth-Moon sub-system. The geocentric frame is to describe rotational motion of Earth and motion of artificial satellites orbiting Earth. The selenocentric frame is introduced to describe rotational motion (physical libration) of Moon and orbital motion of spacecrafts around Moon. The Earth-Moon-barycenter (EMB) frame serves to describe a relative motion of Moon with respect to Earth. The global SSB frame is introduced to describe the orbital motion of the EMB frame with respect to the solar-system barycenter.

Relativistic theory of gravity brings about additional arguments in favor of introduction of the hierarchic structure to the post-Newtonian problem of motion in N-body system. This structure naturally arises on space-time manifold because of its differential structure described in terms of a set of local coordinates and diffeomorphic transformations between them \cite{eisen,mtw}. If one neglected the mutual gravitational interaction between Earth and Moon (test-particles approximation) their relative motion would be governed only by the tidal gravitational field of Sun and other planets. Such relative motion of two test particles is described in metric theories of gravity by the equation of deviation of geodesics \ct{mtw}. This equation is covariant and valid in arbitrary coordinates but it has the most simple form in a local coordinate frame of one of the test particles. The Earth and Moon are not test particles and it complicates their relative motion because one has to account for their mutual gravitational tug in addition to the tidal forces from external bodies. The origin of the local coordinates is also shifted to the center of mass of the Earth-Moon system. These factors bring about rather serious mathematical difficulty to the construction of the local frame in the post-Newtonian approximations since the tidal forces get entangled with the Earth-Moon attraction of gravity and the concept of the Earth-Moon barycenter should be elaborated on more profound basis than in the Newtonian gravity. Any incomprehensive definition used in the construction of the local coordinates in the post-Newtonian approximations leads to appearance of extra terms in equations of relative motion of the Moon-Earth system, which can be removed after transformation to a more appropriate coordinates.

Construction of the set of the local coordinate frames in the Earth-Moon system connected to the global SSB frame by a coordinate transformation helps us to single out and to eliminate a great deal of spurious, gauge-dependent post-Newtonian effects. The spurious radial oscillations of the Earth-Moon distance, expressed in terms of the solar system barycentric coordinates, inhabited earlier relativistic theories of the lunar motion \cite{brm,baier,vab} and researchers assumed that they can be measured \cite{baier,1982A&A...116...75L,1982hper.coll..217L}. The spurious character of the main terms of these coordinate oscillations was recognized in papers \cite{1973PhRvD...7.2347N,1981rcse.conf..283B,sof}. However, the usage of the solar system barycentric coordinates introduces to the Earth-Moon equations of motion (\ref{meih1}), (\ref{meih2}) many other unmeasurable terms, which are explicitly present in the coordinate description of the lunar ephemerides around Earth. Recent paper by Murphy, Nordtvedt and Turyshev \cite{2007PhRvL..98g1102M} intends to interpret some of these coordinate terms as measurable, despite that they depend on the choice of the local coordinates and can be eliminated from the coordinate description after making an appropriate coordinate transformation \cite{k07}. The on-going discussion \ct{2007PhRvL..98v9002M,2008arXiv0809.3392K} stimulates development of the post-Newtonian theory of reference frames in the Earth-Moon system in order to clarify the number of observable relativistic effects, which can be measured with advanced LLR techniques.

In order to understand the theoretical connection between the equations of motion and observables one has to use the light-ray time-delay equation describing the time of flight of a photon from the laser station to the CCR array on Moon. This equation, written down in the SSB frame, has the following form \cite{vab,brum}
\be\l{light}
t_2-t_1=R_{12}+2\sum_{\sss{B}}\frac{G{\rm M}_{\sss{B}}}{c^3}\ln\le[\frac{R_{1B}+R_{2B}+R_{12}}{R_{1B}+R_{2B}-R_{12}}\r]+\sum_{\sss{B}}\lambda_{\sss B}\frac{G{\rm M}_{\sss{B}}}{c^3}\frac{\le(R_{1B}-R_{2B}\r)^2-R^2_{12}}{2R_{1B}R_{2B}R_{12}}\le(R_{1B}+R_{2B}\r)\;,
\en
where $t_1$ is the time of emission of the photon from the laser at point ${\bm x}_1$, $t_2$ is the time of arrival of the photon to the CCR on Moon at point ${\bm x}_2$, $R_{12}=|{\bm x}_2-{\bm x}_1|$, $R_{1B}=|{\bm x}_1-{\bm x}_{\sss{B}}(t_1)|$, $R_{2B}=|{\bm x}_2-{\bm x}_{\sss{B}}(t_2)|$, and $\lambda_{\sss B}$ is the coordinate parameter introduced in equation (\ref{gt2}). Time-delay equation (\ref{light}) does not take into account relativistic perturbations caused by velocities of the gravitating bodies \ct{1999PhRvD..60l4002K,2002PhRvD..65f4025K}. These perturbations may be important in the case of LLR ranging measurements having an accuracy of one millimeter and we shall analyze them in a separate publication. What is important to notice is that the coordinate parameter $\lambda_{\sss B}$ enters both EIH equations of motion of the bodies (\ref{eh1})--(\ref{eh3}) and that of the light ray (\ref{light}). Specifically for this reason, this parameter can not be measured at all by LLR techniques, and must be fixed by observer prior the data processing. Depending on the choice of the coordinates the EIH equations of motion of the bodies can be bring to one or another form. At the same time the equation (\ref{light}) changes synchronously. It makes no sense to fix parameter $\lambda_{\sss B}$ in equation (\ref{light}) and, then, to "measure" it in EIH equations of motion of the bodies (\ref{eh1})--(\ref{eh3}) as discussed in our paper \ct{k07}.

\subsection{The Solar System Barycentric Frame}\label{ssbf}
\subsubsection{Boundary Conditions and Kinematic Properties}\l{zoki}

We assume that the solar system is isolated and there are no masses outside of it.
The number of
bodies in the solar system, which gravitational field must be taken into account in the post-Newtonian theory of reference frames, depends on the accuracy of astronomical observations and the precision of calculation of their ephemerides. We include Sun, Moon, Earth, other planets, and the largest asteroids moving between orbits of Mars and Jupiter. Since we ignore gravitational field of the external astronomical
bodies residing outside of the solar system, the space-time can be considered as
asymptotically-flat at infinity with the metric tensor $g_{\alpha\beta}$ approaching
the Minkowskian metric $\eta_{\alpha\beta}={\rm diag}(-1,+1,+1,+1)$.

The whole space-time manifold associated with the isolated solar system is covered by a single global coordinate frame denoted as $x^\alpha=(x^0,x^i)$, where $x^0=ct$ is coordinate time, and $x^i\equiv{\bm x}$ are spatial coordinates. The global coordinates are used for description of orbital dynamics of the solar system bodies with respect to the solar system barycenter. The coordinate time and spatial coordinates have no physical meaning in those domains of space-time where gravitational field is not negligible. However, when one approaches to infinity the global coordinates approximates the inertial coordinates of observer in the Minkowskian space. For this, reason one can think about the coordinate time $t$ and the spatial coordinates $x^i$ as proper time and proper distance measured by a fictitious observer at infinity, who is at rest with respect to the barycenter of the solar system \ct{fockbook}.

Precise mathematical definition of the global coordinates can be given in terms of the metric tensor, which is a solution of the field equations with a boundary conditions imposed on it at infinity. To formulate the boundary conditions, let us introduce the metric perturbation with respect to the Minkowskian metric (c.f. equation (\ref{exp}))
\be\label{mtp}
h_{\alpha\beta}(t,{\bm x})\equiv g_{\alpha\beta}(t,{\bm x})-\eta_{\alpha\beta}\;.\en
Existence of the global coordinates demand that products $rh_{\alpha\beta}$ and $r^2 h_{\alpha\beta,\gamma}$, where $r=|{\bm x}|$, were bounded, and
\be                                                         \label{12.1}
\lim_{\substack{r\rightarrow\infty\\t+r/c={\rm const.}}}h_{\alpha\beta}(t,{\bm x})=0\;, \en
Additional boundary condition must be imposed on the first
derivatives of the metric tensor to prevent appearance of non-physical radiative solutions associated with gravitational waves incoming to the solar system \cite{fockbook}. This condition is formulated as follows \cite{fockbook,1983grr..proc...58D}
\be                                                         \label{12.2}
\lim_{\substack{r\rightarrow\infty\\t+r/c={\rm const.}}}\left[\left(rh_{\alpha\beta}\right),_r+
\left(rh_{\alpha\beta}\right),_0\right]=0\;.
\en
Though, the first post-Newtonian approximation we are dealing with, does not include terms in the metric tensor, which describe the gravitational waves, the boundary condition tells us to chose the retarded solution of the field equation (\ref{11.33})-(\ref{11.32}).

Similar "no-incoming-radiation" conditions are imposed on the perturbation of the scalar field defined in equation (\ref{10.5})
\be\label{12.3}
\lim_{\substack{r\rightarrow\infty\\t+r/c={\rm const.}}}\zeta(t,{\bm x})=0\;,
\en
\be\label{12.3a}
\lim_{\substack{r\rightarrow\infty\\t+r/c={\rm const.}}}
\left[\left(r\zeta\right),_r+\left(r\zeta\right),_0\right]=0\;.
\en

The global coordinates $x^\alpha$ cover the entire space-time and set up a primary basis for construction of the IAU theory of relativistic reference frames in the
solar system \cite{k89o,iau2000,2007AIPC..886..268K}. The origin of the
global coordinates coincides with the barycenter of the solar system at any
instant of time. This condition can be satisfied after choosing a suitable
definition of the post-Newtonian dipole moment $\mathbb{D}^i$ of the N-body system and equating its numerical
value to zero along with its first and second time derivatives. This requirement can be fulfilled because of the law of conservation of the linear momentum of the solar system  (see equation \ref{13.35b}).

The law of conservation of the angular momentum of the solar system (see equation \ref{13.361}) allows us to make the spatial axes of the global coordinates non-rotating in
space either kinematically or dynamically \cite{bk89,bk-nc}. Coordinates are called kinematically non-rotating
if their spatial orientation does not change with respect to the Minkowskian
coordinates at infinity as time goes on \ct{1989rfag.conf....1K,2004fuas.book.....K}. Such kinematically
non-rotating coordinates are anchored in the sky to a set of distant quasars forming the International Celestial Reference Frame (ICRF) that has the current precision better than $100$ $\mu$arcsec
\cite{1998AJ....116..516M,2004fuas.book.....K}. The ICRF quasars are
uniformly distributed all over the sky and have negligibly small
parallaxes and proper motions. However, temporal stability of the global reference frame can be affected due to internal motions of quasar's jets, which should be constantly monitored in order to maintain ICRF inertial steadiness \cite{2006A&A...452.1107F,2007AstL...33..481T}. Another source of the possible corruption of the inertial properties of the global reference frame is due to the accelerated motion of the solar system with respect to the center of mass of our Galaxy \cite{1933AOTok..36..155H,2006AJ....131.1471K}, influence of the cosmological effects \ct{1970FoPh....1...17B} and ultra-low frequency gravitational waves \ct{1997ApJ...485...87G}. In principle, a more extended post-Newtonian approach to the reference frames, taking into account the fact that the background space-time of the solar system is cosmologically curved, has to be developed \cite{2007AIPC..886..268K,2005ESASP.576..305K} but we do not tackle this problem in the present paper.

Dynamically non-rotating coordinate system is defined by the
condition that equations of motion of test particles moving with respect to these coordinates do not have any
terms that can be interpreted as the Coriolis or centripetal
forces \cite{bk89,1989rfag.conf....1K}. This definition operates only with local properties of the
space-time manifold and does not require observations of distant celestial
objects like stars or quasars. The dynamical definition of non-rotating global coordinates
is used in construction of modern ephemerides of the solar system
bodies which are based primarily on radar and laser ranging measurements to
planets and Moon \cite{2002HiA....12..326S,2008AIPC..977..254S}. Because of the assumption that the solar
system is isolated, one can postulate that the global coordinates do not rotate in any sense. This postulate is firmly supported by observations \ct{1993AdSpR..13..161J,2005HiA....13Q.609S}.

\subsubsection{The Metric Tensor and Scalar Field}

The metric tensor $g_{\alpha\beta}(t,{\bm x})$ and the scalar field $\varphi(t,\bm{x})$ are obtained in the global coordinates by
solving the field equations (\ref{11.29})--(\ref{11.32}) after
imposing the boundary conditions (\ref{12.1})--(\ref{12.3}).
It yields \br
   \nnb{t}{x}&=&U(t,\nnx)\,,                                             \label{12.5}
\\
   h_{00}(t,{\bm x})&=&2\,U(t,\nnx)\,,                                        \label{12.6}
\\
   l_{00}(t,\nnx)&=&2\Psi(t,\nnx)-2(\beta-1)\varphi^2(t,\nnx) -2U^2(t,\nnx)-\frac{\p^2\chi(t,\nnx)}{\p t^2}\,, \label{12.9}
\\
   h_{0i}(t,\nnx)&=& -2(1+\gamma)\,U_i(t,\nnx)\,,                           \label{12.8}
\\
h_{ij}(t,\nnx)&= &2\gamma\delta_{ij}U(t,\nnx)\,,                 \label{12.7}
\er
where the post-Newtonian potential
\be                                                     \label{12.9ex}
\Psi(t,\nnx)\equiv(\gamma+\frac12)\nnff{1}-\frac{1}{6}\nnff{2}+(1+\gamma-2\beta)\nnff{3}
         +\nnff{4}+\gamma\nnff{5}\;,
\en

Gravitational potentials $U,\,U^i,\,\chi$, and
$\Psi_k\,\,$ $(k=1,...,5)$ can be represented as linear
combinations of the gravitational potentials of each body of the solar system
\be\label{12.9a} U=\sum_{A} U_A\,,\qquad U_i=\sum_{A}
U^i_A\,,\qquad\Psi_k=\sum_{A}\Psi_{Ak}\:,\qquad
      \chi=\sum_{A}\chi_A\:,
\en
where the summation index $A$ numerates the bodies of the solar system, which gravitational field contributes to our calculations. In what follows, we shall also use the capital letters $S$, $E$, and $M$ to indicate affiliation of functions to Sun, Earth and Moon respectively.

The gravitational potentials of a body $A$ are defined as integrals taken only over the spatial volume $V_A$ of this body
\be
   U_{{A}}(t,\nnx)=                 \label{12.10}
   G\int\limits_{V_A}\frac{\rho^*(t,{\bm x}')}{
      |{\bm x}-{\bm x}'|}d^3x'\,,
\en\be
   U^i_{{A}}(t,\nnx)=G\int\limits_{V_A}\frac{\rho^*(t,{\bm x}')v^i(t,{\bm x}')}{
      |{\bm x}-{\bm x}'|}d^3x'               \label{12.11}
   \,,
\en\be
   \chi_{{A}}(t,\nnx)=              \label{12.12}
   -\,G\int\limits_{V_A}\rho^*(t,{\bm x}')
      |{\bm x}-{\bm x}'|d^3x'\,,
\en\be
   \Psi_{A1}(t,\nnx)=G\int\limits_{V_A}\frac{\rho^*(t,{\bm x}')v^2(t,{\bm x}')}{
      |{\bm x}-{\bm x}'|}d^3x'    \,,                  \label{12.13}
\en\be
\Psi_{A2}(t,\nnx)=G\int\limits_{V_A}\frac{\rho^*(t,{\bm x}')h(t,{\bm x}')}{
      |{\bm x}-{\bm x}'|} d^3x'                            \label{12.14a}
\en\be
   \Psi_{A3}(t,\nnx)=G\int\limits_{V_A}\frac{\rho^*(t,{\bm x}')\varphi(t,{\bm x}')}{
      |{\bm x}-{\bm x}'|} d^3x'                            \label{12.14z}
\en\be
   \Psi_{A4}(t,\nnx)=G\int\limits_{V_A}\frac{\rho^*(t,{\bm x}')\Pi(t,{\bm x}')}{
      |{\bm x}-{\bm x}'|}d^3x'    \,,                  \label{12.15}
\en\be
\Psi_{A5}(t,\nnx)=G\int\limits_{V_A}\frac{\pi^{kk}(t,{\bm x}')}{
      |{\bm x}-{\bm x}'|}d^3x' \,,                \label{12.16}
\en
where $h(t,{\bm x})=h_{ii}(t,{\bm x})$. Potential $\chi$ is determined as a particular
solution of the inhomogeneous equation \be
   \nabla^2\chi=-2U\,,                                      \label{12.18}
\en with the right side defined in a whole space. Nevertheless, its solution given by equation (\ref{12.12}) has a compact support inside the volumes
of the bodies of the solar system \ct{fockbook,willbook}. It is worthwhile to emphasize that all integrals
defining the metric tensor in the global coordinates are taken over the hypersurface
of constant coordinate time $t$. Any space-time transformation changes the time hypersurface, hence, transforming the corresponding integrals.

Notice that the Newtonian gravitational potential $U(t,{\bm x})$ appears in the solution of the field equations for both the scalar field - equation (\ref{12.5}), and the time-time component of the metric tensor - equation (\ref{12.6}). It does not mean that the time-time component of the metric tensor is a scalar - they are equal only in the Newtonian approximation. This remark is important for making correct post-Newtonian transformation of the functions entering solutions of other field equations since the scalar field and the metric tensor transform differently and should not be confused \ct{kovl}.

\subsubsection{The Post-Newtonian Conservation Laws}\label{pncs}

The laws of conservation allow us to
formulate the post-Newtonian definitions of mass, the center of mass,
the linear and the angular momenta for the whole solar system, which are
crucial in mathematical derivation of equations of motion of the bodies
comprising the system. We employ the mathematical technique relied upon the concept of Landau-Lifshitz pseudo-tensor \cite{lali,mtw} and extended to the scalar-tensor theory by Nutku \cite{1969ApJ...158..991N}. To this end, it is convenient to recast the field equations
(\ref{10.2}) to the form
\br
 \Theta^{\mu\nu} &\equiv& (-g)\frac{\phi}{\phi_0}\Bigl[c^2 T^{\mu\nu}+  t^{\mu\nu}\Bigr]
   =
    \frac{c^4}{16\pi\phi_0}\Bigl[(-g)\phi^2                      \label{13.32}
   (g^{\mu\nu}g^{\alpha\beta}-
    g^{\mu\alpha}g^{\nu\beta})\Bigr]_{,\alpha\beta}\,,
\er where $t^{\mu\nu}$ is an analog of the Landau-Lifshitz
pseudo-tensor of the gravitational field in the scalar-tensor
theory of gravity. This pseudotensor is defined by the following
equation \cite{1969ApJ...158..991N}
\br
\label{13.33}
    t^{\mu\nu}&= & \frac{c^4}{16\pi}\,\frac{\phi^3}{\phi_0^2}\,
    \tau^{\mu\nu}_{LL} +\frac{c^4}{16\pi}
    \frac{2\theta(\phi)+3}{\phi}
    \Bigl(\phi^{,\mu}\phi^{,\nu}-
    \frac{1}{2}g^{\mu\nu}\phi_{,\lambda}\phi^{,\lambda}
    \Bigr)\;,
\er where ${\tau}^{\mu\nu}_{LL}$ is the standard Landau-Lifshitz
pseudotensor \cite{lali,mtw} expressed in terms of the conformal metric
tensor, $\tilde{g}_{\alpha\beta}$, and its first derivatives.
The conservation laws are now obtained from equation (\ref{13.32}) by taking a derivative. One obtains
\be
  \Theta^{\mu\nu}{}_{,\nu}\equiv                           \label{13.34}
  \Bigl[(-g)\frac{\phi}{\phi_0}(c^2 T^{\mu\nu}+t^{\mu\nu})
  \Bigr]_{,\nu}=0\,,
\en
where the right side is zero because of anti-symmetry of the right
side of equation (\ref{13.32}) with respect to indices $\nu$ and
$\alpha$.

We focus on the laws of conservation in the first
post-Newtonian approximation and neglect the energy,
linear and angular momenta taken away from the solar system by
gravitational waves \cite{1992CQGra...9.2093D}. For this reason, the conserved
mass $\mathbb{M}$,
the linear momentum $\mathbb{P}^i$, and spin  $\mathbb{S}^i$ of the solar system are
defined as
\br
  \mathbb{M}&=&\epsilon^2\int_{R^3}\Theta^{00}\,d^3x\;,                       \label{13.35}
\\\nonumber\\
\mathbb{P}^i&=&\epsilon\int_{R^3}\Theta^{0i}\,d^3x\,,                      \label{13.36}
\\\nonumber\\
\mathbb{S}^i&=&\epsilon\int_{R^3}\varepsilon^i_{\;jk}w^j\Theta^{0k}\,d^3x\,,                      \label{13.361}
\er
where the integration is performed over the whole space $R^3$, that is the hypersurface of constant global coordinate time $t$. Let us remark that the integrals
are finite, since in the first post-Newtonian approximation, $\Theta^{00}$ and $\Theta^{0i}$ are of the order of $O(r^{-4})$ for large $r$ \cite{mtw}.
Moreover, in this approximation the domain of integration is reduced to the volume of the bodies comprising the solar system because the functions in the integrands of equations (\ref{13.37})--(\ref{13.38}) have compact support only inside the bodies. Taking into account the asymptotic behavior of $\Theta^{00}$
one can prove that the linear
momentum $\mathbb{P}^i$ can be represented as a time derivative of function
\be
\mathbb{D}^i=\epsilon^2\int_{R^3}\Theta^{00}x^i\,d^3x\;,                                \label{13.35a}
\en
that is
interpreted as the integral of the center of mass. Hence,
\be\label{13.35b} \mathbb{D}^i(t)=\mathbb{P}^i\,t+\mathbb{K}^i\;,\en where $\mathbb{K}^i$ is a constant
vector defining displacement of the solar-system barycenter
from the origin of the global coordinate frame. One can always chose
$\mathbb{K}^i=0$ and $\mathbb{P}^i=0$ making $\mathbb{D}^i=0$ as well.

Direct calculation of the pseudotensor (\ref{13.33}) brings about the post-Newtonian conserved quantities in explicit form
\br
   \mathbb{M}&=&\int_{R^3}\rho^*\left[
   1+\epsilon^2\left(\Pi+\frac{v^2}{2}-\frac{U}{2}\right)\right]\,d^3x    \label{13.37}
   +O(\epsilon^4)\,,
\\\nonumber\\                                                               \label{13.37a}
\mathbb{D}^i&=&\int_{R^3}\rho^*x^i\left[
   1+\epsilon^2\left(\Pi+\frac{v^2}{2}-\frac{U}{2}\right)\right]\,d^3x
   +O(\epsilon^4)\,,
\\\nonumber\\
   \mathbb{P}^i&=&\int_{R^3} \Biggl\{
   \rho^*v^i \left[1+\epsilon^2\left(\Pi+                    \label{13.38}
   \frac{v^2}{2}-\frac{U}{2}\right)\right]+
   \epsilon^2\pi^{ik}v^k-\frac{\epsilon^2}{2}
   \rho^*W^i\Biggr\}\,d^3x+O(\epsilon^4)\,,
\er
where by definition
\be
   W^i(t,\nnx)=G\int_{R^3}\frac{                                 \label{13.39}
   \rho^*(t,\nnx'){\bm v'}\cdot(\nnx-\nnx')(x^i-x'^i)}{|{\bm x}-{\bm x}'|^3}\, d^3x'\,. \en
Position $\mathbb{X}^i$ of the center of mass (barycenter) of the solar system
is defined as $\mathbb{X}^i=\mathbb{D}^i/\mathbb{M}$. Because the total mass, $\mathbb{M}$, of the solar system is constant, position of the solar-system barycenter is fixed in the global coordinates by equating $\mathbb{P}^i=0$, and $\mathbb{D}^i=0$, which makes $\mathbb{X}^i=0$ at any instant of time $t$. This means that the center of mass
of the solar system always coincides with the origin of the
global reference frame as we discussed above.

\subsection{The Earth-Moon Barycentric Frame}\l{ems}

\subsubsection{The Boundary Conditions and Dynamic Properties}\l{pvz1}

Earth and Moon makes a close binary system (a sub-system of the solar system) moving around the barycenter of the solar system. It is convenient to describe orbital motion of Earth and Moon with respect to the local coordinates $w^\a=(cu,w^i)$, which we shall call the Earth-Moon barycentric (EMB) frame.
The EMB frame is constructed in
the neighborhood of the world line of the center of mass of the Earth-Moon sub-system, which
precise definition will be given later along with the derivation
of its equation of motion. The EMB frame was not substantiated explicitly in the numerical construction of ephemerides of Moon. However, the advantage of the EMB frame is that its explicit analytic construction will allow us to clearly decouple the orbital motion of Moon and Earth from the motion of the Earth-Moon barycenter around Sun.

The EMB frame is not
asymptotically Minkowskian as radial distance goes far away from the Earth-Moon sub-system. This  is because solution of the field equations in the EMB frame must smoothly match with the background gravitational field of external bodies - Sun and other planets, which affects the relative motion of Moon around Earth by means of tidal forces. The tidal gravitational field of Sun and other planets makes significant impact on the orbital evolution of the Earth-Moon system, and can not be neglected. Gravitational potential of the tidal force is represented by a Taylor polynomial with respect to the local spatial
coordinates with time-dependent coefficients, which are called the external (tidal)
multipoles \cite{thor,th,k88}. In the Newtonian approximation this polynomial yields a solution of the Laplace equation and starts from the second order (quadratic) term because monopole and dipole external multipoles are usually eliminated by transformation to a freely-falling local coordinates as they are not physically associated with
the tidal force.
In general relativity this monopole-dipole effacing property
of the external gravitational field is retained in the post-Newtonian approximation as a consequence of Einstein's equivalence principle (EEP)
\cite{mtw,lali,willbook}. In particular, EEP suggests that it is always possible to chose local coordinates in such a way that all
first derivatives of the metric tensor (the
Christoffel symbols) vanish along a geodesic world line of a freely falling particle \cite{1978PhRvD..17.1473N}.
In general relativity EEP is also valid for a self-gravitating body moving in external
gravitational field of other bodies \cite{k88,dsx2,breuer}.

In contrast to general relativity the scalar-tensor theory of gravity has a long-range scalar field that can not be eliminated by a coordinate transformation to a locally-inertial frame being at free fall. This is because the scalar field does not change its numerical value under pointwise coordinate transformations and, hence, can not vanish at a point on space-time manifold. The non-vanishing scalar field couples with the intrinsic gravitational field of an extended body and
affects the local characteristics like the gravitational mass of the body. This exemplifies a mechanism of possible violation of EEP discussed by Dicke \ct{1962PhRv..126.1875D,1965AnPhy..31..235D} and Nordtvedt \ct{1973PhRvD...7.2347N}, and known in the PPN formalism as the Nordtvedt effect \cite{willbook}. EEP possible violation is difficult for LLR and other kind of observations because the scalar field does not interacts directly with the measuring instruments but only with the gravitational field. It is hidden in the definition of the universal gravitational constant and reveals itself explicitly only in the first post-Newtonian effects in the equations of motion of the solar system bodies.

We demand that the origin of the local EMB frame
coincides with the center of mass of the Earth-Moon system at any instant of time.
This requires a precise post-Newtonian definition of the center of mass of a sub-system consisting of self-gravitating bodies that is a part of the solar system. Newtonian definition of the Earth-Moon barycenter is straightforward (see, for instance, \cite{brum} and equation (\ref{mse}) of this paper) and used in the analytical theories of the lunar motion in the barycentric coordinates. Post-Newtonian corrections make the concept of the Earth-Moon barycenter more involved and to some extent uncertain because of the gravitational interaction of the Earth-Moon system with the external gravitational field of Sun and other planets. This uncertainty was first noticed by Thorne and Hartle \cite{th}.

The other kind of problem in the scalar-tensor theory of gravity is that mathematically inadequate definition of the center of mass of the Earth-Moon system can bring about fictitious forces that will compel it to move with respect to the origin of the EMB frame. Because the scalar-tensor theory of gravity does not violate the law of conservation of the linear momentum for isolated astronomical systems\ct{1969ApJ...158..991N,willbook}, this motion have no an underlying physical cause and can be removed after making corresponding correction in the post-Newtonian definition of the center of mass. Calculation, which we did in \cite{kovl}, confirms this idea and reveals that the dipole moment $\mathbb{D}^i$ given by equation (\ref{13.37a}), where the integration over $R^3$ is replaced with that over the volumes of Earth and Moon only, gives the definition of the Earth-Moon center of mass that excludes the nonphysical motion of the Earth-Moon barycenter with respect to the origin of the EMB frame.

Another complication in the definition of the origin of the EMB frame is caused by the finite size of the Earth-Moon system making its intrinsic quadrupole moment coupled with the (octupole) tidal gravitational field of Sun and other planets. This coupling exists already in the Newtonian approximation and is a well-established property of gravitationally-interacting extended bodies \ct{dixon1,th}. If one assumes that the origin of the EMB frame moves along a geodesic world line, the quadrupole-octupole coupling makes the second time derivative of the dipole moment of the Earth-Moon system be not equal to zero, that is the Earth-Moon barycenter moves with acceleration.  It means that the assumption of the geodesic motion of the origin of the EMB frame introduces a local gravitational force exerted on the Earth-Moon barycenter, which prevents the Earth-Moon linear momentum (the first time derivative of the dipole moment) to be conserved in such local frame. Nevertheless, mathematics tells us that one can retain the origin of the local EMB frame at the Earth-Moon barycenter if one can make the second time derivative of the dipole moment $\mathbb{D}^i$ equal to zero. This can be done \ct{k88} if the origin of the EMB frame is chosen to move along the accelerated world line of the Earth-Moon barycenter. The acceleration is specified by the external dipole moment $Q^i$ in the multipolar expansion of the homogeneous solution of the field equations (see equation \ref{5.8}) that corresponds physically to the force of inertia in the local EMB frame. The overall procedure of the post-Newtonian determination of the origin of the EMB frame and the Earth-Moon barycenter may look complicated but if it is not applied properly, the equations of the relative motion of Moon around Earth become cluttered with spurious, coordinate-dependent terms having no physical meaning.

We postulate that the local EMB frame is dynamically non-rotating. It
means that orbital equations of motion of Earth and Moon written down in the
local coordinates do not include the Coriolis and centrifugal forces. However, the post-Newtonian nature of the gravitational interaction suggests that spatial axes of the dynamically non-rotating EMB frame must rotate (precess) in the kinematic sense with respect to the spatial axes of the global SSB frame. This kinematic rotational drift of the EMB frame includes three components that are called respectively de-Sitter (geodetic), Lense-Thirring (gravitomagnetic), and Thomas precessions \cite[]{mtw}. The rate of each precession is derived from the law of the Fermi-Walker transport \cite{mtw} of a vector of the orbital momentum of the Earth-Moon system as it moves around Sun. International Astronomical Union (IAU) recommends kinematically non-rotating frames for practical applications. It means that the data processing algorithms that involve the local frames, must have their spatial axes being anchored to ICRF \cite{iau2000}. We notice, however, that any kinematically non-rotating local frame must include the angular speed of its dynamic rotation (the above-mentioned relativistic precession with opposite sign) to the metric tensor. Including such terms to the local EMB frame makes no sense from theoretical point of view as it complicates the post-Newtonian equations of lunar motion. For this reason, we do not follow the IAU resolution in this respect.

\subsubsection{The Metric Tensor and Scalar Field}\label{qop}

We denote the EMB local coordinates by
$w^\alpha=(w^0,w^i)=(cu, w^i)$ where $u$ stands for the local
coordinate time. We are looking for the solution of the field
equations (\ref{11.29})--(\ref{11.32}) inside a world tube surrounding world lines of the centers of mass of Earth and Moon and spreading up to the nearest external bodies. Formally, these are Mars and Venus, but the domain of the EMB local coordinates can be extended later on, if necessary \ct{bk89,k89d}. The matter inside the region covered by the EMB frame, is that of Earth and Moon. Thus, the right side of equations (\ref{11.29})--(\ref{11.32}) should include only the energy-momentum tensor of Earth and Moon.

Functions in a partial solution of the inhomogeneous field equations related to the Earth-Moon system alone, will be labeled by index ({\rm int}) standing for ``internal". These functions describe solution of the internal problem of the post-Newtonian celestial mechanics. Other functions describing a general solution of the homogeneous field equations, characterize the gravitational and scalar field of external bodies and are labeled by index ({\rm ext}) standing for ``external". These functions are given by polynomials in powers of the spatial coordinates. Because the field equations are non-linear, a third group of functions in their solution will appear. These functions describe gravitational mix (coupling) between the internal and external fields, and they will be labeled by index $(\rm mix)$.

Solution of the filed equations
(\ref{11.29})--(\ref{11.32}) is a linear combination of a particular
solution of the inhomogeneous equation and a general
solution of the homogeneous equation. In order to distinguish these solutions from the corresponding solutions of the field equations found in the global SSB frame, we put a hat on any function that is expressed in EMB coordinates. This is because one and the same mathematical function has different forms when expressed in different coordinates. For example, for any scalar function $F(x)$ and a coordinate transformation $x=x(w)$ one has: $F(x)=F[x(w)]\equiv\hat{F}(w)$. It would be mathematically inconsistent to write $F(w)$ instead of $\hat{F}(w)$ because $F(w)$ differs from $F(x)$ \cite{dfn,eisen,arno}.

Accounting for these remarks, solution for a
scalar field in the EMB local coordinates is written as a sum of two terms
\be
\nxi=\,\nnbb{{\rm int}}{u}{w}+\nnbb{\rm ext}{u}{w}\;,                                 \label{1.1}
\en
whereas the EMB perturbation \be\l{nzim}\hat{h}_{\mu\nu}(u,{\bm w})=\hat{g}_{\mu\nu}(u,{\bm w})-\eta_{\mu\nu}\en of the metric tensor $\hat{g}_{\mu\nu}(u,{\bm w})$ is given as a sum of three terms
\be                                                                 \label{1.2}
\nag{\mu\nu}=\nnfa{\rm int}{\mu}{\nu}{u}{w}+
\nnfa{\rm ext}{\mu}{\nu}{u}{w}+\nnfa{\rm mix}{\mu}{\nu}{u}{w}\;,
\en
where terms with index ({\rm int}) refer to the Earth-Moon system and describe the internal solution of the inhomogeneous field equations, terms with index ({\rm ext}) refer to the external bodies (Sun and other planets) and describe the background (external) solution of the homogeneous field equations, and terms with index $(\rm mix)$ are due to the non-linearity of the gravity field equations for the metric tensor. In the first post-Newtonian approximation the mixed terms appear only in $\hat g_{00}(u,\nnw)$ component of the metric tensor.
\\
\paragraph{Internal and External Solutions for the Scalar Field.\\\\}

Equation (\ref{11.29}) gives internal, $\nnbb{\rm int}{u}{w}$, and external, $\nnbb{\rm ext}{u}{w}$, solutions for the scalar field in the following form
\br                                                             \label{1.7}
\nnbb{\rm int}{u}{w}&=&U^{(\mathrm{int})}(u,{\bm w})\;,\\\nonumber\\
  \nnbb{\rm ext}{u}{w}&=& \nn{l=0}\frac1{l!} P^Lw^L\;.        \label{1.7a}
\er
Here $P^L\equiv P^L(u)$ are external symmetric trace-free (STF) multipole moments in the multipolar decomposition of the scalar
field generated by Sun and other planets. These external multipoles
are functions of the local time $u$ only. The internal solution $\nnbb{\rm int}{u}{w}$ describes the scalar field, which is generated by Earth and Moon only. It is expressed in terms of the Newtonian gravitational potential ${U}^{(\mathrm{int})}(u,{\bm w})$ that is defined in next section by equations (\ref{129z}),(\ref{1.11}).

A subtle point of notations should be discussed here. By definition, the scalar field is invariant under coordinate transformation $x^\a=x^\a(w^\b)$, that is the equality $\varphi(x)=\varphi[x(w)]=\hat\varphi(w)$ holds exactly in any post-Newtonian approximation. On the other hand, the Newtonian potentials $U(x)$ and $U(w)$ are not scalar fields, that is $U(x)\not=U(w)$ in all post-Newtonian approximations. For this reason, equations (\ref{12.5}) and (\ref{1.7}) may look self-contradictory because their left sides must be exactly equal by definition, while the right sides are given in terms of the Newtonian potentials $U(x)$ and $U(w)$, which are not exactly equal to each other. This "paradox" is solved if one notices that the transformation from the SSB frame to the EMB frame is represented as a post-Newtonian series $x^\a=w^\a+\epsilon^2 (\mbox{post-Newtonian terms})+O(\epsilon^4)$, which means that if one neglects the post-Newtonian terms in the coordinate transformation, then, $U(x)=U(w)$. The difference between $U(x)$ and $U(w)$ will emerge only if the post-Newtonian terms are taken into account in transformation of the Newtonian potential from one frame to another. This remark also indicates that the scalar filed $\varphi$ can not be expressed only in terms of the Newtonian potential in any post-Newtonian approximation. Hence, equations (\ref{12.5}) and (\ref{1.7}) must be replaced to a more general form in accordance with the exact equation (\ref{10.4}) for the scalar field. This, more exact form of the scalar field is not required in the present paper as the scalar field directly perturbs the metric tensor beyond the Newtonian approximation and equations (\ref{12.5}) and (\ref{1.7}) are sufficient.
\\
\paragraph{Internal Solution for the Metric Tensor.\\\\}

The boundary conditions imposed on the internal solution for the metric tensor in EMB frame are identical with those given in equations (\ref{12.1}), (\ref{12.2}). For this reason the internal solution for the metric tensor has the same form as in the global coordinates, but all quantities must be referred now only to Earth and Moon. We obtain
\br
\label{1.8}
\hat{h}_{00}^{({\rm int})}(u,\nnw)&=&
2 U^{(\mathrm{int})}(u,{\bm w})\;,
\\\nonumber\\                \label{1.8aa}
\hat{l}_{00}^{({\rm int})}(u,\nnw)&=&2{\Psi}^{(\mathrm{int})}(u,\nnw)-
2(\beta-1)\left[\nnbb{\rm int}{u}{w}\right]^2-2\bigg[{U}^{(\mathrm{int})}(u, {\bm w})\bigg]^2-\frac{\partial^2{{\chi}^{(\mathrm{int})}}(u,\nnw)}{\partial u^2}\;,
    \\\nonumber\\                                                           \label{1.9}
\hat{h}_{0i}^{({\rm int})}(u,\nnw)&=& -2(1+\gamma)
{U}^{i}_{(\mathrm{int})}(u,\nnw)\;,
 \\\nonumber\\                                                              \label{1.10}
 \hat h_{ij}^{({\rm int})}(u,\nnw)&=&2\gamma\delta_{ij}{U}^{(\mathrm{int})}(u,{\bm w})\;, \er
where
\begin{eqnarray}
  \label{1.12a}
  {\Psi}^{(\mathrm{int})}(u,\bm{w}) & = & \bigg(\gamma+\frac{1}{2}\bigg){\Psi}^{(\mathrm{int})}_1(u,\bm{w})-\frac{1}{6}{\Psi}^{(\mathrm{int})}_2(u,\bm{w})+(1+\gamma-2\beta){\Psi}^{(\mathrm{int})}_3(u,\bm{w})\nonumber\\
  &&+{\Psi}^{(\mathrm{int})}_4(u,\bm{w})+\gamma{\Psi}^{(\mathrm{int})}_5(u,\bm{w}),
\end{eqnarray}
the internal potentials
\begin{eqnarray}
  \label{129z}
  {U}^{(\mathrm{int})}(u,\bm{w})={U}_{\sss E}(u,\bm{w})+{U}_{\sss M}(u,\bm{w}),& \quad & {U}^{i}_{(\mathrm{int})}(u,\bm{w})={U}^i_{\sss E}(u,\bm{w})+{U}^i_{\sss M}(u,\bm{w}),\nonumber\\
  {\Psi}^{(\mathrm{int})}_k(u,\bm{w})={\Psi}_{Ek}(u,\bm{w})+{\Psi}_{Mk}(u,\bm{w}), & \quad & {\chi}^{(\mathrm{int})}(u,\bm{w})={\chi}_{\sss E}(u,\bm{w})+{\chi}_{\sss M}(u,\bm{w}),
\end{eqnarray}
and the indices $E$ and $M$ represent Earth and Moon respectively.

All these functions are defined as integrals over volumes of Earth and Moon. For example, for Earth (index $E$) one has \br
{U}_{\sss E}(u,\nnw)&=&G\int\limits_{V_{\sss E}}\frac{\rho^*(u,{\bm w}')}{
      |{\bm w}-{\bm w}'|}d^3w'\;,                                    \label{1.11}
  \\\nonumber\\
  {U}^{i}_{\sss E}(u,\nnw)&=&G\int\limits_{V_{\sss E}}\frac{\rho^*(u,{\bm w}')\nu^i(u,{\bm w}')}{
      |{\bm w}-{\bm w}'|}d^3w'\;,
 \\\nonumber\\
 {\chi}_{\sss E}(u,{\bm w})&=&            \label{1.13}
          -G\int\limits_{V_{\sss E}}\rho^*(u,{\bm w}')
      |{\bm w}-{\bm w}'|d^3w'\;,\er
where
\br
{\Psi}_{E1}(u,{\bm w})&=&G\int\limits_{V_{\sss E}}\frac{\rho^*(u,{\bm w}')\nu^2(u,{\bm w}')}{
      |{\bm w}-{\bm w}'|}d^3w'\;,                           \label{1.14}
  \er\br
  {\Psi}_{E2}(u,{\bm w})&=&G\int\limits_{V_{\sss E}}\frac{\rho^*(u,{\bm w}') \hat{h}^{(\mathrm{int})}(u,{\bm w}')}{
      |{\bm w}-{\bm w}'|}d^3w'\;,     \label{1.15}
  \er\br
  {\Psi}_{E3}(u,{\bm w})&=&G\int\limits_{V_{\sss E}}\frac{\rho^*(u,{\bm w}')\nnbb{\rm int}{u}{w'}}{
      |{\bm w}-{\bm w}'|}d^3w'\;,     \label{1.15w}
  \er\br
  {\Psi}_{E4}(u,{\bm w})&=&G\int\limits_{V_{\sss E}}\frac{\rho^*(u,{\bm w}')\Pi(u,{\bm w}')}{
      |{\bm w}-{\bm w}'|}d^3w'\;,                             \label{1.16}
  \er\br
  {\Psi}_{E5}(u,{\bm w})&=&G\int\limits_{V_{\sss E}}\frac{\pi^{kk}(u,{\bm w}')}{
      |{\bm w}-{\bm w}'|}d^3w'\;,                           \label{1.17}
 \er
 $\hat{h}^{(\mathrm{int})}=\hat{h}^{(\mathrm{int})}_{ii}$, the symbol $\nu^i=dw^i/du$ is velocity of Earth's or Moon's matter with respect to the origin of the EMB frame, and all gravitational potentials are taken over Earth's volume denoted respectively as $V_{\sss E}$. Gravitational potentials for Moon are given by similar equations with the integrals performed over the volume of Moon. Formally, one takes equations (\ref{1.14})--(\ref{1.17}) and replace index $E$ for index $M$. It is worth to emphasize that the integrals given in this section, are taken over the hypersurface of the coordinate time $u$. It does not coincide with the hypersurface of the coordinate time $t$, which is used in the integrals defining the gravitational potentials in  the SSB frame. This remark is important for making post-Newtonian transformations of the potentials from one frame to another \cite[section 8.3.1]{kovl}.

The internal terms of the metric tensor in the local EMB frame given by equations (\ref{1.7}), (\ref{1.8})--(\ref{1.10})
must obey the gauge condition (\ref{11.3}) with the external part of the metric being excluded. It yields
\be
\frac{\partial {U}^{(\mathrm{int})}(u,{\bm w})}{\partial u}+               \label{1.19}
         \frac{\partial {U}^{i}_{(\mathrm{int})}(u,{\bm w})}{\partial w^i}=O(\epsilon^2)\,.  \en
This is the only condition, which relates the potentials of the internal (particular) solution of the inhomogeneous field equations in the first post-Newtonian approximation, if one neglects the influence of the external bodies.
We note that equation (\ref{1.19}) is satisfied by the equation of continuity (\ref{11.20}).
\\
\paragraph{External Solution for the Metric Tensor.\\\\}\label{mtex}

Solution of the homogeneous field equations for the metric tensor in EMB frame gives the inertial force exerted on the Earth-Moon barycenter and the tidal gravitational field of Sun and other planets in terms of the external STF tensors \cite{kovl}. Boundary conditions imposed on the external solution should make it convergent on the world line of the origin of the EMB frame, that is at the point with local coordinates ${\bm w}=0$. On the other hand, the external solution should match the tidal gravitational field of Sun and other planets as radial distance $r=|{\bm w}|$  grows outward from the origin of the EMB frame. These boundary conditions are typical for construction of the local frames on curved background space-time \ct{1978PhRvD..17.1473N,th,1986PhRvD..34.3617S,bk89,dsx1,kovl}.

Explicit form of the most general external solution for the linearized metric tensor perturbation in local coordinates is given by \ct{kovl}
\br
\label{1.8a}
\hat{h}_{00}^{({\rm ext})}(u,{\bm w})&=& 2\nn{l=0}
                     \frac{1}{l!}Q^Lw^L+C^2 w^2-C^pC^q w^p w^q  \;,
    \\\nonumber\\\nonumber\\                                            \label{1.8aaa}
\hat h_{0i}^{({\rm ext})}(u,{\bm w})&=&\varepsilon_{ipq}C^pw^q+\sum_{l=2}^{\infty}
          \frac{1}{l!}\varepsilon_{ipq}C^{pL-1}w^{<qL-1>}+\sum\limits_{l=0}^{\infty}\frac{1}{l!}Z^{iL}w^{L}+
          \sum\limits_{l=0}^{\infty}\frac{1}{l!}
           S^Lw^{<iL>}\;,
 \\\nonumber\\\nonumber\\                                               \label{1.10a}
\hat h_{ij}^{({\rm ext})}(u,{\bm w})&=&
          2\delta_{ij}\sum\limits_{l=0}^{\infty}
          \frac{1}{l!}A^Lw^{L}+\sum\limits_{l=0}^{\infty}
          \frac{1}{l!}B^Lw^{<ijL>}+\frac13\left(\delta_{ij}C^2-C^iC^j\right)w^2
                     \\\nonumber
          &&+\sum\limits_{l=1}^{\infty}\frac{1}{l!}
          \Biggl(
          D^{iL-1}w^{<jL-1>}+\varepsilon^{ipq}E^{pL-1}w^{<jqL-1>}
          \Biggr)
          ^{{\rm sym}(ij)}                            \\\nonumber&&
          +\sum\limits_{l=2}^{\infty}\frac{1}{l!}
          \Biggl(
          F^{ijL-2}w^{L-2}+\varepsilon^{pq(i}
          G^{j)pL-2}w^{<qL-2>}
          \Biggr)\;,                       \er
where we use the multi-index notation $L\equiv i_1i_2...i_l$, $L-1\equiv i_1i_2...i_{l-1}$, and so on, symbol `${\rm sym}(ij)$' and the round brackets around indices denote symmetry with respect to the indices, for instance,
$[T_{ijL}]^{{\rm sym}(ij)}\equiv T_{(ij)L}=(1/2)[T_{ijL}+T_{jiL}]$. Function $C^i$ in equations (\ref{1.8a})--(\ref{1.10a}) is the angular velocity of kinematic rotation of the local frame with respect to the global
coordinates, and we keep its contribution only up to terms of the order of $O(C^2)$, which is sufficient for our goal. We also assume that $C^i$ has the post-Newtonian order of magnitude being comparable with the rate of the geodetic precession.

The external solution contains monopole terms $Q$ and $A$ entering $\hat{h}_{00}^{({\rm ext})}$ and $\hat h_{ij}^{({\rm ext})}$ respectively. Function $Q$ defines
the unit of measurement of the coordinate time $u$ at the origin of the EMB frame, and function $A$ defines the
unit of measurement of spatial distances. These functions can be chosen arbitrary in accordance with the practice of astronomical measurements and data processing recommended by the IAU \ct{iau2000}. Current practice is to chose these functions as constants to equate the rate of the solar barycentric and geocentric coordinate times \ct{iau2000,2005USNOC.175.....K}, which would secularly diverge due to the orbital motion of Earth around Sun, if one had chosen $Q=0$. In this paper we do not impose any limitation on functions $Q$ and $A$, which are admitted to depend on time $u$.

Physical meaning of the external multipoles $Q^L$ can be understood if one writes down the Newtonian equation of motion of a test particle being at free fall in the EMB frame. This equation is a time-like geodesic so that after calculation of the Christoffel symbols one gets acceleration of the particle
\br                                                               \label{bcn}
\frac{d^2w^i}{du^2}&=&Q_i-\dot Z^i-2\varepsilon_{ijk}C^j
{\nu}^k-\varepsilon_{ijk}\dot C^jw^k+\left(C^2\delta_{ij}-C_iC_j\right)w^j+Q_{ij}w^j+
\sum_{l=2}^\infty\frac{1}{l!} Q_{iL} w^L+O\left(\epsilon^2\right)\;,
\er
where ${\nu}^i\equiv dw^i/du$. The difference between the first two terms in the right
side of this equation, $Q_i-\dot Z^i$, describe the kinematic acceleration of the particle. It is caused by the force of inertia if the origin of the local frame does not move along geodesic. The third term, $2\varepsilon_{ijk}C^j\nu^k$, in the right side of
equation (\ref{bcn}) is the Coriolis acceleration \cite{arno} caused by the coupling of the particle's velocity to the angular velocity $C^i$ of rotation of the local frame. The forth term, $\varepsilon_{ijk}\dot C^jw^k$,
in the right side of equation (\ref{bcn}) is acceleration due to the non-uniform rotation of the local frame.
The fifth term, $\left(C^2\delta_{ij}-C_iC_j\right)w^j$, describes a centrifugal acceleration
of the particle. The sixth term, $Q_{ij}w^j$, is acceleration due to the presence
of the quadrupole tidal gravitational field of Sun and other planets. The last term in the right side
of equation (\ref{bcn}) is the Newtonian tidal acceleration due to the higher-order external multipoles of the gravitational field of Sun and planets. The centrifugal and quadrupole tidal accelerations have similar structure. The difference, however, is that the matrix of the centrifugal acceleration,
$C^2\delta_{ij}-C_iC_j$,  is not trace-free in contrast to the tidal matrix, $Q_{ij}$. In what follows, we shall keep the angular velocity $C^i$ in the metric tensor but after completion of calculations it will be set equal to zero. This is equivalent to the choice of dynamically non-rotating EMB frame. We also postulate without any limitations that the shift function $Z^i=0$. It means that the time axis of the EMB frame is always locally orthogonal to its spatial axes. However, we shall retain the dipole external moment $Q^i$ in our calculations because it will be associated (see the next section) with acceleration of the origin of the EMB frame due to the gravitational coupling of the internal multipoles of the Earth-Moon system with external gravitational multipoles of Sun and other planets.

A set of eleven external STF multipole moments
$P^L$, $Q^L$, $C^L$, $Z^L$, $S^L$, $A^L$, $B^L$, $D^L$, $E^L$, $G^L$
is defined on the world line
of the origin of the EMB local coordinates so that these multipoles are functions of the local coordinate time $u$ only.
Furthermore, the external multipole moments are symmetric and
trace-free (STF) Cartesian tensors with respect to a pair of any indices \cite{thor,bld1986,kovl}.
Imposing four gauge conditions (\ref{11.11}), (\ref{11.12}) on the metric tensor given by equations (\ref{1.8a})--(\ref{1.10a})
reveals that only 7 from 11 external multipole moments are algebraically independent.
Indeed, after imposing the gauge conditions one can immediately eliminate the external multipole moments $B^L,\,E^L,\,D^L,\,S^L$ from
the local metric \cite{kovl}. The remaining multipoles: $P^L$, $Q^L$, $Z^L$, $C^L$, $A^L$, $F^L$, $G^L$ can be constrained by making use of the residual gauge freedom allowed by differential equation (\ref{11.b}), which excludes four other multipoles $A^L$, $F^L$, $G^L$, $Z^L$ \cite{kovl}. We conclude that only three external moments $P^L$, $Q^L$, and $C^L$ have real physical meaning reflecting one degree of freedom for the scalar field and two degrees of freedom for gravitational field of the external bodies. However, it is convenient to preserve some gauge freedom and do not fix the external multipoles $Z^L$ with $l\ge 2$. They can be chosen later to simplify equations of orbital motion of Moon.

After fixing the gauge freedom as indicated above, the external metric tensor assumes the following
form \cite{kovl}
\br\label{1.24b} \hat{h}_{00}^{({\rm ext})}(u,{\bm w})&=&2\nn{l=0}
                     \frac{1}{l!}Q^Lw^{L}\;,
\\\nonumber\\\nonumber\\                                                                    \label{1.25ba}
\hat h_{0i}^{({\rm ext})}(u,{\bm w})&=&\left(\dot A+\frac{\dot Q}{3}+\frac{1-\gamma}{3}\dot P\right) w^i
+\sum\limits_{l=1}^{\infty}
          \frac{1}{l!}\varepsilon_{ipq}C^{pL-1}w^{<qL-1>}
             \\\nonumber
          &+&
          2\sum\limits_{l=1}^{\infty}\frac{2l+1}{(2l+3)(l+1)!}
          \biggl[2\dot Q^L+(\gamma-1)\dot P^L\biggr]w^{<iL>}+\sum\limits_{l=1}^{\infty}\frac{1}{l!}Z^{iL}w^{L}\;,
 \\\nonumber\\\nonumber\\                                               \label{1.26b}
 \hat h_{ij}^{({\rm ext})}(u,{\bm w})&=&2\delta_{ij}\left\{A+\nn{l=1}\frac{1}{l!}\left[Q^L+(\gamma-1)P^L\right]w^{L}\right\}\;,
\er
where the dot above the external multipoles denotes a derivative with respect to time $u$.

Now we can compute the non-linear part of the
external metric tensor by making
use of equation (\ref{11.33}). It is determined up to
a solution of the homogeneous field (Laplace) equation, which is absorbed to
the post-Newtonian terms  (not shown explicitly) in definition of the external multipoles $Q^L$ in equation (\ref{1.24b}). Thus, for the non-linear terms of the metric tensor one obtains
\br \label{1.24c}
\hat l_{00}^{{({\rm ext})}}(u,{\bm w})&=&
-2\left(\nn{l=0}\frac{1}{l!}Q^Lw^{L}\right)^2-
                 2(\beta-1)\left(\nn{l=0}\frac{1}{l!}P^Lw^{L}\right)^2
		 +\nn{l=0}\frac{1}{(2l+3)l!}\Ddot{Q}^Lw^{L}w^2\,,
\er
where a double dot above $Q_L$ denotes a second time derivative with respect to time $u$. We have included the monopole and dipole terms to the non-linear part of the EMB metric tensor for our convenience. In fact, these terms could be excluded from equation (\ref{1.24c}) by means of re-definition of the monopole $Q$ and the dipole $Q^i$. We could also decompose the product of the two sums in equation (\ref{1.24c}) into irreducible pieces and absorb some of the terms so obtained, to the multipoles $Q^L$ ($l\ge 2$). This procedure is, however, redundant at this stage of calculation as it will be done later on in the procedure of finding the post-Newtonian coordinate transformations between the different frames.
\\
\paragraph{The Mixed Terms.\\\\}

The coupling terms in the metric tensor in the EMB local coordinates are given as a
particular solution of the inhomogeneous field equation (\ref{11.33})
with the right side taken as a product of the internal and
external solutions found on previous step of the approximation procedure. Solving equation (\ref{11.33})
yields
\br \label{1.24o} \hat
l_{00}^{(\rm mix)}(u,{\bm w})&=&-2{U}^{(\mathrm{int})}(u,{\bm w})\left\{A+(2\beta-\gamma-1)P+2\nn{l=0}\frac{1}{l!}
\biggl[Q^L+(\beta-1)P^L\biggr]w^{L}\right\}\\\nonumber&&
-2G\nn{l=1}\frac{1}{l!}\Bigl[Q^L+2(\beta-1)P^L\Bigr]\;\int\limits_{V}\frac{\rho^*(u,{\bm w}')w'^L}{|{\bm w}-{\bm w}'|}d^3w'\;,\er
where $V$ denotes a common volume of Earth's and Moon's matter: $\int_V=\int_{V_{\sss E}}+\int_{V_{\sss M}}$.
This completes derivation of the metric tensor in the EMB local coordinates.

\subsubsection{Internal Multipoles of the Earth-Moon System}\label{mdloc}

Had one ignored the tidal gravitational field of Sun and other planets, the Earth-Moon system would be considered as isolated and its gravitational field would be characterized by the sum of the internal potentials of each body defined in equations (\ref{1.11})--(\ref{1.17}). Multipolar decomposition of the metric tensor of an isolated gravitating system residing in asymptotically flat space-time has been well understood and can be found in papers \cite{thor,bld,di}, where the technique of the irreducible Cartesian tensors has been implemented. The technique has been extended to the case of a self-gravitating system embedded to an asymptotically-curved background space-time in \cite{th,dsx1,kovl}. The Earth-Moon system is not fully isolated from the other bodies of the solar system because it interacts with them gravitationally rather strong -- it suffices to recall the large orbital perturbations in the lunar motion caused by Sun \ct{1971pam..conf....1D}. This mutual interaction brings about the mixed terms to the metric tensor in the EMB frame, which bring uncertainty to the post-Newtonian definition of the internal multipole moments of the Earth-Moon system noted by Thorne and Hartle \ct{th}.

There are two options -- either to include or to exclude the contribution of the mixed terms to the internal multipole moments of the Earth-Moon system. Both options are theoretically admissible but one of them has advantage. Straightforward calculations \cite{kovl} prove that equations of motion of the Earth-Moon system can be significantly simplified if the mixed terms are included to the definition of the internal multipole moments. This is the choice we hold on. There are three classes of the internal multipole moments in the scalar-tensor theory of gravity -- active, conformal, and scalar multipoles  \cite{willbook}. Each class has two physically-different types of mass-induced and current-induced multipoles.

The {\it active} STF mass-type multipoles of the Earth-Moon system are defined by equation \cite{kovl}
\begin{eqnarray}\label{1.31}
   {\cal I}^L&= &\int_{V}\sigma
   (u,{\bm w})w^{<L>}d^3w\\\nonumber\\\nonumber &+&
   \frac{\epsilon^2}{2(2l+3)}\left[\frac{d^2}{du^2}
   \int_{V}\sigma
   (u,{\bm w})w^{<L>}w^2d^3w
   -4(1+\gamma)\,\frac{2l+1}{l+1}
   \frac{d}{du}\int_{V}\sigma^i(u,{\bm w})w^{<iL>}d^3w\right]
   \\\nonumber\\\nonumber &-&\epsilon^2
   \int_{V}d^3w\,\sigma(u,{\bm w})
   \left\{A+(2\beta-\gamma-1)P+\sum_{k=1}^{\infty}\frac{1}{k!}\Bigl[Q^{K}+2(\beta-1) P^{K}\Bigr]w^{K}\right\}w^{<L>}\;,\end{eqnarray}
where $V$ denotes the volume occupied by the matter of the Earth-Moon system, the matter current density
\be\label{pz6}
 \sigma^i(u,{\bm w})=\rho^*(u,{\bm w})\nu^i(u,{\bm w})\;,
 \en
and the {\it active} mass density
\be\label{pz3}
\sigma(u,{\bm w})=\rho^*(u,{\bm w})\left\{1+\epsilon^2\left[(\gamma+\frac12)\nu^2(u,{\bm w})+\Pi(u,{\bm w})-(2\beta-1){U}^{(\mathrm{int})}(u,{\bm w})  \right]\right\}+\epsilon^2\gamma\pi^{kk}(u,{\bm w})\;,\en
where $\hat{U}^{(\mathrm{int})}(u,{\bm w})$ is the gravitational potential of the Earth-Moon system given by equation (\ref{129z}).

The {\it conformal} STF mass-type multipoles of the Earth-Moon system are defined as follows \cite{kovl}
\begin{eqnarray}\label{1.34}
  {I}^L&= &\int_{V}\varrho(u,{\bm w})\left\{1-\epsilon^2\left[A+(1-\gamma)P+\sum_{k=1}^{\infty}\frac{1}{k!}Q_{K}w^{K}\right]\right\}
w^{<L>}\,d^3w\\\nonumber\\\nonumber&+&
\frac{\epsilon^2}{2(2l+3)}\left[\frac{d^2}{du^2}
\int_{V}\varrho
(u,{\bm w})w^{<L>}w^2\,d^3w
-\frac{8(2l+1)}{l+1}
\frac{d}{du}\int_{V}\sigma^i(u,{\bm w})w^{<iL>}\,d^3w\right]\;,\end{eqnarray}
with the {\it conformal} mass density of matter
\begin{equation}\label{pz5}
 \varrho=\rho^*(u,{\bm w})\left[1+\epsilon^2\left(\frac32\nu^2(u,{\bm w})+\Pi(u,{\bm w})-{U}^{(\mathrm{int})}(u,{\bm w})\right)\right]+\epsilon^2\pi^{kk}(u,{\bm w})\;.\end{equation}
The {\it conformal} mass density does not depend on the PPN parameters $\beta$ and $\gamma$ as opposed to the definition of the {\it active}mass density.

The {\it scalar} field multipoles, $\bar{I}^L$,
are not independent and are related to the {\it active} and {\it conformal} multipoles with the aid of a linear relationship \cite{kovl}
\br\label{wh1}
\bar{I}^L=2{\cal I}^L-(1+\g){I}^L\;.
\er
The hypersurface of the integration in equations (\ref{1.31}), (\ref{1.34}) is that of
the constant local EMB coordinate time $u$, which does not coincide with the hypersurface of the constant time $t$ in the global SSB coordinates.

In addition to the gravitational mass-type multipoles, ${\cal I}^L$ and ${I}^L$, there is a set of other multipoles, which are called spin multipoles. In the post-Newtonian approximation they are defined by equation \cite{thor,bld,kovl}
\be
   {\cal S}^L= \int_{V}\varepsilon^{pq<a_l} w^{L-1>p}             \label{1.32}
   \sigma^q(u,{\bm w})\, d^3w\,,
\en
where the matter current density $\sigma^q$ is defined in equation (\ref{pz6}). The spin multipoles of the Earth-Moon system are different from zero because both Earth and Moon move with respect to the EMB frame making the matter current density $\sigma^i\not=0$.

\subsection{The Geocentric Frame}\label{grf}

\subsubsection{The Boundary Conditions and Dynamic Properties}\label{bcdp}

The Earth is a self-gravitating, extended body moving around the barycenter of the Earth-Moon system, which in its own turn moves around the barycenter of the solar system. It is convenient to introduce local geocentric frame $X^\a=(cT,X^i)$ to describe rotational motion of Earth and orbital motion of artificial satellites orbiting Earth. It will also help us to decouple the orbital motion of Earth around the Earth-Moon system barycenter from the rotational harmonics (nutation, precession, polar motion) of Earth.

Mathematical principles of the construction of the geocentric reference frame (GRF) are similar to those, which were used in the construction of the local Earth-Moon barycentric frame. However, in case of the GRF the internal problem is solved only for Earth. The external bodies are Sun, Moon, and other planets.

The origin of the GRF is located at Earth's center of mass, and the domain of space covered by the frame incloses the world line of the center of mass of Earth with a spatial cross-section that extends to the orbit of Moon. The GRF is not
asymptotically Minkowskian because of the tidal gravitational field of Sun, Moon, and other planets. This means that the GRF metric tensor diverges as distance from the origin of the GRF grows. At the same time, if one neglects gravitational field of Earth, the remaining part of the GRF metric tensor is reduced to the Minkowskian tensor $\eta_{\a\b}$ on the world-line of the origin of the geocentric frame.

We postulate that the GRF is dynamically non-rotating. It
means that orbital equations of motion of Earth's artificial satellites, referred to the GRF, do not include the Coriolis and centrifugal forces. However, the post-Newtonian force causes spatial axes of the GRF rotate slowly in kinematic sense with respect to spatial axes of the EMB frame, which undergoes relativistic precession with respect to the global SSB frame as well.

\subsubsection{The Metric Tensor and the Scalar Field}\label{qopq}

We denote the geocentric coordinates by
$X^\alpha=(X^0,X^i)=(cT, X^i)$ where $T$ stands for the geocentric
coordinate time. We are looking for the non-vacuum (internal) solution of the field
equations (\ref{11.29})--(\ref{11.33}) inside a domain spreading up from the geocenter to Moon. Thus, the right side of equations (\ref{11.29})--(\ref{11.32}) includes only the energy-momentum tensor of Earth. The scalar field and the metric tensor in the internal solution of the field equations related to Earth only, will be labeled by index ({\rm int}) and a capital letter $E$. Other functions describing the metric tensor and the scalar field of the external bodies are labeled by index ({\rm ext}), and the coupling between the internal and external terms will be labeled by index $(\rm mix)$. The solution of the field equations
(\ref{11.29})--(\ref{11.33}) is a linear combination of a general
solution of the homogeneous equation and a particular
solution of the inhomogeneous equation.

Solution for a
scalar field in the GRF coordinates is denoted as $\Phi$, and has the following structure
\be
{\Phi}(T,{\bm X})={\Phi}^{({\rm int})}(T,{\bm X})+{\Phi}^{({\rm ext})}(T,{\bm X})\;.                                 \label{1.1a}
\en
The GRF metric tensor is denoted as $G_{\mu\nu}(T,{\bm X})$, and its perturbation \be\l{mtp8}
H_{\mu\nu}(T,{\bm X})={G}_{\mu\nu}(T,{\bm X})-\eta_{\mu\nu}\;,
\en
is given in the following form
\be                                                                 \label{1.2a}
H_{\mu\nu}(T,{\bm X})=H^{({\rm int})}_{\mu\nu}(T,{\bm X})+H^{({\rm ext})}_{\mu\nu}(T,{\bm X})+H^{(\rm mix)}_{\mu\nu}(T,{\bm X})\;,
\en
where terms with index ({\rm int}) refer to Earth and describe the internal solution of the inhomogeneous field equations, terms with index ({\rm ext}) refer to the external bodies (Sun, Moon, and other planets) and describe the external (background) solution of the homogeneous part of the field equations, and terms with index $(\rm mix)$ are due to the gravitational coupling of the internal and external solutions.
\\
\paragraph{Internal and External Solutions for the Scalar Field.\\\\}

Equation (\ref{11.29}) gives internal, ${\Phi}^{(\rm int)}$, and external, ${\Phi}^{(\rm ext)}$, solutions for the scalar field in the following form
\br                                                             \label{1.7y}
 {\Phi}^{({\rm int})}(T,{\bm X})&=&U_{\sss E}(T,{\bm X})\;,\\\nonumber\\
 {\Phi}^{({\rm ext})}(T,{\bm X})&=& \nn{l=0}\frac1{l!} P_{\sss E}^LX^L\;.        \label{1.7aa}
\er
Here $P_{\sss E}^L\equiv P_{\sss E}^L(T)$ are external STF multipoles entering the irreducible Cartesian-tensor decomposition of the scalar field generated by Sun, Moon, and other planets. These external multipoles are functions of the local time $T$ only, and are attached to the world line of the origin of the GRF. The internal solution ${\Phi}^{(\rm int)}$ describes the scalar field, which is generated by Earth. It is expressed in terms of the Newtonian gravitational potential of Earth $U_{\sss E}(T, {\bm X})$ that is defined in equation (\ref{1.11a}). Notice again that the laws of the post-Newtonian coordinate transformation for the scalar field ${\Phi}^{({\rm int})}(T,{\bm X})$, the metric tensor $H_{00}^{({\rm int})}(T,{\bm X})$, and the Newtonian potential $U_{\sss E}(T,{\bm X})$ are different because they represent different mathematical objects. This point will be taken into account in the next section.
\\
\paragraph{Internal Solution for the Metric Tensor.\\\\}

The boundary conditions imposed on the internal solution for the metric tensor
are similar with equations (\ref{12.1})--(\ref{12.2}). Solving the inhomogeneous field equations with Earth as a source of the field, yields:
\br
\label{1.8w}
H_{00}^{({\rm int})}(T,{\bm X})&=&
2U_{\sss E}(T,{\bm X})\;,
\\\nonumber\\                \label{1.8s}
L_{00}^{({\rm int})}(T,{\bm X})&=&2\Psi_{\sss E}(T,{\bm X})-
    2(\beta-1)\left[{\Phi}^{({\rm int})}(T,{\bm X})\right]^2-2 U_{\sss E}^2(T,{\bm X})-\frac{\partial^2{\chi_{\sss E}}(T,{\bm X})}{\partial T^2}\;,
    \\\nonumber\\                                                           \label{1.9q}
H_{0i}^{({\rm int})}(T,{\bm X})&=& -2(1+\gamma)
          U_{\sss E}^i(T,{\bm X})\;,
 \\\nonumber\\                                                              \label{1.10q}
 H_{ij}^{({\rm int})}(T,{\bm X})&=&2\gamma\delta_{ij}U_{\sss E}(T,{\bm X})\;, \er
 where all gravitational potentials are taken over Earth's volume denoted in the following equations as $V_{\sss E}$. Specifically, \br
  U_{\sss E}(T,{\bm X})&=&G\int\limits_{V_{\sss E}}\frac{\rho^*(T,{\bm X}')}{
      |{\bm X}-{\bm X}'|}d^3X'\;,                                    \label{1.11a}
  \\\nonumber\\
  U^i_{\sss E}(T,{\bm X})&=&G\int\limits_{V_{\sss E}}\frac{\rho^*(T,{\bm X}')\nu^i_{\sss E}(T,{\bm X}')}{
      |{\bm X}-{\bm X}'|}d^3X'\;,
 \\\nonumber\\
\chi_{\sss E}(T,{\bm X})&=&            \label{1.13a}
          -G\int\limits_{V_{\sss E}}\rho^*(T,{\bm X}')
      |{\bm X}-{\bm X}'|d^3X'\;,\er
\br
\label{1.12aa}
 \Psi_{\sss E}(T,{\bm X})&=&
 (\gamma+\frac12)\Psi_{E1}(T,{\bm X})-\frac{1}{6}\Psi_{E2}(T,{\bm X})\\\nonumber&&+
 (1+\gamma-2\beta)\Psi_{E3}(T,{\bm X})
    +\Psi_{E4}(T,{\bm X}) +\gamma\Psi_{E5}(T,{\bm X})\;,
\er
where
\br
\Psi_{E1}(T,{\bm X})&=&G\int\limits_{V_{\sss E}}\frac{\rho^*(T,{\bm X}')\nu^2_{\sss E}(T,{\bm X}')}{
      |{\bm X}-{\bm X}'|}d^3X'\;,                           \label{1.14a}
  \er\br
  \Psi_{E2}(T,{\bm X})&=&G\int\limits_{V_{\sss E}}\frac{\rho^*(T,{\bm X}')H^{(\mathrm{int})}(T,{\bm X}')}{
      |{\bm X}-{\bm X}'|}d^3X'\;,     \label{1.15a}
  \er\br
\Psi_{E3}(T,{\bm X})&=&G\int\limits_{V_{\sss E}}\frac{\rho^*(T,{\bm X}'){\Phi}^{({\rm int})}(T,{\bm X}')}{
      |{\bm X}-{\bm X}'|}d^3X'\;,     \label{1.15u}
  \er\br
\Psi_{E4}(T,{\bm X})&=&G\int\limits_{V_{\sss E}}\frac{\rho^*(T,{\bm X}')\Pi(T,{\bm X}')}{
      |{\bm X}-{\bm X}'|}d^3X'\;,                             \label{1.16a}
  \er\br
  \Psi_{E5}(T,{\bm X})&=&G\int\limits_{V_{\sss E}}\frac{\pi^{kk}(T,{\bm X}')}{
      |{\bm X}-{\bm X}'|}d^3X'\;,                           \label{1.17a}
 \er
$H^{(\mathrm{int})}=H^{(\mathrm{int})}_{ii}$, the symbol $\nu^i_{\sss E}(T,{\bm X})=dX^i/dT$ is the velocity of Earth's matter with respect to the origin of GRF, and all integrals given in this section, are taken over the hypersurface of a constant value of the coordinate time $T$.

The internal part of the local metric given by equations (\ref{1.7y}), (\ref{1.8w})--(\ref{1.10q})
must obey the gauge condition (\ref{11.3}), which yields
\be
         \frac{\partial U_{\sss E}}{\partial T}+               \label{1.19a}
         \frac{\partial U_{\sss E}^i}{\partial X^i}=O(\epsilon^2)\,,  \en
         where we have neglected contribution of the external part of the metric tensor.
Equation (\ref{1.19a}) is satisfied under this assumption because of the equation of continuity (\ref{11.20}) for Earth's matter.
\\
\paragraph{External Solution for the Metric Tensor.\\\\}\label{mtexa}

The boundary conditions imposed on the external solution of the field equations tells us that it must be convergent on the world line of the origin of the GRF, where ${\bm X}=0$. However, the external solution diverges as the radial distance from the origin of the GRF increases because it should match with the tidal gravitational field of Sun, Moon, and other planets. By making use of the gauge condition and the residual gauge freedom, the external metric tensor can be brought to the following simple form \cite{kovl}
\br\label{1.24ba} H_{00}^{({\rm ext})}(T,{\bm X})&=&2\nn{l=0}
                     \frac{1}{l!}Q_{\sss E}^LX^{L}\;,
\\\nonumber\\\nonumber\\                                                                    \label{1.25i}
H_{0i}^{({\rm ext})}(T,{\bm X})&=&\left(\dot A_{\sss E}+\frac{\dot Q_{\sss E}}{3}+\frac{1-\gamma}{3}\dot P_{\sss E}\right) X^i
          +\sum\limits_{l=1}^{\infty}
          \frac{1}{l!}\varepsilon_{ipq}C_{\sss E}^{pL-1}X^{<qL-1>}
             \\\nonumber
          &+&
          2\sum\limits_{l=1}^{\infty}\frac{2l+1}{(2l+3)(l+1)!}
          \biggl[2\dot Q_{\sss E}^L+(\gamma-1)\dot P_{\sss E}^L\biggr]X^{<iL>}+\sum\limits_{l=1}^{\infty}\frac{1}{l!}Z_{\sss E}^{iL}X^{L}\;,
 \\\nonumber\\\nonumber\\                                               \label{1.26ba}
H_{ij}^{({\rm ext})}(T,{\bm X})&=&2\delta_{ij}\left\{A_{\sss E}+\nn{l=1}\frac{1}{l!}\left[Q_{\sss E}^L+(\gamma-1)P_{\sss E}^L\right]X^{L}\right\}\;,
\er
where $P_{\sss E}^L=P_{\sss E}^L(T)$, $Q_{\sss E}^L=Q_{\sss E}^L(T)$, $C_{\sss E}^L=C_{\sss E}^L(T)$ are STF Cartesian tensors depending only on time $T$ and characterizing gravitational multipoles of the external gravitational field by Sun, Moon, and other planets on the world line of the origin of the GRF, and dot above the multipoles denotes a time derivative with respect to time $T$. External multipoles $Z_{\sss E}^L=Z_{\sss E}^L(T)$ are left free to accommodate the residual gauge freedom remained in the description of the external gravitational field.

The non-linear part of the GRF metric tensor is
\br \label{1.24ca}
L_{00}^{({\rm ext})}(T,{\bm X})&=&
-2\left(\nn{l=0}\frac{1}{l!}Q_{\sss E}^LX^{L}\right)^2-
                 2(\beta-1)\left(\nn{l=0}\frac{1}{l!}P_{\sss E}^LX^{L}\right)^2
                 +\nn{l=0}\frac{1}{(2l+3)l!}\Ddot{Q}_{\sss E}^LX^{L}X^2\,,
\er
where dots above $Q_{\sss E}^L$ denotes time derivative with respect to time $T$.

We have included the monopole $(l=0)$ and dipole $(l=1)$ terms to the non-linear part of the GRF metric tensor for our convenience. These terms could be excluded from equation (\ref{1.24ca}) by means of re-definition of the monopole $Q_{\sss E}$ and the dipole $Q^i_{\sss E}$ in equation (\ref{1.24ba}). We could also decompose the product of the two sums in equation (\ref{1.24ca}) into a single series consisting of irreducible pieces, and absorb some of the terms to the multipoles $Q_{\sss E}^L$ ($l\ge 2$) in equation (\ref{1.24ba}). However, it is more simple to do calculations directly with expression (\ref{1.24ca}).
\\
\paragraph{The Mixed Terms.\\\\}

The coupling terms in the GRF metric tensor are given as a
particular solution of the inhomogeneous field equation (\ref{11.33})
with the right side taken as a product of the internal and
external solutions found on the previous step. It reads
\br \label{1.24oa}
L_{00}^{(\rm mix)}(T,{\bm X})&=&-2U_{\sss E}(T,{\bm X})\left\{A_{\sss E}+(2\beta-\gamma-1)P_{\sss E}+2\nn{l=0}\frac{1}{l!}
\biggl[Q_{\sss E}^L+(\beta-1)P_{\sss E}^L\biggr]X^{L}\right\}\\\nonumber&&
-2G\nn{l=1}\frac{1}{l!}\Bigl[Q_{\sss E}^L+2(\beta-1)P_{\sss E}^L\Bigr]\;\int\limits_{V_{\sss E}}\frac{\rho^*(T,{\bm X}')X'^L}{|{\bm X}-{\bm X}'|}d^3X'\;.\er
This completes derivation of the metric tensor in the local GRF coordinates.

\subsubsection{Gravitational Multipoles of Earth}\label{meoc}

Gravitational field of Earth is described in the GRF in terms of the internal potentials $U_{\sss E}$, $ U^i_{\sss E}$, $ \Psi_{\sss E}$, etc., defined in equations (\ref{1.11a})--(\ref{1.17a}). Multipolar decomposition of the metric tensor of Earth is given in terms of the irreducible STF Cartesian tensors.
The {\it active} STF mass-type multipole moments of Earth are \cite{kovl}
\begin{eqnarray}\label{g1.31}
   {\cal I}^L_{\sss E}&= &\int_{V_{\sss E}}\sigma_{\sss E}
   (T,{\bm X})X^{<L>}d^3X\\\nonumber\\\nonumber &+&
   \frac{\epsilon^2}{2(2l+3)}\left[\frac{d^2}{dT^2}
   \int_{V_{\sss E}}\sigma_{\sss E}
   (T,{\bm X})X^{<L>}X^2d^3X
   -4(1+\gamma)\,\frac{2l+1}{l+1}
   \frac{d}{dT}\int_{V_{\sss E}}\sigma^i_{\sss E}(T,{\bm X})X^{<iL>}d^3X\right]
   \\\nonumber\\\nonumber &-&\epsilon^2
   \int_{V_{\sss E}}d^3X\,\sigma_{\sss E}(T,{\bm X})
   \left\{A_{\sss E}+(2\beta-\gamma-1)P_{\sss E}+\sum_{k=1}^{\infty}\frac{1}{k!}\Bigl[Q_{\sss E}^{K}+2(\beta-1) P_{\sss E}^{K}\Bigr]X^{K}\right\}X^{<L>}\;,\end{eqnarray}
where $V_{\sss E}$ denotes the volume occupied by the matter of Earth, the matter current density
\be\label{gpz6}
 \sigma^i_{\sss E}(T,{\bm X})=\rho^*(T,{\bm X})\nu^i_{\sss E}(T,{\bm X})\;,
 \en
and the {\it active} mass density is defined as
\be\label{gpz3}
 \sigma_{\sss E}(T,{\bm X})=\rho^*(T,{\bm X})\left\{1+\epsilon^2\left[(\gamma+\frac12)\nu^2_{\sss E}(T,{\bm X})+\Pi(T,{\bm X})-(2\beta-1) U_{\sss E}(T,{\bm X})  \right]\right\}+\epsilon^2\gamma\pi^{kk}(T,{\bm X})\;,\en
where $ U_{\sss E}(T,{\bm X})$ is the gravitational potential of Earth given by equation (\ref{1.11a}).
The {\it conformal} STF mass-type multipole moments of Earth is
\begin{eqnarray}\label{g1.34}
  {I}^L_{\sss E}&= &\int_{V_{\sss E}}\varrho_{\sss E}(T,{\bm X})\left\{1-\epsilon^2\left[A_{\sss E}+(1-\gamma)P_{\sss E}+\sum_{k=1}^{\infty}\frac{1}{k!}Q_{\sss E}^{K}X^{K}\right]\right\}
X^{<L>}\,d^3X\\\nonumber\\\nonumber&+&
\frac{\epsilon^2}{2(2l+3)}\left[\frac{d^2}{dT^2}
\int_{V_{\sss E}}\sigma_{\sss E}
(T,{\bm X})X^{<L>}X^2\,d^3X
-\frac{8(2l+1)}{l+1}
\frac{d}{dT}\int_{V_{\sss E}}\sigma^i_{\sss E}(T,{\bm X})X^{<iL>}\,d^3X\right]\;,\end{eqnarray}
with the {\it conformal} mass density of Earth's matter defined as
\begin{equation}\label{gpz5}
 \varrho_{\sss E}=\rho^*(T,{\bm X})\left[1+\epsilon^2\left(\frac32\nu^2_{\sss E}(T,{\bm X})+\Pi(T,{\bm X})-U_{\sss E}(T,{\bm X})\right)\right]+\epsilon^2\pi^{kk}(T,{\bm X})\;.\end{equation}
The {\it conformal} density does not depend on the PPN parameters $\beta$ and $\gamma$.
Scalar multipoles, $\bar{I}_{\sss E}^L$,
are related to the active and conformal multipoles via linear relationship
\br\label{gwh1}
\bar{I}_{\sss E}^L=2{\cal I}_{\sss E}^L-(1+\g){I}_{\sss E}^L\;.
\er
The hypersurface of the integration in equations (\ref{g1.31}), (\ref{g1.34}) is that of
a constant value of the GRF coordinate time $T$, which does not coincide either with the hypersurface of the constant time $u$ in the EMB frame or that of the constant time $t$ of the SSB frame.

The spin multipoles of Earth are defined by equation
\be
   S^L_{\sss E}= \int_{V_{\sss E}}\varepsilon^{pq<a_l} X^{L-1>p}             \label{g1.32}
   \sigma^q_{\sss E}(T,{\bm X})\, d^3X\,,
\en
where the matter current density $\sigma^q_{\sss E}$ is defined in equation (\ref{gpz6}). From their definitions, it is clear that the current-type multipoles are different from zero, if and only if, velocity of matter with respect to the GRF is not zero. Since Earth rotates around its axis, the spin moments for Earth are well-defined.

\subsection{The Selenocentric Frame}\label{lrf}

\subsubsection{The Boundary Conditions and Dynamic Properties}\label{bgro}

From the point of view of space research Moon is a self-gravitating, extended body moving in space around the barycenter of the Earth-Moon system, which, in its own turn, orbits the barycenter of the Solar system. Previous post-Newtonian theories of the lunar motion neglected its mass \ct{brm,vab}. However, the millimeter ranging accuracy of LLR measurements may be sensitive to some post-Newtonian effects associated with the finite value of mass of Moon. For this reason, we are taking it into account. It is convenient to introduce a local selenocentric reference frame (SRF) to describe rotational motion of Moon, motion of a CCR on Moon around center of mass of Moon, and orbital motion of spacecrafts around Moon.
The origin of the SRF should be located at Moon's center of mass -- this will be achieved later by making use of the law of conservation of the linear momentum of Moon in local selenocentric coordinates.  The spatial domain covered by the SRF encloses the world line of the center of mass of Moon and extends to the orbit of Earth.
The SRF is not
asymptotically Minkowskian because the external part of the SRF metric tensor must match with the tidal gravitational field of Sun, Earth, and other planets. This means that the SRF metric tensor diverges as distance from Moon grows. If one neglects the internal part of the SRF metric tensor that describes gravitational field of Moon, the external part of the SRF metric tensor must approach the Minkowskian metric on the world-line of the origin of the SRF.
We postulate that the SRF is dynamically non-rotating. It
means that orbital equations of motion of Moon's artificial satellites, written down in the local selenocentric coordinates, do not include the Coriolis and centrifugal forces. However, the post-Newtonian gravitational interaction of Moon with Earth and the other solar system bodies makes the spatial axes of the SRF slowly rotating in the kinematic sense with respect to the spatial axes of the EMB frame. Dynamically non-rotating SRF is useful for doing the post-Newtonian calculations. However, one has to remember that Moon is tidally locked so that its orbital and rotational motions are synchronized and obey the Cassini laws. For this reason, the post-Newtonian precession of the spatial axes of the SRF should superimpose on physical libration of Moon.

Currently, it is not quite clear whether the Newtonian theory
is sufficient for complete interpretation of the high-precision rotational
data of Moon (and Earth), or the post-Newtonian corrections
should be earnestly taken into account. Several papers \cite{bbf,vok,vokr,brg,ksr1}
pointed out that the relativistic corrections might be important in the rotational
theory of Moon (and Earth). For example, it seems likely that apart from the well-known geodetic precession of the lunar orbit
\cite{wnd}, the SRF undergoes an additional
precession of 28.9 milli-arcsec/century \cite{vok}. This value
is theoretically within the range of
LLR technique attaining precision of 1 millimeter, but the question is how to de-correlate it from other secular selenophysical effects. Existence of the orbit co-rotation 1:1
resonance in the lunar dynamics imposes a specific constrain on
the relativistic libration of Moon making
it hard to observe \cite{vokr}. M\"uller \cite{muelphd} made a significant effort
to incorporate relativistic effects to the rotational-orbital
dynamics of Moon by making use of Thorne-Hartle's paper \cite{th}
and Brumberg-Kopeikin formalism \cite{bk-nc,bk89} and tested
their presence in his LLR software. He had showed that their impact was
fairly small at that time, and removed those terms, because their computation
was time consuming.
However, the millimeter LLR demands to reconsider this
problem at a new theoretical level. Certain progress towards this direction
has been achieved by German-Chinese research group in a series of papers \cite{xws1,xwsk,xws2}. It would be interesting to extend this line of research and to apply it to real LLR data processing.

\subsubsection{The Metric Tensor and the Scalar Field}\label{olas}

We denote the selenocentric coordinates by
$Y^\a=(Y^0,Y^i)=(c\Sigma,Y^i)$ where $\Sigma$ stands for the selenocentric
coordinate time. We are looking for the internal solution of the field
equations (\ref{11.29})--(\ref{11.32}) inside the spatial domain spreading up from the center of mass of Moon to Earth. Thus, the right side of equations (\ref{11.29})--(\ref{11.32}) includes only the energy-momentum tensor of Moon.

 The internal solution of the field equations for the scalar field and the metric tensor relates only to Moon and is labeled by index ({\rm int}). The external solution of the field equations describes the gravitational and scalar field of the external bodies (Earth, Sun, other planets) and is labeled by index ({\rm ext}). The non-linear solution of the field equations describing the gravitational mixing (coupling) of the internal and external solutions will be labeled by index $(\rm mix)$. Solution of the field equations
(\ref{11.29})--(\ref{11.32}) is a linear combination of a general
solution of the homogeneous (the external field) equation and a particular
solution of the inhomogeneous (the internal field) equation. In order to distinguish these solutions from the corresponding solutions of the field equations in the geocentric frame, we put a hat above any function expressed in the SRF coordinates.

Solution for a
scalar field in the SRF coordinates is
\be
\hat{\Phi}(\Sigma,{\bm Y})=\hat{\Phi}^{({\rm int})}(\Sigma,{\bm Y})+\hat{\Phi}^{({\rm ext})}(\Sigma,{\bm Y})\;,                                 \label{1.1ax}
\en
and the SRF perturbation
\be\l{srfp}
\hat{H}_{\mu\nu}(\Sigma,{\bm Y})=\hat{G}_{\mu\nu}(\Sigma,{\bm Y})-\eta_{\mu\nu}\;,
 \en
of the metric tensor $\hat{G}_{\mu\nu}(\Sigma,{\bm Y})$ is given in the form
\be                                                                 \label{1.2ax}
\hat{H}_{\mu\nu}(\Sigma,{\bm Y})=\hat{H}^{({\rm int})}_{\mu\nu}(\Sigma,{\bm Y})+\hat{H}^{({\rm ext})}_{\mu\nu}(\Sigma,{\bm Y})+\hat{H}^{(\rm mix)}_{\mu\nu}(\Sigma,{\bm Y})\;,
\en
where the terms with index ({\rm int}) refer to Moon and describe the internal solution of the inhomogeneous field equations, the terms with index ({\rm ext}) refer to the external bodies (Sun, Earth, and other planets) and describe the external (background) solution of the homogeneous field equations, and the terms with index $(\rm mix)$ are due to the gravitational coupling of the internal and external solutions.
\\
\paragraph{Internal and External Solutions for the Scalar Field.\\\\}

Non-homogeneous equation (\ref{11.29}) yields the internal solution $\hat{\Phi}^{(\rm int)}$ for the scalar field.  Homogeneous part of this equation gives rise to the external solution, $\hat{\Phi}^{(\rm ext)}$,  for the scalar field. They are
\br                                                             \label{1.7yx}
 \hat{\Phi}^{({\rm int})}(\Sigma,{\bm Y})&=&U_{\sss M}(\Sigma,{\bm Y})\;,\\\nonumber\\
  \hat{\Phi}^{({\rm ext})}(\Sigma,{\bm Y})&=& \nn{l=0}\frac1{l!} P_{\sss M}^LY^L\;.        \label{1.7aax}
\er
Here $P_{\sss M}^L\equiv P_{\sss M}^L(\Sigma)$ are the external STF multipoles of the scalar field generated by Sun, Earth, and other planets. These external multipoles are functions of the coordinate time $\Sigma$ only. The internal solution $\hat{\Phi}^{({\rm int})}$ describes the scalar field, which is generated by Moon. It is expressed in terms of the Newtonian gravitational potential of Moon, $U_{\sss M}(\Sigma,{\bm Y})$, that is defined explicitly by equation (\ref{1.11ax}).
\\
\paragraph{Internal Solution for the Metric Tensor.\\\\}\label{obqm}

The boundary conditions imposed on the internal solution for the metric tensor in the SRF
are similar with equations (\ref{12.1}), (\ref{12.2}). Solving the inhomogeneous field equations yields:
\br
\label{1.8wx}
\hat H_{00}^{({\rm int})}(\Sigma,{\bm Y})&=&
2U_{\sss M}(\Sigma,{\bm Y})\;,
\\\nonumber\\                \label{1.8sx}
\hat H_{0i}^{({\rm int})}(\Sigma,{\bm Y})&=&2\Psi_{\sss M}(\Sigma,{\bm Y})-
    2(\beta-1)\left[\hat{\Phi}^{({\rm int})}(\Sigma,{\bm Y})\right]^2-2 U_{\sss M}^2(\Sigma,{\bm Y})-\frac{\partial^2{\chi}_{\sss M}(\Sigma,{\bm Y})}{\partial\Sigma^2}\;,
    \\\nonumber\\                                                           \label{1.9qx}
\hat H_{0i}^{({\rm int})}(\Sigma,{\bm Y})&=& -2(1+\gamma)
          U_{\sss M}^i(\Sigma,{\bm Y})\;,
 \\\nonumber\\                                                              \label{1.10qx}
 \hat H_{ij}^{({\rm int})}(\Sigma,{\bm Y})&=&2\gamma\delta_{ij}U_{\sss M}(\Sigma,{\bm Y})\;, \er
 where all functions in the right side of equations (\ref{1.8wx})--(\ref{1.10qx}) are taken over Moon's volume denoted as $V_{\sss M}$. More specifically, \br
   U_{\sss M}(\Sigma,{\bm Y})&=&G\int\limits_{V_{\sss M}}\frac{\rho^*(\Sigma,{\bm Y}')}{
      |{\bm Y}-{\bm Y}'|}d^3Y'\;,                                    \label{1.11ax}
  \\\nonumber\\
  {U}_{\sss M}^i(\Sigma,{\bm Y})&=&G\int\limits_{V_{\sss M}}\frac{\rho^*(\Sigma,{\bm Y}'){\nu_{\sss M}^i}(\Sigma,{\bm Y}')}{
      |{\bm Y}-{\bm Y}'|}d^3Y'\;,
 \\\nonumber\\
\chi_{\sss M}(\Sigma,{\bm Y})&=&            \label{1.13ax}
          -G\int\limits_{V_{\sss M}}\rho^*(\Sigma,{\bm Y}')
      |{\bm X}-{\bm X}'|d^3X'\;,\er
\br
\label{1.12aax}
 \Psi_{\sss M}(\Sigma,{\bm Y})&=&
 (\gamma+\frac12)\Psi_{M1}(\Sigma,{\bm Y})-\frac{1}{6}\Psi_{M2}(\Sigma,{\bm Y})+
    (1+\gamma-2\beta)\Psi_{M3}(\Sigma,{\bm Y})
    +\Psi_{M4}(\Sigma,{\bm Y}) +\gamma\Psi_{M5}(\Sigma,{\bm Y})\;,
\er
where
\br
\Psi_{M1}(\Sigma,{\bm Y})&=&G\int\limits_{V_{\sss M}}\frac{\rho^*(\Sigma,{\bm Y}')\nu_{\sss M}^2(\Sigma,{\bm Y}')}{
      |{\bm Y}-{\bm Y}'|}d^3Y'\;,                           \label{1.14ax}
  \er\br
  \Psi_{M2}(\Sigma,{\bm Y})&=&G\int\limits_{V_{\sss M}}\frac{\rho^*(\Sigma,{\bm Y}') \hat{H}^{(\mathrm{int})}(\Sigma,{\bm Y}')}{
      |{\bm Y}-{\bm Y}'|}d^3Y'\;,     \label{1.15ax}
  \er\br
\Psi_{M3}(\Sigma,{\bm Y})&=&G\int\limits_{V_{\sss M}}\frac{\rho^*(\Sigma,{\bm Y}') \hat{\Phi}^{({\rm int})}(\Sigma,{\bm Y}')}{
      |{\bm Y}-{\bm Y}'|}d^3Y'\;,     \label{1.15axa}
  \er\br
\Psi_{M4}(\Sigma,{\bm Y})&=&G\int\limits_{V_{\sss M}}\frac{\rho^*(\Sigma,{\bm Y}')\Pi(\Sigma,{\bm Y}')}{
      |{\bm Y}-{\bm Y}'|}d^3Y'\;,                             \label{1.16ax}
  \er\br
  \Psi_{M5}(\Sigma,{\bm Y})&=&G\int\limits_{V_{\sss M}}\frac{\pi^{kk}(\Sigma,{\bm Y}')}{
      |{\bm Y}-{\bm Y}'|}d^3Y'\;,                           \label{1.17ax}
 \er
the symbol $\nu^i_{\sss M}(\Sigma,{\bm Y})=dY^i/d\Sigma$ is the velocity of Moon's matter with respect to the origin of the SRF, and all integrals given in this section, are taken over the hypersurface of a constant coordinate time $\Sigma$ passing through Moon's volume $V_{\sss M}$.

The local SRF metric given by equations (\ref{1.8wx})--(\ref{1.10qx})
must obey the gauge condition (\ref{11.3}), which yields
\be
         \frac{\partial U_{\sss M}(\Sigma,{\bm Y})}{\partial S}+               \label{1.19ax}
         \frac{\partial U^i_{\sss M}(\Sigma,{\bm Y})}{\partial Y^i}=O(\epsilon^2)\,,  \en
         where we have neglected the contribution of the gravitational field of the external bodies.
Under this assumption equation (\ref{1.19ax}) is satisfied because of the equation of continuity (\ref{11.20}) for Moon's matter.
\\
\paragraph{External Solution for the Metric Tensor.\\\\}\label{mibx}

The boundary conditions imposed on the external solution tell us that it must be convergent on the world line of the SRF origin, where ${\bm Y}=0$. On the other hand, the external solution should match the tidal gravitational field of Sun, Earth, and other planets as the radial distance from the Moon grows. The procedure of finding the external solution for the metric tensor in the SRF is identical with that used for construction of the external solution in the EMB frame and in the GRF. For this reason, we do not describe all its details here. After solving the homogeneous field equations, making use of the gauge conditions and the residual gauge freedom, the external metric tensor acquires the following form
\br\label{1.24bax} \hat H_{00}^{{({\rm ext})}}(\Sigma,{\bm Y})&=&2\nn{l=0}
                     \frac{1}{l!}Q_{\sss M}^LY^{L}\;,
\\\nonumber\\\nonumber\\                                                                    \label{1.25ix}
\hat H_{0i}^{({\rm ext})}(\Sigma,{\bm Y})&=&\left(\dot A_{\sss M}+\frac{\dot Q_{\sss M}}{3}+\frac{1-\gamma}{3}\dot P_{\sss M}\right) Y^i
          +\sum\limits_{l=1}^{\infty}
          \frac{1}{l!}\varepsilon_{ipq}C_{\sss M}^{pL-1}Y^{<qL-1>}
             \\\nonumber
          &+&
          2\sum\limits_{l=1}^{\infty}\frac{2l+1}{(2l+3)(l+1)!}
          \biggl[2\dot{Q}_{\sss M}^L+(\gamma-1)\dot{P}_{\sss M}^L\biggr]Y^{<iL>}+\sum\limits_{l=1}^{\infty}\frac{1}{l!}Z_{\sss M}^{iL}Y^{L}\;,
 \\\nonumber\\\nonumber\\                                               \label{1.26bax}
\hat H_{ij}^{({\rm ext})}(\Sigma,{\bm Y})&=&2\delta_{ij}\left\{A_{\sss E}+\nn{l=1}\frac{1}{l!}\left[Q_{\sss M}^L+(\gamma-1)P_{\sss M}^L\right]Y^{L}\right\}\;,
\er
where $P_{\sss M}^L=P_{\sss M}^L(\Sigma)$, $Q_{\sss M}^L=Q_{\sss M}^L(\Sigma)$, $C_{\sss M}^L=C_{\sss M}^L(\Sigma)$ are STF Cartesian tensors depending only on time $\Sigma$ characterizing gravitational multipolar structure of the external gravitational field by Earth, Sun, and other planets, the dot above the multipoles denotes a time derivative with respect to time $\Sigma$. Multipoles $Z_{\sss M}^L=Z_{\sss M}^L(\Sigma)$ are left free in the external solution as they are associated with the residual gauge freedom, which will be fixed later in derivation of equations of motion of Moon.

The non-linear part of the background SRF metric tensor is
\br \label{1.24cax}
\hat L_{00}^{({\rm ext})}(\Sigma,{\bm Y})&=&
-2\left(\nn{l=0}\frac{1}{l!}Q_{\sss M}^LY^{L}\right)^2-
                 2(\beta-1)\left(\nn{l=0}\frac{1}{l!}P_{\sss M}^LY^{L}\right)^2
                 +\nn{l=0}\frac{1}{(2l+3)l!}\Ddot{Q}_{\sss M}^LY^{L}Y^2\,,
\er
where the double dot above $Q_{\sss M}^L$ denotes the second time derivative with respect to time $\Sigma$.

We have included the monopole $(l=0)$ and dipole $(l=1)$ terms to the non-linear part of the SRF metric tensor for our convenience. These terms could be excluded from equation (\ref{1.24cax}) by means of re-definition of the monopole $Q_{\sss M}$ and the dipole $Q^i_{\sss M}$ in equation (\ref{1.24bax}). We could also decompose the product of the two sums in equation (\ref{1.24cax}) into a single series consisting of irreducible pieces, and absorb some of the terms to the multipoles $Q_{\sss M}^L$ ($l\ge 2$) in equation (\ref{1.24bax}). However, it is more simple to do calculations directly with expression (\ref{1.24cax}).
\\
\paragraph{The Mixed Terms.\\\\}

The non-linear coupling terms in the SRF metric tensor are obtained as a
particular solution of the inhomogeneous field equation (\ref{11.33})
with the right side being a product of the internal and
external solutions found at the previous step. It reads
\br \label{1.24oax} \hat
L_{00}^{(\rm mix)}(\Sigma,{\bm Y})&=&-2U_{\sss M}(\Sigma,{\bm Y})\left\{A_{\sss M}+(2\beta-\gamma-1)P_{\sss M}+2\nn{l=0}\frac{1}{l!}
\biggl[Q_{\sss M}^L+(\beta-1)P_{\sss M}^L\biggr]Y^{L}\right\}\\\nonumber&&
-2G\nn{l=1}\frac{1}{l!}\Bigl[Q_{\sss M}^L+2(\beta-1)P_{\sss M}^L\Bigr]\;\int\limits_{V_{\sss M}}\frac{\rho^*(\Sigma,{\bm Y}')Y'^L}{|{\bm Y}-{\bm Y}'|}d^3Y'\;.\er
This equation completes the derivation of the metric tensor in the local SRF coordinates.

\subsubsection{Gravitational Multipoles of Moon}\label{mluc}

Gravitational field of Moon is described in the SRF in terms of the internal potentials $U_{\sss M}$, $ U^i_{\sss M}$, $ \Psi_{\sss M}$, etc., defined in equations (\ref{1.11ax})--(\ref{1.17ax}). We have found \ct{kovl} that the mixed terms given by equation (\ref{1.24oax}) should be also taken into account to describe the multipolar structure of the internal gravitational field. Multipolar decomposition of the metric tensor of Moon is given in terms of the STF Cartesian tensors of two types.
The {\it active} STF mass-type multipoles of Moon are
\begin{eqnarray}\label{q1.31}
   {\cal I}^L_{\sss M}&= &\int_{V_{\sss M}}\sigma_{\sss M}
   (\Sigma,{\bm Y})Y^{<L>}d^3Y\\\nonumber\\\nonumber &+&
   \frac{\epsilon^2}{2(2l+3)}\left[\frac{d^2}{d\Sigma^2}
   \int_{V_{\sss M}}\sigma_{\sss M}
   (\Sigma,{\bm Y})Y^{<L>}Y^2d^3Y
   -4(1+\gamma)\,\frac{2l+1}{l+1}
   \frac{d}{d\Sigma}\int_{V_{\sss M}}\sigma^i_{\sss M}(\Sigma,{\bm Y})Y^{<iL>}d^3Y\right]
   \\\nonumber\\\nonumber &-&\epsilon^2
   \int_{V_{\sss M}}d^3Y\,\sigma_{\sss M}(\Sigma,{\bm Y})
   \left\{A_{\sss M}+(2\beta-\gamma-1)P_{\sss M}+\sum_{k=1}^{\infty}\frac{1}{k!}\Bigl[Q_{\sss M}^{K}+2(\beta-1) P_{\sss M}^{K}\Bigr]Y^{K}\right\}Y^{<L>}\;,\end{eqnarray}
where $V_{\sss M}$ denotes the volume occupied by the matter of Moon, the density of Moon's matter current
\be\label{qpz6}
 \sigma^i_{\sss M}(\Sigma,{\bm Y})=\rho^*(\Sigma,{\bm Y})\nu^i_{\sss M}(\Sigma,{\bm Y})\;,
 \en
and the {\it active} mass density is defined by
\be\label{qpz3}
 \sigma_{\sss M}(\Sigma,{\bm Y})=\rho^*(\Sigma,{\bm Y})\left\{1+\epsilon^2\left[(\gamma+\frac12)\nu^2_{\sss M}(\Sigma,{\bm Y})+\Pi(\Sigma,{\bm Y})-(2\beta-1) U_{\sss M}(\Sigma,{\bm Y})  \right]\right\}+\epsilon^2\gamma\pi^{kk}(\Sigma,{\bm Y})\;,\en
where $ U_{\sss M}(\Sigma,{\bm Y})$ is the gravitational potential of Moon given by equation (\ref{1.11ax}).

The {\it conformal} STF mass-type multipoles of Moon are defined by equation
\begin{eqnarray}\label{q1.34}
  {I}^L_{\sss M}&= &\int_{V_{\sss M}}\varrho(\Sigma,{\bm Y})\left\{1-\epsilon^2\left[A_{\sss M}+(1-\gamma)P_{\sss M}+\sum_{k=1}^{\infty}\frac{1}{k!}Q_{\sss M}^{K}Y^{K}\right]\right\}
Y^{<L>}\,d^3Y\\\nonumber\\\nonumber&+&
\frac{\epsilon^2}{2(2l+3)}\left[\frac{d^2}{d\Sigma^2}
\int_{V_{\sss M}}\varrho
(\Sigma,{\bm Y})Y^{<L>}Y^2\,d^3Y
-\frac{8(2l+1)}{l+1}
\frac{d}{d\Sigma}\int_{V_{\sss M}}\sigma^i_{\sss M}(\Sigma,{\bm Y})Y^{<iL>}\,d^3Y\right]\;,\end{eqnarray}
with the {\it conformal} mass density of matter defined as
\begin{equation}\label{qpz5}
 \varrho_{\sss M}=\rho^*(\Sigma,{\bm Y})\left[1+\epsilon^2\left(\frac32\nu^2_{\sss M}(\Sigma,{\bm Y})+\Pi(\Sigma,{\bm Y})-U_{\sss M}(\Sigma,{\bm Y})\right)\right]+\epsilon^2\pi^{kk}(\Sigma,{\bm Y})\;.\end{equation}
The {\it conformal} density does not depend on the PPN parameters $\beta$ and $\gamma$.

Scalar multipoles, $\bar{I}_{\sss M}^L$,
are related to the {\it active} and {\it conformal} multipoles by means of a linear relationship
\br\label{qwh1}
\bar{I}_{\sss M}^L=2{\cal I}_{\sss M}^L-(1+\g){I}_{\sss M}^L\;.
\er
The hypersurface of the integration in equations (\ref{q1.31}), (\ref{q1.34}) is that of
the constant coordinate time $\Sigma$, which does not coincide with the hypersurface of the constant time $u$ in the EMB frame or that of the constant time $t$ of the SSB frame.

The spin multipoles of Moon are defined by equation
\be
   S^L_{\sss M}= \int_{V_{\sss M}}\varepsilon^{pq<a_l} Y^{L-1>p}             \label{q1.32}
   \sigma^q_{\sss M}(\Sigma,{\bm Y})\, d^3Y\,,
\en
where the matter current density $\sigma^q_{\sss M}$ is defined in equation (\ref{gpz6}). From their definitions, it is clear that the spin multipoles are different from zero, if and only if, the velocity of matter with respect to the SRF is not zero.

\section{Post-Newtonian Transformations Between the Reference Frames}\label{pntb}
\subsection{Transformation from the Earth-Moon to the Solar-System Frame}\label{embssb}
\subsubsection{General Structure of the Transformation}\label{gzm}
The PPN coordinate transformations between various reference frames in the advanced theory of the lunar motion can be split in three basic categories:\begin{enumerate}
\item post-Newtonian transformation from the local EMB frame to the global SSB frame,
\item post-Newtonian transformation from the local GRF to the local EMB frame,
\item post-Newtonian transformation from the local SRF to the local EMB frame.
\end{enumerate}
However, in order to model LLR data processing we have to add two more post-Newtonian transformations - a transformation from the proper reference frame of the laser station on Earth to the GRF coordinates, and a transformation from the proper reference frame of a CCR on Moon to the SRF coordinates. These two transformations are required for linking the observer's proper time, $\tau$, with the coordinate time $T$ of the GRF, and for connecting the CCR proper time $\lambda$ with the coordinate time $\Sigma$ of the SRF. They are also necessary for description of the small adjustments of the origin and orientation of the proper reference frames with the exact position and orientation of the laser and the CCR plate array. The present paper will describe only the basic post-Newtonian transformations. The post-Newtonian transformations from the proper reference frames to the GRF and the SRF will be discussed elsewhere in connection with the LLR data processing model. The reader, who is interested in principles of derivation of observer-related post-Newtonian transformations can find further details in \ct{1981rcse.conf..283B,kop1991b,brum}.

We draw attention of the reader that one can construct post-Newtonian transformations from the geocentric and selenocentric frames directly to the global SSB frame without the intermediate EMB frame. This approach, however, does not reflect the hierarchic structure of the local coordinates associated with the Earth-Moon system and does not allow us to make a complete decoupling of the relative motion of Moon around Earth from the orbital motion of the Earth-Moon barycenter around Sun. Moreover, once one knows the transformations from the geocentric and selenocentric frame to the EMB frame, and that from the EMB frame to the SSB frame - the post-Newtonian transformation from the geocentric and selenocentric frame directly to the SSB frame can be derived by means of successive application of the post-Newtonian transformations between the hierarchic gravitating systems. This kind of procedure may lead to some difficulties in formulation of the post-Newtonian conventions about the dynamically and kinematically non-rotating frames \ct{klioner93}.

Post-Newtonian transformations between the frames are derived by making use of the mathematical technique known as asymptotic matching of the post-Newtonian expansions of the scalar field and the metric tensor. This technique was originally proposed in relativity by D'Eath \cite{das1,das2} as a tool for derivation of equations of motion of black holes. Other researchers had proved its efficiency in the post-Newtonian theory of reference frames in the solar system \cite{k85,k88,k89d,bk89,ashb2,kovl,dsx1,dsx2}.
The metric tensor and the scalar field are given as a solution of the field equations in each particular frame of reference. This solution is expressed in the form of different functions depending on the choice of the coordinates associated with the reference frame. However, these functions describe one and the same physical situation, which means that they must match smoothly in the space-time domain where two coordinate charts overlap. The matching assumes that the tensor transformation law is applied to the post-Newtonian metric tensor and the scalar field. The matching domain is bounded by the radius of convergence of the post-Newtonian series. After the matching is finished and the post-Newtonian transformations between the reference frames are found, the local solutions of the field equations can be analytically continued to a much larger spatial domain, if it is required by practical applications \ct{kop1991b,bk-nc,klv,kovl}.

The post-Newtonian coordinate transformation between the frames must comply with the gauge condition (\ref{11.3}). Therefore, one begins with finding the most general structure of the coordinate transformation that obeys equation (\ref{11.3}). After this structure is established, it is further specialized by making use of the residual gauge freedom admitted by equation (\ref{11.b}). Technically, post-Newtonian transformation between the global and local frames is found as a general solution of the homogeneous equation (\ref{11.b}) describing the residual gauge freedom of the metric tensor. The solution is given by a post-Newtonian series of harmonic polynomial expanded in powers of the spatial coordinates of the local frame with the polynomial coefficients being functions of time that are STF Cartesian tensors defined on the world line of the origin of the local coordinates \cite{k88,dsx2}. This solution is substituted to the matching equations between the solutions of the field equations expressed in the global and local coordinates. The matching of the post-Newtonian expansions of the scalar field and the metric tensor allows us to fix all degrees of the gauge freedom in the final form of the post-Newtonian coordinate transformation. Notice that we have partially used this gauge freedom in sections \ref{mtex}, \ref{mtexa}, \ref{mibx} to remove non-physical multipoles in the external solution for the metric tensor.

The post-Newtonian transformation between the coordinate times of the two frames describes the integral Lorentz (velocity-dependent) and Einstein (gravitational field-dependent) time delays associated with the different definition of simultaneity of events in the two frames \cite{k88,ashb2}. They also include a series of complicated polynomial terms \ct{1990CeMDA..48...23B}. The post-Newtonian transformation between the space coordinates of the two frames consists of linear and non-linear parts. The linear part of the transformation includes the Lorentz and Einstein contractions as well as a matrix of relativistic rotation describing the post-Newtonian precession of the spatial axes of one frame with respect to another due to the orbital motion of the local frame and gravitational fields of the solar system bodies, which are external with respect to the local frame  \cite{k85,d89}.
The Lorentz contraction takes into account the kinematic
aspects of the post-Newtonian transformation that depends on the relative velocity of motion of the local frame with respect to the global one. The Einstein gravitational contraction accounts for static effects of the scalar and gravitational fields \ct{kovl}. The non-linear part of the spatial transformation depends on the orbital acceleration of the local frame and accounts for the effects of the derivatives of the gravitational field associated with the Christoffel symbols.

Let us discuss the mathematical structure of the post-Newtonian transformation taking as an example
the transformation between the EMB local frame, $w^\alpha=(w^0,w^i)=(cu,{\bm w})$, and the SSB global frame,
$x^\alpha=(x^0,x^i)=(ct,{\bm x})$. This coordinate transformation must be compatible with the weak-field and slow-motion approximation used as a cornerstone of the post-Newtonian iteration procedure. Hence, the transformation is also given as a post-Newtonian series by two equations - one for time and another one for space coordinates:
\br u&=&t +\epsilon^2\xi^0(t,{\bm x})\;,
\label{2.2} \\\nonumber\\
 w^i&=&\nnr{i}+\epsilon^2\xi^i(t,{\bm x})\;,      \label{2.3} \er
 where $\xi^0$ and $\xi^i$ are the post-Newtonian corrections to the Galilean translation,
$\nnr{i}=x^i-x^i_{{ B}}(t)$, and $x^i_{{ B}}(t)$ is the position of
the origin of the local frame at time $t$ with respect to the origin of the global coordinates. The origin of the EMB frame can be always chosen at any instant of time at
the barycenter of the Earth-Moon system as we shall demonstrate later (see also \ct{kovl}).
In what
follows, we denote velocity and acceleration
of the origin of the local coordinates as $\nnv{i}\equiv\dot x^i_{{ B}}$ and
$\nnaa{i}\equiv\Ddot x^i_{{ B}}$ respectively, where here and everywhere else the dot above a function
must be understood as a total time derivative with respect to time $t$.

Pointwise matching equations for the scalar field, the metric tensor, and
the Christoffel symbols are given by the general law of coordinate transformations of these objects \cite{dfn}
\br
\varphi(t,{\bm x})&=&\nxi\;, \label{2.5}
      \\\nonumber\\
      g_{\mu\nu}(t,{\bm x})&=&                         \label{2.6}
          \hat g_{\alpha\beta}(u, {\bm w})\frac{\partial w^{\alpha}}{\partial x^{\mu}}
          \frac{\partial w^{\beta}}{\partial x^{\nu}}\;,
          \\\nonumber\\
   \Gamma^{\mu}_{\alpha\beta}(t,{\bm x})&=&
    \hat{\Gamma}^{\nu}_{\rho\sigma}(u,\bm w)
     \frac{\partial x^{\mu}}{\partial w^{\nu}}
      \frac{\partial w^{\rho}}{\partial x^{\alpha}}
       \frac{\partial w^{\sigma}}{\partial x^{\beta}}+           \label{2.7}
        \frac{\partial x^{\mu}}{\partial w^{\nu}}
         \frac{{\partial}^2 w^{\nu}}{\partial x^{\alpha}
          \partial x^{\beta}}\,,
\er
where
\br\label{oi1}
{\Gamma}^{\mu}_{\alpha\beta}(t,{\bm x})&=&\frac12 g^{\mu\nu}\left(\frac{\p g_{\nu\alpha}}{\p x^\beta}+\frac{\p g_{\nu\beta}}{\p x^\alpha}-\frac{\p g_{\alpha\beta}}{\p x^\nu}\right)\;,
\\\label{oi2}
\hat{\Gamma}^{\mu}_{\alpha\beta}(u,{\bm w})&=&\frac12\hat g^{\mu\nu}\left(\frac{\p\hat g_{\nu\alpha}}{\p w^\beta}+\frac{\p\hat g_{\nu\beta}}{\p w^\alpha}-\frac{\p\hat g_{\alpha\beta}}{\p w^\nu}\right)\;,
\er
are the Christoffel symbols expressed in the SSB and EMB frames respectively.

It is worth noticing that the matching equations (\ref{2.5})--(\ref{2.7}) are valid in a 4-dimensional space-time volume, which includes the world tube with a space-like cross-section covered by spatial coordinates of the local EMB frame. The scalar field, the metric tensor and their first derivatives are continuously differentiated functions in this volume. This point, in fact, means that equations (\ref{2.5}), (\ref{2.6}) are sufficient for the purposes of matching procedure because equation (\ref{2.7}) does not bear any new physical information. The matching equations are not identities, which are automatically satisfied. The left side of these equations contain known functions which are defined as integrals over the volumes of the bodies of the solar system. The right side of the matching equations contain yet unknown functions, which are the external multipoles of the metric tensor in the EMB frame as well as functions $\xi^\a$ entering the post-Newtonian transformations (\ref{2.2}), (\ref{2.3}). These functions are determined by solving the matching equations.

The starting point in this iterative post-Newtonian procedure is the $g_{0i}$ component of the
metric tensor. One notices that it does not contain terms of the order of $O(\epsilon)$ because one have assumed that both the
global and the local frames are not dynamically rotating, which cancels the angular and linear velocity terms \cite{kovl,dsx1}. This fact having been used in equation (\ref{2.6}), implies that
function $\xi^0(t,{\bm x})$ from time-transformation equation (\ref{2.2}) must be subject to
the following restriction:
\be\l{ha+}
\xi^0,_{k}=-\nnv{i}+O(\epsilon^2)\;.
\en
This is a partial differential equation which
can be integrated so that function $\xi^0$ can be the most generally represented
as \be \xi^0(t,\nnx)=-{\cal A}(t)-
       \nnv{k}\nnr{k}                                           \label{2.8}
          +\epsilon^2 \kappa(t,\nnx)+O(\epsilon^4)\;,     \en
where ${\cal A}(t)$ and
$\kappa(t,\nnx)$ are analytic, but otherwise unspecified functions. Notice that
function ${\cal A}(t)$ depends only on time $t$.

Let us now use the gauge conditions (\ref{gau}) in order to
impose further restrictions of the post-Newtonian functions $\xi^0$ and $\xi^i$
from equations (\ref{2.2}) and (\ref{2.3}).
The law of transformation of the Christoffel connection, equation
(\ref{2.7}), being substituted to equation (\ref{gau}) yields a partial
differential equation of the second order
\be
   g^{\alpha\beta}(t,\nnx)\,
    \frac{{\partial}^2 w^{\mu}}{\partial x^{\alpha}             \label{2.9}
     \partial x^{\beta}}=0\;,
\en
which describes any possible freedom in the post-Newtonian transformations from the EMB to SSB coordinates.
Let us now substitute functions $w^0=cu$ and $w^i$ from equations
(\ref{2.2}) and (\ref{2.3}), and $\xi^0$ from equation (\ref{2.8}) to
equation (\ref{2.9}). One obtains
 \br
   \nabla^2 \kappa(t,\nnx)&=&3\nnv{k}\nnaa{k} -\Ddot {\cal A}- \label{2.10}
        \dot a^k_{{ B}}\nnr{k}+ O(\epsilon^2)\;,
\\\nonumber\\
   \nabla^2\xi^i(t,\nnx)&=&-\nnaa{i}+ O(\epsilon^2)\;.              \label{2.11}
\er
General solution of these elliptic-type equations can be written in the form of
the Taylor series expansion in terms of the irreducible Cartesian tensors. Furthermore, solution for functions
$\kappa(t,\nnx)$ and $\xi^i(t,\nnx)$ in equations (\ref{2.10})
and (\ref{2.11}) consists of two parts -- a fundamental solution of the
homogeneous Laplace equation and a particular solution of the inhomogeneous Poisson
equation. We discard the part of the fundamental solution that has a singularity at the origin of the local coordinates, where $w^i=0$. This is because the singular part does not present in the internal solution of the field equations for the metric tensor and the scalar field, which are represented by integrals over the continuous distribution of matter. However, had we worked in the region outside of the gravitating bodies, we would have to include the singular part of the fundamental solution of the Laplace equation to the coordinate transformation between the frames. In this case the singular part of the transformation is responsible for the gauge freedom in the definition of the multipole moments of the gravitating bodies \ct{bld1986,thor,bld}. This freedom has been fixed in the present paper by picking up the post-Newtonian definition of the multipole moments in the form proposed by Blanchet and Damour \ct{bld}.  For this reason the singular terms characterizing the residual gauge freedom have no matching counterparts and must be equated to zero.

Integrating equations (\ref{2.10}) and (\ref{2.11})
results in
\br  \kappa&=& \Bigl(\frac{1}{2}
        \nnv{k}\nnaa{k}-\frac{1}{6}
        \Ddot {\cal A}\Bigr)\nnr{2}-\frac{1}{10}             \label{2.12}
        \dot a^k_{{ B}}\nnr{k}\nnr{2}+
        \Xi(t,\nnx)\;,\\\nonumber\\                             \label{2.13}
\xi^i&=&-\frac{1}{6} \nnaa{i} \nnr{2}+\Xi^i(t,\nnx)\,,
\er
where functions $\Xi$ and $\Xi^i$ are the fundamental solutions of the homogeneous Laplace equation, which is convergent at the origin of the EMB frame. These solutions can be
written in the form of scalar and vector harmonic polynomials
\br\label{eq1}
\Xi(t,\nnx)&=&\sum\limits_{l=0}^\infty\frac{1}{l!}{\cal B}^{L}
        \nnr{L}\;,\\\nonumber\\\label{eq2}
\Xi^i(t,\nnx)&=&\sum\limits_{l=1}^{\infty}
        \frac{1}{l!}{\cal D}^{iL}\nnr{L}
        +\sum\limits_{l=0}^{\infty}\frac{\varepsilon_{ipq}}{(l+1)!}
        {\cal F}^{pL}\nnr{<qL>}+
        \sum\limits_{l=0}^{\infty}
        \frac{1}{l!}{\cal E}^{L}
        \nnr{<iL>}\;,
        \er
where coefficients ${\cal B}^{L}\equiv{\cal B}^{<L>}(t)$, ${\cal D}^{L}\equiv{\cal D}^{<L>}(t)$, ${\cal F}^{L}\equiv{\cal F}^{<L>}(t)$, and ${\cal E}^{L}\equiv{\cal E}^{<L>}(t)$ are the STF Cartesian tensors \ct{kovl}. These coefficients are defined on the world line of the
origin of the local coordinates and depend only on the coordinate time $t$ of the global SSB frame. Explicit form of these functions
will be obtained in the process of matching of the metric tensor and the scalar field in accordance with equations (\ref{2.5})--(\ref{2.7}).

Formulas (\ref{2.12})--(\ref{eq2}) allow us to evaluate the size of the spatial domain
of applicability of the EMB local coordinates. It is determined by the condition that determinant of the matrix $\Lambda^{\alpha}_{\;\beta}$ of the four-dimensional coordinate transformation, is zero \ct{1980gmmp.book.....S}. Calculating the
determinant yields
\br\label{ma5}
{\rm
det}\left(\Lambda^{\alpha}_{\;\beta}\right)&=&1+\epsilon^2\left[-\dot{\cal
A}+3\,{\cal E}-\frac43\left(\nnaa{k}-\frac52{\cal
E}_k\right)\nnr{k}+\sum_{l=2}^\infty\frac{(l+1)(2l+3)}{(2l+1)l!}{\cal
E}^{L}\nnr{L}\right]+O\left(\epsilon^4\right)\;.
\er
Radius of
convergence of the polynomial in the right side of equation (\ref{ma5})
crucially depends on the choice of functions ${\cal E}^L$. We have proved \cite{bk-nc,kovl} that it is possible to make function
${\cal E}^i=\nnaa{i}$, and all other functions ${\cal E}^L=0$ for any $l\ge2$. Thus,
determinant (\ref{ma5}) vanishes when distance $\nnr{}\approx
c^2/(2\nnaa{})$. In case of the EMB frame, moving around Sun
with acceleration $\nnaa{}\simeq 0.6$ $cm/s^2$, this
distance $\nnr{}$ is about $10^{21}$ cm or about 300 parsec.  Hence, the EMB frame covers a spatial region, which includes the entire solar system and its neighborhood.
This
consideration suggests that the metric tensor defined originally in the EMB
local coordinates only in the domain restricted by the distance to the nearest external gravitating body (Venus and Mars in case of the EMB frame) can
be re-formulated in terms of some other functions and extrapolated beyond this boundary. Such extrapolation of the local coordinates and the corresponding metric tensor was considered in papers \cite{klv,1992A&A...257..777B}.

\subsubsection{Matching of the Post-Newtonian Expansions}

Method of the matched post-Newtonian expansions is a powerful mathematical tool to find the law of transformation from one celestial frame to another and to determine the external multipoles in the post-Newtonian expansions of the background metric tensor.
These post-Newtonian expansions, which are used in the matching procedure,
are solutions of the
gravity field equations for the metric tensor and the scalar field found respectively in the global and local coordinates.
These solutions are shown in equations (\ref{12.5})--(\ref{12.9}) for the SSB frame, and (\ref{1.7}), (\ref{1.7a}), (\ref{1.24b})--(\ref{1.26b}) for the EMB frame. The solution for the metric tensor and the scalar field in the SSB frame is valid
everywhere inside and outside of the solar system up to infinity.
One may think that the global
coordinates alone are sufficient to describe the post-Newtonian celestial dynamics of the solar system bodies. Indeed, the original Fock-Papapetrou approach  \cite{fock1,pap1,pap2,fockbook} assumed that only one coordinate chart is used to solve the internal and external problems of motion.
However, the single coordinate-chart approach is not satisfactory in doing the post-Newtonian approximations for two reasons.

First, a gravitating body (or a sub-system of bodies) that is a member of N-body system, has its own (internal) gravitational field that is characterized outside the body by gravitational multipoles like mass -- monopole, center-of-mass -- dipole, oblateness -- quadrupole, etc. Local coordinates attached to the body are required to give physically meaningful, post-Newtonian definition of the internal multipoles. One must know how
this definition of the multipoles given in the local frame of reference conforms to the definition of the same multipoles given in the global coordinates. Post-Newtonian relationship between the frame-dependent definitions of the internal multipoles plays a key role in derivation of the orbital equations of motion of extended bodies \cite{bk-nc,dsx2,kovl}.

Second,
the global SSB frame is inappropriate for the gauge-independent description of the orbital motion of Moon and artificial Earth's satellites. This is because the Earth-Moon system is moving in the external gravitational field of Sun and other planets of the solar system. The most simple, Galilean translation of the origin of the global SSB coordinates to the barycenter of the Earth-Moon system (to the geocenter in case of the artificial satellite) that is currently used for construction of the lunar ephemeris \cite{chatc,sta,epm,pmoe}, denies the post-Newtonian aspects of the coordinate transformation in relativistic theory of gravity and brings up the gauge-dependent terms to the description of the orbital motion of Moon. These terms are unobservable and should be discarded by making appropriate choice of the global and local coordinate frames.
The adequate post-Newtonian transformation accounting for the gauge freedom should make description of the lunar motion essentially simpler by suppressing all the spurious orbital harmonics \cite{bk-nc,sof,dsx4,k07}.

The internal solution for the metric tensor and the scalar field in the local EMB coordinates contains the external multipoles $Q^L$, $C^L$, $P^L$ describing gravitational field of Sun and other planets. Their explicit functional dependence on the gravitational potentials of the external bodies can not be determined by solving the field equations in the local EMB coordinates alone -- the matching with the solution of the field equations in the global SSB coordinates is required. The external multipoles are found simultaneously with the post-Newtonian coordinate transformation between the SSB and EMB coordinates. Matching also helps to understand better the physical foundation underlying the principle of equivalence for self-gravitating bodies.

Matching the metric tensor and the scalar field in the local and global coordinates is based on equations (\ref{2.5}) and (\ref{2.6}), and consists of the following steps (for exhaustive mathematical details of this procedure the reader is referred to \cite[Section 8]{kovl})
\begin{enumerate}
\item[\it Step 1.] One re-writes the local metric tensor $\hat g_{\a\b}(u,{\bm w})$ and the scalar field $\nxi$ in the right side of equations (\ref{2.5}) and (\ref{2.6}) in terms of the global coordinates $(t,{\bm x})$. This is achieved by making use of a Taylor expansion of the internal and external potentials defining $\nxi$ and $\hat{g}_{\alpha\beta}(u,{\bm w})$ around the point $x^\alpha=(ct,{\bm x})$. The concept of the Lie transfer \ct{1980gmmp.book.....S,mtw} must be applied in order to change the integration in the integrals of the internal potentials from the hypersurface of the local coordinate time $u$ to that of the global coordinate time $t$.
\item[\it Step 2.] One calculates the partial derivatives of the local coordinates, $w^\a$, with respect to the global coordinates, $x^\b$, that is the matrix of transformation of the coordinate bases, $\Lambda^\a_{\;\b}=\p w^\a/\p x^\b$.
\item[\it Step 3.] One separates the gravitational potentials in the left side of equations (\ref{2.5}) and (\ref{2.6}) to the internal part, relating to the Earth-Moon system, and to the external part generated by Sun and other planets:
    \br
    \varphi(t,\nnx)&=&\varphi_{E}(t,\nnx)+\varphi_{M}(t,\nnx)+\bar{\varphi}(t,\nnx)\,,     \label{3.1s}
\\\nonumber\\
   U(t,\nnx)&=&U_{E}(t,\nnx)+U_{M}(t,\nnx)+\bar{U}(t,\nnx)\,,                 \label{3.1}
\\\nonumber\\
   U^i(t,\nnx)&=&U^i_{\sss E}(t,\nnx)+U^i_{\sss M}(t,\nnx)+    				 \label{3.2}
        {\bar{U}}^i(t,\nnx)\,,
\\\nonumber\\
   \chi(t,\nnx)&=&\chi_{\sss E}(t,\nnx)+\chi_{\sss M}(t,\nnx)+  				 \label{3.3}
         \bar{\chi}(t,\nnx)\,,
\\\nonumber\\
   \nnff{k}&=&\Psi_{Ek}(t,\nnx)+\Psi_{Mk}(t,\nnx)+ \bar{\Psi}_k(t,\nnx)\;,       \label{3.4}
   \qquad\qquad(k=1,\,...\,,5)\,,\er
   where functions with indices $(E)$ and $(M)$ are given by integrals (\ref{12.10})--(\ref{12.16}) taken over the volume of Earth and Moon respectively, and the bar above functions indicates, here and everywhere else, that the corresponding sum in the definitions (\ref{12.9a}) of these functions excludes Earth and Moon, that is the sum takes into account only external bodies which are Sun and other planets
\be\label{omap}
\bar{\varphi}=\sum_{A\not=E,M} \varphi_A\,,\quad
\bar{U}=\sum_{A\not=E,M} U_A\,,\quad\bar{U}_i=\sum_{A\not=E,M}
U^i_A\,,\quad\bar{\Psi}_k=\sum_{A\not=E,M} \Psi_{Ak}\:,\quad
      \bar{\chi}=\sum_{A\not=E,M} \chi_A\:.
\en
\item[\it Step 4.] One expands the gravitational potentials of the external masses, that is functions with bars in equations (\ref{3.1s})--(\ref{omap}), in the Taylor series in powers of $R^i_{B}=x^i-x^i_{B}$ in the neighborhood of the origin of the local EMB frame, which image in the global coordinates is at the point $x^i_{B}=x^i_{B}(t)$.
\item[\it Step 5.] One equates similar terms depending on the internal structure of the Earth-Moon system. It turns out that all the internal potentials cancel out in equations (\ref{2.5}),(\ref{2.6}). It proves that the internal structure of the bodies is compatible with the differential structure of space-time manifold. Because we did not set any limitations on the distribution of matter's density and stress inside the bodies, it means that the {\it effacing principle} \ct{1983grr..proc...58D,d89,2006gr.qc....12017K} is satisfied in the scalar-tensor theory of gravity.
\item[\it Step 6.] One equates similar terms of the Taylor expansions from the left side of the matching equations (\ref{2.5}) and (\ref{2.6}) with the corresponding terms of the Taylor expansions entering the right side of these equations. This reduces the original matching equations to the set of algebraic and ordinary differential equations expressing the yet unknown external multipoles and the coefficients of the harmonic polynomials in the post-Newtonian coordinate transformations in terms of the external gravitational potentials (the over-barred functions) and their derivatives.
\item[\it Step 7.] One separates the algebraic equations into irreducible pieces and, finally, determine the external multipoles as well as the coefficients of the harmonic polynomials. It fixes the residual gauge freedom and brings about the laws of orbital and precessional motion of the local EMB frame with respect to the global SSB frame
\end{enumerate}
Final results of the matching procedure are given below.

\subsubsection{Post-Newtonian Coordinate Transformation}\label{ffemb}

The $\b-\g$ parameterized post-Newtonian coordinate
transformation from the EMB frame to the SSB frame is given by two equations:
\br                                                 \label{5.12}
u&=&t-\epsilon^2\left({\cal
A}+\nnv{k}\nnr{k}\right)
+\epsilon^4\left[{\cal B}+\frac{1}{6}
\biggl(\dot{Q}-
\dot{\bar U}\nnxe+2\nnv{k}\nnaa{k}\biggr)\nnr{2}-
   \frac{1}{10}\Dot{a}^{k}_{\scriptstyle B}\nnr{k}\nnr{2}+
   \sum\limits_{l=1}^{\infty}\frac{1}{l!}
   {\cal B}^{L}\nnr{L}\right]+O(\epsilon^5)\;,
   \\\nonumber\\                                         \label{5.13}
   w^i&=&\nnr{i}+\epsilon^2\left[
   \biggl(\frac{1}{2}\nnv{i}\nnv{k}+
   \gamma\delta^{ik}\bar U\nnxe-\delta^{ik}A+{F}^{ik}\biggr)\nnr{k}+
   \nnaa{k}\nnr{i}\nnr{k}-\frac{1}{2}\nnaa{i}\nnr{2}
   \right]+O(\epsilon^4)\;.
   \er
Here functions ${\cal A}$ and ${\cal B}$ depends on the global coordinate time $t$ only and are solutions of the ordinary differential equations
\br
\frac{d{\cal A}}{dt}&=&+\frac12\nnv{2}+\bar U\nnxe-Q\,,             \label{5.14}
\\\nonumber\\
   \frac{d{\cal B}}{dt}&=&-                                         \label{5.7a}
        \frac{1}{8}\nnv{4}-
	(\gamma+\frac{1}{2})\nnv{2}\bar U\nnxe+
        \frac{1}{2}\bar U^2\nnxe+2(1+\gamma)\nnv{k}\bar U^k\nnxe
        -\bar\Psi\nnxe+\frac12\bar\chi_{,tt}\nnxe\nonumber\\
	 &&+Q\bigg[-\frac{1}{2}v_{\sss{B}}^2+\frac{1}{2}Q-\bar{U}\nnxe\bigg]\;,
\er
that describe the post-Newtonian transformation between the coordinate time $u$ of the EMB frame and the coordinate time $t$ of the SSB frame.
The other functions are defined by algebraic relationships as follows
\br
{\cal B}^i&=&2(1+\gamma)\bar U^i\nnxe-                   \label{5.15}
(1+2\gamma)\nnv{i}\bar U\nnxe-\frac12
\nnv{i}\nnv{2}-Qv^i_{\sss{B}}\,,
\\\nonumber\\
       {\cal B}^{ik}&=&\!\!
       Z^{ik}+
        2(1+\gamma)\bar U^{<i,k>}\nnxe
	-2(1+\gamma)\nnv{<i}\bar U^{,k>}\nnxe              \label{5.16}
	+2\nnv{<i}a^{k>}_{\sss{B}}\,,
           \\\nonumber\\
        {\cal B}^{iL}&=&\!\!
       Z^{iL}+
        2(1+\gamma)\bar U^{<i,L>}\nnxe
	-2(1+\gamma)\nnv{<i}\bar U^{,L>}\nnxe\;               \label{5.17}
          , \qquad\qquad\qquad (l\ge 2)
\er
where some residual gauge freedom parameterized by STF Cartesian tensors $Z^L$ is explicitly shown.

The anti-symmetric rotational matrix ${F}^{ik}$ couples algebraically with the dipole moment $C^i$, which describes the post-Newtonian precession of the spatial axes of the EMB frame,
\br\label{5.18}
\varepsilon_{ipk}C^p+\frac{dF^{ik}}{dt}&=&-2(1+\gamma)\bar U^{[i,k]}\nnxe
+(1+2\gamma)\nnv{[i}\bar U^{,k]}\nnxe
    +\nnv{[i}Q^{k]}\;.
\er
The first term in the right side of equation (\ref{5.18})
describes the Lense-Thirring (gravitomagnetic) precession, the
second term describes the de Sitter (geodetic) precession in the scalar-tensor theory of gravity, and the third term
describes the Thomas precession \cite{mtw} depending on the local (non-geodesic) acceleration $Q^i$ of the origin of the EMB frame. In the
scalar-tensor theory both the Lense-Thirring and the de Sitter
precessions depend on the PPN parameter $\gamma$ while the Thomas
precession does not. The reason is that the Thomas precession is generically a special relativistic effect \cite{mtw} that can not depend on
any particular choice of a specific gravitational theory. The presence of matrix ${F}^{ik}$ in the spatial part of the post-Newtonian transformation means that spatial axes of the local
EMB frame rotates kinematically with respect to axes of the SSB frame which is anchored to distant quasars. At the same time the dipole moment $C^i$ is the angular velocity of the dynamic rotation of the spatial axes of the EMB frame. IAU recommends \ct{iau2000} to adopt ${F}^{ik}=0$ making the EMB frame dynamically rotating with angular velocity $C^i$ defined by equation (\ref{5.18}). In the present paper we prefer another choice, namely $C^i=0$, making the EMB kinematically rotating with the precessional matrix $F^{ik}$ defined by equation (\ref{5.18}). The advantage of our choice is that it eliminates the Coriolis and centrifugal forces from the equations of motion of Moon with respect to Earth written in the local EMB frame.

\subsubsection{The External Multipoles}

Matching determines the external multipoles in terms of the derivatives of gravitational potentials of the external bodies that are Sun and other planets \cite{kovl}. The external multipoles of the scalar field are
\be
       P^L=\bar\varphi_{,L}\nnxe+O(\epsilon^2)\,,                     \label{3.13}
\en
where the external scalar field $\bar\varphi$ coincides in this approximation with the external Newtonian potential $\bar{U}$, as defined in equation (\ref{omap}), and is computed at the origin of the
local EMB coordinates, $x^i_{ B}(t)$, at the instant of time $t$. We remind that the scalar field perturbation is coupled with the factor $\gamma-1$, so that all physically-observed scalar-field effects must be proportional to this factor. We also notice that the lower-order ($l=0,1$) external multipoles of the scalar field can not be removed from the observable gravitational effects by making coordinate transformation because if the scalar field presents in one coordinate frame, it must be present in any other as coordinate transformations do not change the numerical value of the scalar field.

The matching equation determines the external dipole moment of the EMB metric tensor as follows \cite{kovl}
\br\nonumber
 Q^i&=&
           \bar U_{,i}\nnxe-\nnaa{i}\\\nonumber\\\nonumber  &
           &+\epsilon^2
           \bigg\{\bar\Psi_{,i}\nnxe-\frac{1}{2}\bar{\chi}_{,itt}\nnxe
           +2(1+\gamma)\dot{\bar{U}}^i\nnxe-2(1+\gamma)
           \nnv{k}\bar U^{k,i}\nnxe\\\nonumber\\\nonumber&&
	   -(1+2\gamma)\nnv{i}\dot{\bar U}\nnxe +(2-\gamma) \bar U\nnxe \bar U_{,i}\nnxe+(2+\gamma)\nnv{2}
           \bar U_{,i}\nnxe-
           \frac{1}{2}\nnv{i}\nnv{k}\bar U_{,k}\nnxe\\\nonumber\\\nonumber  &
           &
	   -\frac{1}{2}\nnv{i}\nnv{k}\nnaa{k}-2\nnv{2}\nnaa{i}-(4+\gamma)
           \nnaa{i}\bar U\nnxe+F^{ik}\bar U_{,k}\nnxe-F^{ik}\nnaa{k}
           \\\nonumber\\  &
	   &+Q_i[A-v^2_{\sss{B}}-2\bar{U}\nnxe]
	   \bigg\}+O(\epsilon^4)\;.\label{5.8}
\er
The external dipole, $Q^i$, is explicitly expressed in terms of the external gravitational potentials and the barycentric acceleration $\nnaa{i}$ of the origin of the local EMB frame with respect to the global SSB coordinates. It is remarkable that $Q^i$ is not limited by the gauge conditions and can be chosen arbitrary because it determines the magnitude and direction of the inertial force acting in the local EMB frame on a test particle being in a free fall. It means, that equation (\ref{5.8}) must be effectively understood as the law of the orbital motion of the origin of the EMB frame in the global SSB coordinates, which is governed by a particular choice of the dipole moment $Q^i$. Only after the choice of $Q^i$ is made, the coordinate acceleration $\nnaa{i}$ of the origin of the EMB frame with respect to the global SSB coordinates can be fully defined.

The most simple choice of $Q^i=0$ means that the origin of the EMB frame moves along a geodesic world line in the background space-time defined by the external part of the EMB metric tensor. However, it does not allow us to keep the origin of the local coordinates at the barycenter of the Earth-Moon system. This is because the Earth-Moon system has an internal quadrupole moment interacting with the tidal gravitational field of Sun and other planets, and forcing the barycenter of the Earth-Moon system to move along an accelerated (non-geodesic) world line \cite{k88}. Thus, $Q^i$ must be defined in such a way that the Earth-Moon barycenter and the origin of the EMB local frame would coincide at any instant of time. This is equivalent to solving the internal problem of motion of the Earth-Moon barycenter with respect to the local EMB frame that will be discussed elsewhere (some details can be found in \ct{k88,kovl}).

External mass-type multipoles $Q^L$ for $l\ge2$ are defined by the following equation
\br\nonumber
    Q^L&=&\bar U^{,<L>}\nnxe\\\nonumber\\\nonumber  && +\epsilon^2\left[\bar\Psi^{,<L>}\nnxe-\frac12\bar\chi^{,tt<L>}\nnxe
           +2(1+\gamma)\dot{\bar U}^{<i_l,L-1>}\nnxe-2(1+\gamma)
	   \nnv{k}\bar U^{k,<L>}\nnxe\right.\\\nonumber\\\nonumber  && \left.  +(l-2\gamma-2)
	   \nnv{<i_l}\dot{\bar U}^{,L-1>}\nnxe+(1+\gamma)\nnv{2}
           \bar U^{,<L>}\nnxe
	   -\frac{l}{2}\nnv{k}\nnv{<i_l}\bar U^{,L-1>k}\nnxe-l\gamma\,\bar U\nnxe\bar U^{,L}\nnxe\right.
           \\\nonumber\\ & &\left.-
	   (l^2-l+2\gamma+2)\nnaa{<i_l}\bar U^{,L-1>}\nnxe
           -lF^{k<i_l}\bar U^{,L-1>k}\nnxe+K^{L}+
	   \dot Z^L+lA\bar{U}^{,<L>}\nnxe\right]+O\left(\epsilon^4\right)\;,\label{5.9}
\er
where we have used notations
\br
    K^{ij}&\equiv&3\nnaa{<i}\nnaa{j>}\,,                      \label{5.10}
\\\nonumber\\
    K^{L}&\equiv&0\,,\qquad\qquad\qquad (l\ge3).              \label{5.11}
\er
External current-type multipoles $C^L$ for $l\ge2$ are given by
\br\label{3.29}
        \varepsilon^{ipj}C^{pL-1}&=&
	\frac{4l(1+\gamma)}{l+1}\left\{
          \nnv{[i}\bar U^{,j]L-1}\nnxe
	  -\bar U^{[i,j]L-1}\nnxe-\frac{l-1}{l}\delta^{i_{l-1}[i}\dot{\bar{U}}^{,j]L-2}\nnxe
             \right\}+O\left(\epsilon^2\right)\,,
\er
where the dot means the time derivative with respect to time $t$.
\subsection{Transformation from the Geocentric to the Earth-Moon Frame}\l{tgem}
\subsubsection{Matching Procedure}\l{mtpr}
Matching of the geocentric and the Earth-Moon barycentric solutions of the metric tensor and the scalar field follows the same steps as in the matching of the post-Newtonian expansions written in the EMB and SSB frames. We recall that the geocentric coordinates of the GRF are denoted by $X^\a=(X^0,X^i)=(cT,{\bm X})$, and the EMB local coordinates are $w^\a=(w^0,w^i)=(cu,{\bm w})$. In doing this matching, the GRF is "more local" than the EMB frame in the sense of the hierarchy of the astronomical frames discussed in the introduction of the present paper. Indeed, the local GRF coordinates cover the space from Earth to Moon, while the EMB coordinates cover much larger region spreading out to the orbits of Mars and Venus.

The matching equations are as follows:
\br
\nxi&=&\Phi(T,{\bm X})\;, \label{2.5x}
      \\\nonumber\\
      \hat g_{\mu\nu}(u, {\bm w})&=&                         \label{2.6x}
      G_{\alpha\beta}(T, {\bm X})\frac{\partial X^{\alpha}}{\partial w^{\mu}}
          \frac{\partial X^{\beta}}{\partial w^{\nu}}\;,
          \\\nonumber\\
   \hat{\Gamma}^{\mu}_{\alpha\beta}(u,\bm w)&=&
    C^{\nu}_{\rho\sigma}(T,{\bm X})
     \frac{\partial w^{\mu}}{\partial X^{\nu}}
      \frac{\partial X^{\rho}}{\partial w^{\alpha}}
       \frac{\partial X^{\sigma}}{\partial w^{\beta}}+           \label{2.7x}
        \frac{\partial w^{\mu}}{\partial X^{\nu}}
         \frac{{\partial}^2 X^{\nu}}{\partial w^{\alpha}
          \partial w^{\beta}}\,,
\er
where $\hat{\Gamma}^{\mu}_{\alpha\beta}(u,\bm w)$ is given by equation (\ref{oi2}), and
\br\label{oim1}
C^{\nu}_{\rho\sigma}(T,{\bm X})&=&\frac12G^{\mu\nu}\left(\frac{\p G_{\mu\rho}}{\p w^\sigma}+\frac{\p G_{\mu\sigma}}{\p w^\rho}-\frac{\p G_{\rho\sigma}}{\p w^\mu}\right)\;,
\er
is the Christoffel symbol expressed in the geocentric coordinates. The law of transformation of the Christoffel connection, equation
(\ref{2.7x}), being substituted to the gauge condition (\ref{gau}) yields a partial
differential equation of the second order
\be
   \hat g^{\alpha\beta}(u,{\bm w})\,
    \frac{{\partial}^2 X^{\mu}}{\partial w^{\alpha}             \label{2.9x}
     \partial w^{\beta}}=0\;,
\en
which describes the residual gauge freedom that remains in the post-Newtonian transformations from the GRF to EMB coordinates.

Matching is done in accordance with the following procedure (compare with the matching of the EMB and SSB coordinates):
\begin{enumerate}
\item[\it Step 1.] One writes down the most general solution admitted by the homogeneous equation (\ref{2.9x}). In the first post-Newtonian approximation this equation is reduced to two wave equations
    \br\label{le1}
-\frac{1}{c^2}\frac{\p T(u,{\bm w})}{\p u^2}+\nabla^2 T(u,{\bm w})&=&0\;,\\\label{le2}
-\frac{1}{c^2}\frac{\p {\bm X}(u,{\bm w})}{\p u^2}+\nabla^2{\bm X}(u,{\bm w})&=&0\;.
\er
Solution of these equations is given in the form of functions $T=T(u,{\bm w})$ and ${\bm X}={\bm X}(u,{\bm w})$ which are harmonic polynomials with respect to the EMB spatial coordinates ${\bm w}$ with the coefficients which are STF Cartesian tensors depending on time $u$.
\item[\it Step 2.] One re-writes the GRF metric tensor $G_{\alpha\beta}(T, {\bm X})$ and the scalar field $\Phi(T,{\bm X})$ in the right side of equations (\ref{2.5x})
and (\ref{2.6x}) in terms of the EMB coordinates $(u,{\bm w})$. This is achieved by making use of a Taylor
expansion of $\Phi(T,{\bm X})$ and $G_{\alpha\beta}(T,{\bm X})$ around the point $w^\alpha=(cu,{\bm w})$.
\item[\it Step 3.] One calculates the partial derivatives of the GRF coordinates with respect to the EMB coordinates, that is the matrix of transformation of the coordinate bases: $\p X^\alpha/\p w^\beta$.
\item[\it Step 4.] One separates the internal (Earth-related) scalar field and the metric tensor in the left side of equations (\ref{2.5x}) and (\ref{2.6x}) from the external part generated by Moon, Sun, and other planets:
\br
\hat\varphi(u,{\bm w})&=&\hat\varphi_{E}(u,{\bm w})+{\hat\varphi}_{\sss M}(u,{\bm w})+{\hat\varphi}^{\mathrm{(ext)}}(u,{\bm w})\,,                     \label{3.1xa}
\\\nonumber\\
\hat{h}_{00}(u,{\bm w})&=&\hat{h}^E_{00}(u,{\bm w})+{\hat{h}}^M_{00}(u,{\bm w})+{\hat{h}}^{\mathrm{(ext)}}_{00}(u,{\bm w})\,,                     \label{3.1x}
\\\nonumber\\
\hat{h}_{0i}(u,{\bm w})&=&\hat{h}^E_{0i}(u,{\bm w})+{\hat{h}}^M_{0i}(u,{\bm w})+{\hat{h}}^{\mathrm{(ext)}}_{0i}(u,{\bm w})\,,                      \label{3.1qop}
\\\nonumber\\
\hat{h}_{ij}(u,{\bm w})&=&\hat{h}^E_{ij}(u,{\bm w})+{\hat{h}}^M_{ij}(u,{\bm w})+{\hat{h}}^{\mathrm{(ext)}}_{ij}(u,{\bm w})\,,                      \label{3.4x}\,
\\\nonumber\\
{\hat{l}}_{00}(u,{\bm w})&=&{\hat{l}}^{E}_{00}(u,{\bm w})+{\hat{l}}^{M}_{00}(u,{\bm w})+{\hat{l}}^{\mathrm{(ext)}}_{00}(u,{\bm w})+{\hat{l}}^{\mathrm{(mix)}}_{00}(u,{\bm w})
\er
where functions with index $E$ are integrals taken over the
volume $V_{\sss E}$ of Earth, functions with index $M$ are integrals taken over the
volume $V_{\sss M}$ of Moon, functions with index ${\mathrm{(ext)}}$ are given by equations (\ref{1.24b})--(\ref{1.24c}), and functions with index ${\mathrm{(mix)}}$ are given by equations (\ref{1.24o}).
\item[\it Step 5.] One expands the gravitational potentials of the external masses (that is Moon, Sun, and other planets) in Taylor's series in powers of the spatial distance $r^i_{E}=w^i-w^i_{E}$
in the neighborhood  of the origin of the GRF, that is the point with spatial coordinates $w^i_{E}=w^i_{E}(u)$.
\item[\it Step 6.] One equates similar terms in the left and right sides of the matching equations (\ref{2.5x}) and (\ref{2.6x}) depending on internal structure of Earth and on the Taylor expansions from {\it step 5} of this procedure. It is remarkable that all terms depending on the internal structure of Earth as well as all mixed (coupling) terms will cancel out. What  remains is a set of algebraic and ordinary differential equations, which contain the coefficients of the post-Newtonian coordinate transformation and the external multipoles.
\item[\it Step 7.] One separates the algebraic equations into irreducible parts and
determine the external multipoles $P^L_{\sss E},$ $Q^L_{\sss E},$ $C^L_{\sss E}$ in the local GRF metric tensor and the scalar field
as well as the time-dependent coefficients in the coordinate transformations. This fixes the residual gauge freedom and brings about the laws of translational and precessional motion of the GRF -- geocentric reference frame.
\end{enumerate}

Technically, the matching procedure is identical with that executed in section \ref{embssb}. Therefore, it yields equations, which have the same functional structure as in matching between the EMB and SSB coordinates. Because of this similarity we can get post-Newtonian transformation from the GRF to the EMB frame by making use of the following replacements of coordinates, velocities, and accelerations in the coordinate transformations (\ref{5.12}), (\ref{5.13}) from the EMB to the SSB frame
\begin{eqnarray}
      \label{babka}
      \bm{x} & \rightarrow & \bm{w},\\\nonumber
      \bm{x}_{\sss{B}} & \rightarrow & \bm{w}_{\sss E},\\\nonumber
      \bm{R}_{\sss{B}} & \rightarrow & \bm{r}_{\sss E},\\\nonumber
      \bm{v}_{\sss{B}} & \rightarrow & \bm{\nu}_{\sss E},\\\nonumber
      \bm{a}_{\sss{B}} & \rightarrow & \bm{\alpha}_{\sss E}.\\\nonumber
      \end{eqnarray}
One has to make replacement of the differential equations (\ref{5.14}), (\ref{5.7a}), (\ref{5.18})
\begin{eqnarray}
      \label{babka1}
      \frac{d\mathcal{A}}{dt} & \rightarrow & \frac{d\mathcal{A}_{\sss E}}{du}-\frac{1}{2}\hat{h}^{\mathrm{(ext)}}_{00}(\bm{w}_{\sss E}),\\\label{babka2}
      \frac{d\mathcal{B}}{dt} & \rightarrow & \frac{d\mathcal{B}_{\sss E}}{du}+\hat{\mathfrak{f}}_{\mathcal{B}}(\bm{w}_{\sss E}),\\\label{babka3}
      \varepsilon_{ipk}C^p+\frac{dF^{ik}}{dt} & \rightarrow &\varepsilon_{ipk}C_{\sss E}^p+\frac{dF_{\sss E}^{ik}}{du}-\hat{\mathfrak{f}}^{[i,k]},
\er
where the time derivatives of functions $\mathcal{A}_{\sss E}$, $\mathcal{B}_{\sss E}$, and $F_{\sss E}^{ik}$ are given by the right sides of those equations after replacing
\begin{eqnarray}
      \label{babka4}
      \bar{U} & \rightarrow & \hat{U}_{\sss M},\\\nonumber
      \bar{U}^i & \rightarrow & \hat{U}^i_{\sss M},\nonumber
    \end{eqnarray}
and we have introduced two auxiliary functions
\begin{eqnarray}
      \label{babka5}
      \hat{\mathfrak{f}}_{\mathcal{B}} & \equiv& \frac{1}{2}\hat{l}_{00}^{(\mathrm{ext})}+
      \frac{d\mathcal{A}_{\sss E}}{du}\hat{h}^{(\mathrm{ext})}_{00}+\nu^k_{\sss E}\hat{h}^{(\mathrm{ext})}_{0k}
      +\frac{1}{6}\hat{h}^{(\mathrm{ext})}_{kk}\;,\\\l{babka6}
      \hat{\mathfrak{f}}^{i} & \equiv & \hat{h}^{(\mathrm{ext})}_{0i}+\nu_{\sss E}^i\hat{h}^{(\mathrm{ext})}_{00}+\nu^k_{\sss E}\hat{h}^{(\mathrm{ext})}_{ik}\;.
      \er
The coordinate functions $\mathcal{B}^{<iL>}$ are replaced as follows
\be\l{babka7}
\mathcal{B}^{<iL>} \rightarrow  \mathcal{B}^{<iL>}_{\sss E}+\hat{\mathfrak{f}}^{<i,L>}(\bm{w}_{\sss E}),\;.
\en
The external multipole moments are replaced according to the following rules:
\br\l{babka8}
P^L & \rightarrow & P^L_{\sss E}-\hat{\varphi}^{(\mathrm{ext})}_{,<L>}(\bm{w}_{\sss E})\;,\\\l{babka9}
	Q^L & \rightarrow & Q^L_{\sss E}-\hat{\mathfrak{f}}_Q^{<L>}(\bm{w}_{\sss E})\qquad(l\ge1)\;,\\\l{babka10}
      \varepsilon_{ipi_l}C^{pL-1} & \rightarrow & \varepsilon_{ipi_l}C_{\sss E}^{pL-1}-\frac{2l}{l+1}\hat{\mathfrak{f}}^{[i,i_l]L-1}(\bm{w}_{\sss E})
      -\frac{4l}{(l+1)(2l+1)}\delta_{i_{l-1}[i}\hat{\mathfrak{f}}^k_{\phantom{k},i_l]kL-2}(\bm{w}_{\sss E})\qquad(l\ge2),\\
	\end{eqnarray}
where
\be\l{babka11}
      \hat{\mathfrak{f}}^{<L>}_Q \equiv\frac{1}{2}\hat{h}^{(\mathrm{ext})}_{00,L}
      +\epsilon^2\left[\frac{1}{2}\hat{l}^{(\mathrm{ext})}_{00,L}+\frac{d\mathcal{A}_{\sss E}}{du}\hat{h}_{00,L}^{(\mathrm{ext})}+\alpha_{\sss E}^{<i_l}\hat{h}_{00}^{(\mathrm{ext}),L-1>}
      +\nu^k_{\sss E}\hat{h}_{0k,L}^{(\mathrm{ext})}+\frac{1}{6}\nu^2_{\sss E}\hat{h}_{kk,L}^{(\mathrm{ext})}\right]\qquad(l\ge1)\;.
    \en
We have confirmed these rules of replacement by straightforward calculations. Next two subsections give explicitly the post-Newtonian coordinate transformations from the GRF to the EMB frame and the external multipoles of the GRF metric tensor.

\subsubsection{Post-Newtonian Coordinate Transformation}\label{ctram}

The parameterized post-Newtonian coordinate transformations from the geocentric to the EMB frame are given by two equations:
\br                                                 \label{grf1}
T & = & u-\epsilon^2(\mathcal{A}_{\sss E}+\nu^k_{\sss E}r_{\sss E}^k)\nonumber\\
&&+\epsilon^4\bigg[\mathcal{B}_{\sss E}+\bigg(\frac{1}{3}\nu^k_{\sss E}\alpha^k_{\sss E}-\frac{1}{6}\frac{d}{du}{\hat{U}}_{\sss M}({\bm{w}}_{\sss E})
+\frac{1}{6}\frac{d{Q}_{\sss E}}{du}\bigg)r_{\sss E}^2-\frac{1}{10}\dot{\alpha}^k_{\sss E}r^k_{\sss E}r^2_{\sss E}+\sum_{l=1}^{\infty}\frac{1}{l!}\mathcal{B}_{\sss E}^Lr_{\sss E}^L\bigg]+O(\epsilon^5)\;,
   \\\nonumber\\                                         \label{grf2}
   X^i&=&r^i_{\sss E}+\epsilon^2\bigg[\bigg(\frac{1}{2}\nu^i_{\sss E}\nu_{\sss E}^k+\gamma\delta^{ik}\hat{U}_{\sss M}({\bm{w}}_{\sss E})
   -\delta^{ik}A_{\sss E}+F_{\sss E}^{ik}\bigg)r_{\sss E}^k+\alpha^k_{\sss E}r^i_{\sss E}r^k_{\sss E}-\frac{1}{2}\alpha^i_{\sss E}r_{\sss E}^2\bigg]+O(\epsilon^4)\;.
   \er
Here functions ${\cal A}_{\sss E}$ and ${\cal B}_{\sss E}$ are solutions of the ordinary differential equations
\br
\frac{d\mathcal{A}_{\sss E}}{du} & = &+\frac{1}{2}\nu_{\sss E}^2+\hat{U}_{\sss M}(\bm{w}_{\sss E})-Q_{\sss E}+\sum_{k=0}^{\infty}\frac{1}{k!}Q^Kw_{\sss E}^K\,,             \label{grf3}
\er
and
\br
\frac{d\mathcal{B}_{\sss E}}{du} & = &                                         \label{grf4}
        -\frac{1}{8}\nu^4_{\sss E}-\bigg(\gamma+\frac{1}{2}\bigg)\nu_{\sss E}^2\hat{U}_{\sss M}(\bm{w}_{\sss E})+\frac{1}{2}\hat{U}_{\sss M}^2(\bm{w}_{\sss E})+Q_{\sss E}\bigg[-\frac{1}{2}\nu^2_{\sss E}+\frac{1}{2}Q_{\sss E}-\hat{U}_{\sss M}(\bm{w}_{\sss E})\bigg]\nonumber\\
  &&+2(1+\gamma)\nu_{\sss E}^k\hat{U}_{\sss M}^k(\bm{w}_{\sss E})-\hat{\Psi}_{\sss M}(\bm{w}_{\sss E})+\frac{1}{2}\hat{\chi}_{M,uu}(\bm{w}_{\sss E})\nonumber\\
  &&-A\nu_{\sss E}^2-Q_{\sss E}\sum_{k=0}^{\infty}\frac{1}{k!}Q^Kw_{\sss E}^K+Q\bigg[-\frac{1}{2}\nu_{\sss E}^2+\frac{1}{2}Q-\hat{U}_{\sss M}(\bm{w}_{\sss E})+\sum_{k=1}^{\infty}\frac{1}{k!}Q^Kw_{\sss E}^K\bigg]\nonumber\\
  &&-2(\beta-1)P\hat{U}_{\sss M}(\bm{w}_{\sss E})-\bigg(\frac{dA}{du}+\frac{1}{3}\frac{dQ}{du}+\frac{1-\gamma}{3}\frac{dP}{du}\bigg)\nu_{\sss E}^kw_{\sss E}^k\nonumber\\
  &&-\bigg[(\gamma-1)\nu_{\sss E}^2+2(\beta-1)\hat{U}_{\sss M}(\bm{w}_{\sss E})\bigg]\sum_{k=1}^{\infty}\frac{1}{k!}P^Kw_{\sss E}^K+\frac{1}{2}\bigg(\sum_{k=1}^{\infty}\frac{1}{k!}Q^Kw_{\sss E}^K\bigg)^2\nonumber\\
  &&-\bigg[\frac{3}{2}\nu^2_{\sss E}+\hat{U}_{\sss M}(\bm{w}_{\sss E})\bigg]\sum_{k=1}^{\infty}\frac{1}{k!}Q^Kw_{\sss E}^K-\sum_{k=1}^{\infty}\frac{1}{k!}\nu_{\sss E}^m\varepsilon_{mpi_k}C^{pK-1}w_{\sss E}^{K}\nonumber\\
  &&-2\sum_{k=1}^{\infty}\frac{2k+1}{(2k+3)(k+1)!}\nu_{\sss E}^mw_{\sss E}^m\bigg[2\frac{dQ}{du}+(\gamma-1)\frac{dP}{du}\bigg]w_{\sss E}^K\nonumber\\
  &&+2\sum_{k=1}^{\infty}\frac{k}{(2k+3)(k+1)!}w_{\sss E}^2\nu_{\sss E}^{i_k}\bigg[2\frac{dQ^{K}}{du}+(\gamma-1)\frac{dP^{K}}{du}\bigg]w_{\sss E}^{<K-1>}\nonumber\\
  &&-\sum_{k=1}^{\infty}\frac{1}{k!}\nu_{\sss E}^iZ^{iK}w_{\sss E}^K-\frac{1}{2}\sum_{k=0}^{\infty}\frac{1}{(2k+3)k!}\frac{d^2Q^K}{du^2}w_{\sss E}^Kw_{\sss E}^2\;,
\er
that describe the post-Newtonian transformation between the coordinate time $T$ of the geocentric frame and the coordinate time $u$ of the EMB frame.
Other functions entering equations (\ref{grf1}), (\ref{grf2}) are defined by algebraic relationships as follows
\br
{\cal B}^i_{\sss E} & = &2(1+\gamma)\hat{U}_{\sss M}^i(\bm{w}_{\sss E})-(1+2\gamma)\nu^i_{\sss E}\hat{U}_{\sss M}(\bm{w}_{\sss E})-\frac{1}{2}\nu^i_{\sss E}\nu^2_{\sss E}-\nu^i_{\sss E}Q_{\sss E}\nonumber\\
  &&-\nu^i_{\sss E}\bigg[+2A+Q+3\sum_{k=1}^{\infty}\frac{1}{k!}Q^Kw_{\sss E}^K+2(\gamma-1)\sum_{k=1}^{\infty}\frac{1}{k!}P^Kw_{\sss E}^K\bigg]\nonumber\\
  &&-w_{\sss E}^i\bigg(\frac{dA}{du}+\frac{1}{3}\frac{dQ}{du}+\frac{1-\gamma}{3}\frac{dP}{du}\bigg)-\sum_{k=1}^{\infty}\frac{1}{k!}\varepsilon_{ipi_k}C^{pK-1}w_{\sss E}^K\nonumber\\
  &&-2\mathfrak{G}^{i}(\bm{w}_{\sss E})-\sum_{k=1}^{\infty}\frac{1}{k!}Z^{iK}w_{\sss E}^K\,,\label{grf5}\\
{\cal B}^{ij}_{\sss E} & = &
       2(1+\gamma)[\hat{U}_{\sss M}^{<i,j>}(\bm{w}_{\sss E})-\nu_{\sss E}^{<i}\hat{U}_{\sss M}^{,j>}(\bm{w}_{\sss E})]+2\alpha_{\sss E}^{<i}\alpha_{\sss E}^{j>}+Z_{\sss E}^{ij}\nonumber\\
  &&-2\sum_{k=0}^{\infty}\frac{1}{k!}[2\nu_{\sss E}^{<i}Q^{j>K}+(\gamma-1)\nu_{\sss E}^{<i}P^{j>K}]w_{\sss E}^K-\sum_{k=0}^{\infty}\frac{1}{k!}Z^{ijK}w_{\sss E}^K\nonumber\\
  &&-\sum_{k=0}^{\infty}\frac{1}{(k+2)k!}\varepsilon_{ipi_{k+1}}C^{jpK}w_{\sss E}^{K+1}-2\mathfrak{G}^{<i,j>}(\bm{w}_{\sss E})\,,\label{grf6}\\
  {\cal B}^{iL}_{\sss E} & = &
       2(1+\gamma)[\hat{U}_{\sss M}^{<i,L>}(\bm{w}_{\sss E})-\nu_{\sss E}^{<i}\hat{U}_{\sss M}^{,L>}(\bm{w}_{\sss E})]+Z_{\sss E}^{iL}\nonumber\\
  &&-2\sum_{k=0}^{\infty}\frac{1}{k!}[2\nu_{\sss E}^{<i}Q^{L>K}+(\gamma-1)\nu_{\sss E}^{<i}P^{L>K}]w_{\sss E}^K-\sum_{k=0}^{\infty}\frac{1}{k!}Z^{iLK}w_{\sss E}^K\nonumber\\
  &&-\sum_{k=0}^{\infty}\frac{1}{(k+l+1)k!}\varepsilon_{ipi_{k+1}}C^{pLK}w_{\sss E}^{K+1}-2\mathfrak{G}^{<i,L>}(\bm{w}_{\sss E})\;\label{grf7}
          , \qquad\qquad\quad (l\ge 2)
\er
where we have left some residual gauge freedom parameterized by STF Cartesian tensors $Z^L$ and $Z^L_{\sss E}$ ($l\ge 2$), and introduced a shorthand notation for an auxiliary function
\begin{equation}
  \label{}
  \mathfrak{G}^i\equiv\sum_{k=1}^{\infty}\frac{2k+1}{(2k+3)(k+1)!}\bigg[2\frac{dQ^K}{du}+(\gamma-1)\frac{dP^K}{du}\bigg]w^{<iK>}.
\end{equation}

The anti-symmetric matrix ${F}^{ik}_{\sss E}$ describes kinematic rotation of the GRF frame with respect to the EMB frame. It is represented as an algebraic sum of terms produced by the gravitational field of Moon alone and the terms associated with the external bodies and the dynamic rotation of the EMB frame. The matrix ${F}^{ik}_{\sss E}$ of the kinematic rotation couples with the matrix of dynamic rotation $\varepsilon_{ipj}C_{\sss E}^p$ of the GRF yielding the following result
\br\label{5.18x}
\varepsilon_{ipj}C_{\sss E}^p+\frac{dF^{ij}_{\sss E}}{du} & = & -2(1+\gamma)\hat{U}_{\sss M}^{[i,j]}(\bm{w}_{\sss E})+(1+2\gamma)\nu_{\sss E}^{[i}\hat{U}^{,j]}_{\sss M}(\bm{w}_{\sss E})+\nu_{\sss E}^{[i}Q_{\sss E}^{j]}\nonumber\\
  &&+3\sum_{k=0}^{\infty}\frac{1}{k!}\nu_{\sss E}^{[i}Q^{j]K}w_{\sss E}^K+2(\gamma-1)\sum_{k=0}^{\infty}\nu_{\sss E}^{[i}P^{j]K}w_{\sss E}^K\nonumber\\
  &&+\sum_{k=0}^{\infty}\frac{1}{(k+1)!}\varepsilon_{ipj}C^{pK}w_{\sss E}^K+2\mathfrak{G}^{[i,j]}(\bm{w}_{\sss E})\;,
\er
The first two terms in the right side of  equation (\ref{5.18x})
describe the Lense-Thirring (gravitomagnetic) and the de Sitter (geodetic) precessions caused by Moon. The third term is the Thomas precession due to the non-geodesic motion of the origin of the GRF. The forth and the fifth terms describe the de Sitter
precessions due to the external mass-type multipoles of Sun and other planets. The sixth term is the Lense-Thirring precession caused by the external current-type multipoles of Sun and other planets. The last term is the relativistic precession caused by the time evolution of the external multipoles as the origin of the GRF moves along its world line. We should make a choice of the rotation of the GRF axes. Our preference is to make it dynamically non-rotating so that the Coriolis and centrifugal forces are eliminated from the equations of motion of artificial satellites of Earth. This preference is realized with equating $C^i_{\sss E}=0$. With this choice equation (\ref{5.18x}) defines the matrix of the kinematic rotation.

\subsubsection{The External Multipoles}

Matching determines the external geocentric multipoles in terms of the derivatives of gravitational potentials of external bodies that are Moon, Sun and other planets. The external GRF multipoles of the scalar field are
\be
       P_{\sss E}^L=U_{\sss M}^{,L}({\bm w}_{\sss E})+\sum_{k=0}\frac{1}{k!}P^{LK}w^K_{\sss E}+O(\epsilon^2)\,,                     \label{3.13x}
\en
and are taken at the origin of the GRF, $w^i_{E}=w^i_{E}(u)$, at the instant of time $u$. We emphasize that the lower-order ($l=0,1$) external multipoles of the scalar field can not be removed by making coordinate transformation as the scalar field does not depend on the choice of coordinates.

The matching equation for the metric tensor determines the external dipole moment of the GRF metric tensor as follows
\br\label{5.8x}
Q_{\sss E}^i & = & \hat{U}_{\sss M}^{,i}(\bm{w}_{\sss E})-\alpha^i_{\sss E}+\sum_{k=0}^{\infty}\frac{1}{k!}Q^{iK}w_{\sss E}^K\nonumber\\
  &&+\epsilon^2\bigg\{\hat{\Psi}_{\sss M}^{,i}(\bm{w}_{\sss E})-\frac{1}{2}\hat{\chi}_{\sss M}^{,iuu}(\bm{w}_{\sss E})+Q_{\sss E}^i[A_{\sss E}-\nu_{\sss E}^2-2\hat{U}_{\sss M}(\bm{w}_{\sss E})]+2(1+\gamma)\frac{d\hat{U}_{\sss M}^i(\bm{w}_{\sss E})}{du}\nonumber\\
  &&\phantom{+\epsilon^2\bigg\{}-2(1+\gamma)\nu^k_{\sss E}\hat{U}_{\sss M}^{k,i}(\bm{w}_{\sss E})-(1+2\gamma)\nu^i_{\sss E}\frac{d\hat{U}_{\sss M}(\bm{w}_{\sss E})}{du}+(2-\gamma)\hat{U}_{\sss M}(\bm{w}_{\sss E})\hat{U}_{\sss M}^{,i}(\bm{w}_{\sss E})\nonumber\\
  &&\phantom{+\epsilon^2\bigg\{}+(2+\gamma)\nu^2_{\sss E}\hat{U}_{\sss M}^{,i}(\bm{w}_{\sss E})-\frac{1}{2}\nu^i_{\sss E}\nu^k_{\sss E}\hat{U}_{\sss M}^{,k}(\bm{w}_{\sss E})-\frac{1}{2}\nu^i_{\sss E}\nu^k_{\sss E}\alpha^k_{\sss E}-2\nu^2_{\sss E}\alpha^i_{\sss E}\nonumber\\
  &&\phantom{+\epsilon^2\bigg\{}-(4+\gamma)\alpha^i_{\sss E}\hat{U}_{\sss M}(\bm{w}_{\sss E})+F_{\sss E}^{ik}\hat{U}_{\sss M}^{,k}(\bm{w}_{\sss E})-F_{\sss E}^{ik}\alpha^k_{\sss E}+\sum_{k=0}^{\infty}\frac{1}{k!}F_{\sss E}^{im}Q^{mK}w_{\sss E}^K\nonumber\\
  &&\phantom{+\epsilon^2\bigg\{} -w_{\sss E}^i\bigg(\frac{d^2A}{du^2}+\frac{1-\gamma}{3}\frac{d^2P}{du^2}\bigg)-\nu_{\sss E}^i\bigg(\frac{dA}{du}+\frac{2}{3}\frac{dQ}{du}-\frac{1-\gamma}{3}\frac{dP}{du}\bigg)-\nu_{\sss E}^i\bigg[\frac{1}{2}\sum_{k=0}^{\infty}\nu_{\sss E}^mQ^{mK}w_{\sss E}^K\nonumber\\
  &&\phantom{+\epsilon^2\bigg\{} +\sum_{k=1}^{\infty}\frac{6k^2+7k+5}{(2k+3)(k+1)!}\frac{dQ^K}{du}w_{\sss E}^K+2(\gamma-1)\sum_{k=1}\frac{2k^2+3k+2}{(2k+3)(k+1)!}\frac{dP^K}{du}w_{\sss E}^K\bigg]\nonumber\\
  &&\phantom{+\epsilon^2\bigg\{} -\alpha_{\sss E}^i\bigg[2A+2Q+4\sum_{k=1}^{\infty}\frac{1}{k}Q^Kw_{\sss E}^K+2(\gamma-1)\sum_{k=1}^{\infty}\frac{1}{k!}P^Kw_{\sss E}^K\bigg]\nonumber\\
  &&\phantom{+\epsilon^2\bigg\{} +2\hat{U}_{\sss M}^{,i}(\bm{w}_{\sss E})\sum_{k=0}^{\infty}\frac{1}{k!}\bigg[Q^K+(\beta-1)P^K\bigg]w_{\sss E}^K-AQ^i_{\sss E}-Q_{\sss E}^i\sum_{k=0}^{\infty}\frac{1}{k!}\bigg[Q^K+(\gamma-1)P^K\bigg]w_{\sss E}^K\nonumber\\
  &&\phantom{+\epsilon^2\bigg\{} +\bigg[3\nu_{\sss E}^2+(4-\gamma)\hat{U}_{M}(\bm{w}_{\sss E})\bigg]\sum_{k=0}^{\infty}\frac{1}{k!}Q^{iK}w_{\sss E}^K+\bigg[(\gamma-1)\nu_{\sss E}^2+2(\beta-1)\hat{U}_{M}(\bm{w}_{\sss E})\bigg]\sum_{k=0}^{\infty}\frac{1}{k!}P^{iK}w_{\sss E}^K\nonumber\\
  &&\phantom{+\epsilon^2\bigg\{} -\sum_{k=1}^{\infty}\frac{1}{k!}\varepsilon_{ipi_k}\frac{dC^{pK-1}}{du}w_{\sss E}^K+\sum_{k=1}^{\infty}\frac{1}{k!}\varepsilon_{ipi_k}\nu_{\sss E}^pC^{K}w_{\sss E}^{<K-1>}+\sum_{k=0}^{\infty}\frac{1}{(k+2)k!}\varepsilon_{mpq}\nu_{\sss E}^mC^{ipK}w_{\sss E}^{<qK>}\nonumber\\
  &&\phantom{+\epsilon^2\bigg\{} +\sum_{k=0}^{\infty}\frac{1}{k!}\nu_{\sss E}^mZ^{imK}w_{\sss E}^K-\sum_{k=1}^{\infty}\frac{1}{k!}\frac{dZ^{iK}}{du}w_{\sss E}^K+\frac{1}{2}w_{\sss E}^2\sum_{k=0}^{\infty}\frac{1}{(2k+3)k!}\frac{d^2Q^{iK}}{du^2}w_{\sss E}^K\nonumber\\
  &&\phantom{+\epsilon^2\bigg\{} -\sum_{k=1}^{\infty}\frac{7k+3}{(2k+3)(k+1)!}\frac{d^2Q^K}{du^2}w_{\sss E}^{<iK>}-2(\gamma-1)\sum_{k=1}^{\infty}\frac{2k+1}{(2k+3)(k+1)!}\frac{d^2P^K}{du^2}w_{\sss E}^{<iK>}\nonumber\\
  &&\phantom{+\epsilon^2\bigg\{} +2\nu_{\sss E}^mw_{\sss E}^m\sum_{k=1}^{\infty}\frac{k(2k+1)}{(2k+3)(k+1)!}\bigg[2\frac{d}{du}Q^{iK-1}+(\gamma-1)\frac{d}{du}P^{iK-1}\bigg]w_{\sss E}^{<K-1>}\nonumber\\
  &&\phantom{+\epsilon^2\bigg\{} -4\sum_{k=1}^{\infty}\frac{k}{(2k+3)(k+1)!}\nu_{\sss E}^{i_k}\bigg[2\frac{d}{du}Q^{K}+(\gamma-1)\frac{d}{du}P^{K}\bigg]w_{\sss E}^{<iK-1>}\nonumber\\
  &&\phantom{+\epsilon^2\bigg\{} -2w_{\sss E}^2\sum_{k=2}^{\infty}\frac{(2k+1)k(k-1)}{(2k+3)(2k-1)(k+1)!}\nu_{\sss E}^m\bigg[2\frac{d}{du}Q^{imK-2}+(\gamma-1)\frac{d}{du}P^{imK-2}\bigg]w_{\sss E}^{<K-2>}\bigg\}+O(\epsilon^4)\;.
\er
The external dipole, $Q_{\sss E}^i$, is explicitly expressed in terms of the external gravitational potentials, velocity $\nu^i_{\sss E}$, and acceleration $\alpha^i_{\sss E}$ of the origin of the GRF with respect to the EMB frame. The dipole $Q_{\sss E}^i$ is not subject any limitation from the gauge condition and can be chosen arbitrary, because it determines the magnitude and direction of the inertial force acting on a test particle being in a free fall with respect to the GRF. Thus, equation (\ref{5.8x}) should be understood as the law of orbital motion of the origin of the GRF with respect to the EMB frame, which is specified by a particular choice of the dipole moment $Q_{\sss E}^i$. As soon as $Q_{\sss E}^i$ is chosen, the coordinate acceleration $\alpha_{\sss E}^i$ of the origin of the GRF with respect to the EMB frame can be determined. Notice that $Q_{\sss E}^i=0$ corresponds to motion of the origin of the GRF along a geodesic world line. This choice, however, does not keep the origin of the GRF at the geocenter because Earth has an intrinsic quadrupole moment interacting with the tidal gravitational field of Moon, Sun, and other planets. Therefore, the geocenter moves along a non-geodesic world line. To keep the origin of the GRF at the geocenter at any instant of time, the dipole $Q_{\sss E}^i$ must be determined from the solution of the internal problem of motion of the geocenter with respect to the GRF. We shall discuss this issue somewhere else.

The mass-type external multipoles $Q^L_{\sss E}$ for the case $l\ge2$ are derived from the matching equations and are defined by the following equation
\br\label{5.9x}
Q_{\sss E}^L & = & \hat{U}_{\sss M}^{,<L>}(\bm{w}_{\sss E})+\sum_{k=0}^{\infty}\frac{1}{k!}Q^{LK}w_{\sss E}^K\nonumber\\
  & & +\epsilon^2\bigg\{\hat{\Psi}_{\sss M}^{,<L>}(\bm{w}_{\sss E})-\frac{1}{2}\hat{\chi}_{\sss M}^{,uu<L>}(\bm{w}_{\sss E})+K_{\sss E}^L+\frac{dZ_{\sss E}^L}{du}+lA_{\sss E}\hat{U}_{\sss M}^{,<L>}(\bm{w}_{\sss E})\nonumber\\
  & & \phantom{+\epsilon^2\bigg\{} +2(1+\gamma)\frac{d}{du}\hat{U}_{\sss M}^{<i_l,L-1>}(\bm{w}_{\sss E})-2(1+\gamma)\nu_{\sss E}^k\hat{U}_{\sss M}^{k,<L>}(\bm{w}_{\sss E})\nonumber\\
  & & \phantom{+\epsilon^2\bigg\{} +(l-2\gamma-2)\nu_{\sss E}^{<i_l}\frac{d}{du}\hat{U}_{\sss M}^{,L-1>}(\bm{w}_{\sss E})+(1+\gamma)\nu_{\sss E}^2\hat{U}_{\sss M}^{,<L>}(\bm{w}_{\sss E})\nonumber\\
  & & \phantom{+\epsilon^2\bigg\{} -\frac{l}{2}\nu^k_{\sss E}\nu_{\sss E}^{<i_l}\hat{U}_{\sss M}^{,L-1>k}(\bm{w}_{\sss E})-l\gamma\hat{U}_{\sss M}(\bm{w}_{\sss E})\hat{U}_{\sss M}^{,<L>}(\bm{w}_{\sss E})\nonumber\\
  & & \phantom{+\epsilon^2\bigg\{} -(l^2-l+2\gamma+2)\alpha_{\sss E}^{<i_l}\hat{U}_{\sss M}^{,L-1>}(\bm{w}_{\sss E})-lF_{\sss E}^{k<i_l}\hat{U}_{\sss M}^{,L-1>k}(\bm{w}_{\sss E})\nonumber\\
  & & \phantom{+\epsilon^2\bigg\{} +\sum_{k=0}\frac{1}{k!}\nu_{\sss E}^m\varepsilon^{mp<i_l}C^{KL-1>p}w_{\sss E}^K-\sum_{k=0}^{\infty}\frac{1}{(k+l)k!}\varepsilon_{i_lpq}\frac{dC^{pKL-1}}{du}w_{\sss E}^{<qK>}\nonumber\\
  & & \phantom{+\epsilon^2\bigg\{} +\sum_{k=0}^{\infty}\frac{1}{k!}\nu_{\sss E}^mZ^{mLK}w_{\sss E}^K-\sum_{k=0}^{\infty}\frac{1}{k!}\frac{dZ^{LK}}{du}w_{\sss E}^K-lA\hat{U}_{\sss M}^{,<L>}(\bm{w}_{\sss E})\nonumber\\
  & & \phantom{+\epsilon^2\bigg\{} -lQ_{\sss E}^L\sum_{k=1}^{\infty}\frac{1}{k!}\bigg[Q^K+(\gamma-1)P^K\bigg]w_{\sss E}^K-l\sum_{k=0}^{\infty}\frac{1}{k!}F_{\sss E}^{m<i_l}Q^{L-1>mK}w_{\sss E}^K\nonumber\\
  & & \phantom{+\epsilon^2\bigg\{} +\bigg[l\bigg(A_{\sss E}-A-\gamma\hat{U}_{\sss M}(\bm{w}_{\sss E})\bigg)+2\nu_{\sss E}^2\bigg]\sum_{k=0}^{\infty}\frac{1}{k!}Q^{LK}w_{\sss E}^K\nonumber\\
  & & \phantom{+\epsilon^2\bigg\{} -\frac{l}{2}\sum_{k=0}^{\infty}\frac{1}{k!}\nu_{\sss E}^m\nu_{\sss E}^{<i_l}Q^{L-1>mK}w_{\sss E}^K+2\sum_{k=0}^l\sum_{r=0}^{\infty}\frac{l!}{(l-k)!k!r!}Q^{R<L-K}\hat{U}_{\sss M}^{,K>}(\bm{w}_{\sss E})w_{\sss E}^R\nonumber\\
  & & \phantom{+\epsilon^2\bigg\{} +2(\beta-1)\sum_{k=0}^l\sum_{r=0}^{\infty}\frac{l!}{(l-k)!k!r!}P^{R<L-K}\hat{U}_{\sss M}^{,K>}(\bm{w}_{\sss E})w_{\sss E}^R-2\frac{d}{du}\mathfrak{G}^{<i_l,L-1>}(\bm{w}_{\sss E})\nonumber\\
  & & \phantom{+\epsilon^2\bigg\{} -(l^2-l+4)\sum_{k=0}^{\infty}\alpha_{\sss E}^{<i_l}Q^{L-1>K}w_{\sss E}^K+(l-4)\sum_{k=0}^{\infty}\frac{1}{k!}\nu_{\sss E}^{<i_l}\frac{d}{du}Q^{L-1>K}w_{\sss E}^K\nonumber\\
  & & \phantom{+\epsilon^2\bigg\{} -2(\gamma-1)\sum_{k=0}^{\infty}\frac{1}{k!}\nu_{\sss E}^{<i_l}\frac{d}{du}P^{L-1>K}w_{\sss E}^K-2(\gamma-1)\sum_{k=0}^{\infty}\frac{1}{k!}\alpha_{\sss E}^{<i_l}P^{L-1>K}w_{\sss E}^K\nonumber\\
  & & \phantom{+\epsilon^2\bigg\{} +(\gamma-1)\nu_{\sss E}^2\sum_{k=0}^{\infty}\frac{1}{k!}P^{LK}w_{\sss E}^K+\frac{1}{2}w_{\sss E}^2\sum_{k=0}^{\infty}\frac{1}{(2k+2l+3)k!}\frac{d^2}{du^2}Q^{LK}w_{\sss E}^K\nonumber\\
  & & \phantom{+\epsilon^2\bigg\{} +\sum_{k=0}^{\infty}\frac{l}{(2k+2l+1)k!}w_{\sss E}^{<i_l}\frac{d^2}{du^2}Q^{L-1>K}w_{\sss E}^K\nonumber\\
  & & \phantom{+\epsilon^2\bigg\{} +2\sum_{k=0}^{\infty}\frac{2k+2l-1}{(2k+2l+1)k!}\bigg[2\nu_{\sss E}^{<i_l}\frac{d}{du}Q^{KL-1>}+(\gamma-1)\nu_{\sss E}^{<i_l}\frac{d}{du}P^{KL-1>}\bigg]w_{\sss E}^K\bigg\}+O\left(\epsilon^4\right)\;,
\er
where we have used notations
\br
    K_{\sss E}^{ij}&\equiv&3\alpha_{\sss E}^{<i}\alpha_{\sss E}^{j>}\,,                      \label{5.10x}
\\\nonumber\\
    K_{\sss E}^{L}&\equiv&0\,,\qquad\qquad\qquad (l\ge3).                          \label{5.11x}
\er

The current-type external multipoles $C_{\sss E}^L$ $(l\ge 1)$ are given by
\br\label{3.29x}
\varepsilon^{ipi_l}C_{\sss E}^{pL-1} & =& \frac{4l(1+\gamma)}{l+1}\bigg[\nu_{\sss E}^{[i}\hat{U}_{\sss M}^{,i_l]L-1}(\bm{w}_{\sss E})-\hat{U}_{\sss M}^{[i,i_l]L-1}(\bm{w}_{\sss E})-\frac{l-1}{l}\delta^{i_{l-1}[i}\dot{\hat{U}}_{\sss M}^{,i_l]L-2}(\bm{w}_{\sss E})\bigg]\nonumber\\
  &&+\sum_{k=0}^{\infty}\frac{l}{(k+l)k!}\varepsilon^{ipi_l}C^{pKL-1}w_{\sss E}^K\nonumber\\
  &&+\frac{4l}{l+1}\bigg[\mathfrak{G}^{[i,i_l]L-1}(\bm{w}_{\sss E})+\frac{2}{2l+1}\delta_{i_{l-1}[i}\mathfrak{G}^k_{\phantom{K},i_l]kL-2}(\bm{w}_{\sss E})\bigg]\nonumber\\
  &&+\frac{4l}{l+1}\sum_{k=0}^{\infty}\frac{1}{k!}\bigg[2\nu_{\sss E}^{[i}Q^{i_l]KL-1}+(\gamma-1)\nu_{\sss E}^{[i}P^{i_l]KL-1}\bigg]w_{\sss E}^K\nonumber\\
  &&+\frac{8l}{(l+1)(2l+1)}\sum_{k=0}^{\infty}\frac{1}{k!}\nu_{\sss E}^m\bigg[2\delta^{i_{l-1}[i}Q^{i_l]mKL-2}+(\gamma-1)\delta^{i_{l-1}[i}P^{i_l]mKL-2}\bigg]w_{\sss E}^K+O\left(\epsilon^2\right)\,,
\er
where the dot above any function denotes a total time derivative with respect to time $u$.

\subsection{Transformation from the Selenocentric to the Earth-Moon Frame}\l{tsem}
\subsubsection{Matching Procedure}\l{mapr}
Matching solutions of the field equations for the metric tensor and the scalar field in the SRF and EMB frame repeats exactly the same steps as the matching of the GRF and the EMB frames. The only change is that Moon is now the internal object and Earth is an external one. All matching equations remain the same as in the previous subsection except that the indices belonging to Moon and to Earth should be exchanged: $E M$. The results of the matching are given below.

\subsubsection{Post-Newtonian Coordinate Transformation}\label{ytram}
The parameterized post-Newtonian coordinate transformations from the SRF to the EMB frame are given by two equations:
\br                                                 \label{lurf1}
\Sigma & = & u-\epsilon^2(\mathcal{A}_{\sss M}+\nu^k_{\sss M}r_{\sss M}^k)\nonumber\\
  &&+\epsilon^4\bigg[\mathcal{B}_{\sss M}+\bigg(\frac{1}{3}\nu^k_{\sss M}\alpha^k_{\sss M}-\frac{1}{6}\frac{d}{du}{\hat{U}}_{\sss E}({\bm{w}}_{\sss M})+\frac{1}{6}\frac{d{Q}_{\sss M}}{du}\bigg)r_{\sss M}^2-\frac{1}{10}\dot{\alpha}^k_{\sss M}r^k_{\sss M}r^2_{\sss M}+\sum_{l=1}^{\infty}\frac{1}{l!}\mathcal{B}_{\sss M}^Lr_{\sss M}^L\bigg]+O(\epsilon^5)\;,
   \\\nonumber\\                                         \label{lurf2}
   Y^i & = & r^i_{\sss M}+\epsilon^2\bigg[\bigg(\frac{1}{2}\nu^i_{\sss M}\nu_{\sss M}^k+\delta^{ik}\gamma\hat{U}_{\sss E}({\bm{w}}_{\sss M})-\delta^{ik}A_{\sss M}+F_{\sss M}^{ik}\bigg)r_{\sss M}^k+\alpha^k_{\sss M}r^i_{\sss M}r^k_{\sss M}-\frac{1}{2}\alpha^i_{\sss M}r_{\sss M}^2\bigg]+O(\epsilon^4)\;.
   \er
Here functions ${\cal A}_{\sss M}$ and ${\cal B}_{\sss M}$ are solutions of the ordinary differential equations
\br
\frac{d\mathcal{A}_{\sss M}}{du} & = & +\frac{1}{2}\nu_{\sss M}^2+\hat{U}_{\sss E}(\bm{w}_{\sss M})-Q_{\sss M}+\sum_{k=0}^{\infty}\frac{1}{k!}Q^Kw_{\sss M}^K\,,             \label{lurf3}
\\\nonumber\\
   \label{lurf4}
   \frac{d\mathcal{B}_{\sss M}}{du} & = & -\frac{1}{8}\nu^4_{\sss M}-\bigg(\gamma+\frac{1}{2}\bigg)\nu_{\sss M}^2\hat{U}_{\sss E}(\bm{w}_{\sss M})+\frac{1}{2}\hat{U}_{\sss E}^2(\bm{w}_{\sss M})+Q_{\sss M}\bigg[-\frac{1}{2}\nu^2_{\sss M}+\frac{1}{2}Q_{\sss M}-\hat{U}_{\sss E}(\bm{w}_{\sss M})\bigg]\nonumber\\
  &&+2(1+\gamma)\nu_{\sss M}^k\hat{U}_{\sss E}^k(\bm{w}_{\sss M})-\hat{\Psi}_{\sss E}(\bm{w}_{\sss M})+\frac{1}{2}\hat{\chi}_{E,uu}(\bm{w}_{\sss M})\nonumber\\
  &&-A\nu_{\sss M}^2-Q_{\sss M}\sum_{k=0}^{\infty}\frac{1}{k!}Q^Kw_{\sss M}^K+Q\bigg[-\frac{1}{2}\nu_{\sss M}^2+\frac{1}{2}Q-\hat{U}_{\sss E}(\bm{w}_{\sss M})+\sum_{k=1}^{\infty}\frac{1}{k!}Q^Kw_{\sss M}^K\bigg]\nonumber\\
  &&-2(\beta-1)P\hat{U}_{\sss E}(\bm{w}_{\sss M})-\bigg(\frac{dA}{du}+\frac{1}{3}\frac{dQ}{du}+\frac{1-\gamma}{3}\frac{dP}{du}\bigg)\nu_{\sss M}^kw_{\sss M}^k\nonumber\\
  &&-\bigg[(\gamma-1)\nu_{\sss M}^2+2(\beta-1)\hat{U}_{\sss E}(\bm{w}_{\sss M})\bigg]\sum_{k=1}^{\infty}\frac{1}{k!}P^Kw_{\sss M}^K+\frac{1}{2}\bigg(\sum_{k=1}^{\infty}\frac{1}{k!}Q^Kw_{\sss M}^K\bigg)^2\nonumber\\
  &&-\bigg[\frac{3}{2}\nu^2_{\sss M}+\hat{U}_{\sss E}(\bm{w}_{\sss M})\bigg]\sum_{k=1}^{\infty}\frac{1}{k!}Q^Kw_{\sss M}^K-\sum_{k=1}^{\infty}\frac{1}{k!}\nu_{\sss M}^m\varepsilon_{mpi_k}C^{pK-1}w_{\sss M}^{K}\nonumber\\
  &&-2\sum_{k=1}^{\infty}\frac{2k+1}{(2k+3)(k+1)!}\nu_{\sss M}^mw_{\sss M}^m\bigg[2\frac{dQ}{du}+(\gamma-1)\frac{dP}{du}\bigg]w_{\sss M}^K\;,
\er
that describe the post-Newtonian transformation between the coordinate time $\Sigma$ of the selenocentric frame and the coordinate time $u$ of the EMB frame.
The other functions are defined by algebraic relationships as follows
\br
\label{lurf5}
\mathcal{B}_{\sss M}^i & = & 2(1+\gamma)\hat{U}_{\sss E}^i(\bm{w}_{\sss M})-(1+2\gamma)\nu^i_{\sss M}\hat{U}_{\sss E}(\bm{w}_{\sss M})-\frac{1}{2}\nu^i_{\sss M}\nu^2_{\sss M}-\nu^i_{\sss E}Q_{\sss M}\nonumber\\
  &&-\nu^i_{\sss M}\bigg[+2A+Q+3\sum_{k=1}^{\infty}\frac{1}{k!}Q^Kw_{\sss M}^K+2(\gamma-1)\sum_{k=1}^{\infty}\frac{1}{k!}P^Kw_{\sss M}^K\bigg]\nonumber\\
  &&-w_{\sss M}^i\bigg(\frac{dA}{du}+\frac{1}{3}\frac{dQ}{du}+\frac{1-\gamma}{3}\frac{dP}{du}\bigg)-\sum_{k=1}^{\infty}\frac{1}{k!}\varepsilon_{ipi_k}C^{pK-1}w_{\sss M}^K\nonumber\\
  &&-2\mathfrak{G}^i(\bm{w}_{\sss M})-\sum_{k=1}^{\infty}\frac{1}{k!}Z^{iK}w_{\sss M}^K\,,
\\\nonumber\\
\label{lurf6}
\mathcal{B}_{\sss M}^{ij} & = & 2(1+\gamma)[\hat{U}_{\sss E}^{<i,j>}(\bm{w}_{\sss M})-\nu_{\sss M}^{<i}\hat{U}_{\sss E}^{,j>}(\bm{w}_{\sss M})]+2\alpha_{\sss M}^{<i}\alpha_{\sss M}^{j>}+Z_{\sss M}^{ij}\nonumber\\
  &&-2\sum_{k=0}^{\infty}\frac{1}{k!}[2\nu_{\sss M}^{<i}Q^{j>K}+(\gamma-1)\nu_{\sss M}^{<i}P^{j>K}]w_{\sss M}^K-\sum_{k=0}^{\infty}\frac{1}{k!}Z^{ijK}w_{\sss M}^K\nonumber\\
  &&-\sum_{k=0}^{\infty}\frac{1}{(k+2)k!}\varepsilon_{ipi_{k+1}}C^{jpK}w_{\sss M}^{K+1}-2\mathfrak{G}^{<i,j>}(\bm{w}_{\sss M})\,,
           \\\nonumber\\\label{lurf7}
	   \mathcal{B}_{\sss M}^{<iL>} & = & 2(1+\gamma)[\hat{U}_{\sss E}^{<i,L>}(\bm{w}_{\sss M})-\nu_{\sss M}^{<i}\hat{U}_{\sss E}^{,L>}(\bm{w}_{\sss M})]+Z_{\sss M}^{iL}\nonumber\\
  &&-2\sum_{k=0}^{\infty}\frac{1}{k!}[2\nu_{\sss M}^{<i}Q^{L>K}+(\gamma-1)\nu_{\sss M}^{<i}P^{L>K}]w_{\sss M}^K-\sum_{k=0}^{\infty}\frac{1}{k!}Z^{iLK}w_{\sss M}^K\nonumber\\
  &&-\sum_{k=0}^{\infty}\frac{1}{(k+l+1)k!}\varepsilon_{ipi_{k+1}}C^{pLK}w_{\sss M}^{K+1}-2\mathfrak{G}^{<i,L>}(\bm{w}_{\sss M})\;
          , \qquad\qquad\quad (l\ge 2)
\er
where we have left some residual gauge freedom parameterized by STF Cartesian tensors $Z^L$.

The anti-symmetric rotational matrix ${F}^{ik}_{\sss M}$ is a linear combination of three terms
\br\label{5.18yt}
\varepsilon_{ipj}C_{\sss M}^p+\frac{dF^{ij}_{\sss M}}{du} & = & -2(1+\gamma)\hat{U}_{\sss E}^{[i,j]}(\bm{w}_{\sss M})+(1+2\gamma)\nu_{\sss M}^{[i}\hat{U}^{,j]}_{\sss E}(\bm{w}_{\sss M})+\nu_{\sss M}^{[i}Q_{\sss M}^{j]}\nonumber\\
  &&+3\sum_{k=0}^{\infty}\frac{1}{k!}\nu_{\sss M}^{[i}Q^{j]K}w_{\sss M}^K+2(\gamma-1)\sum_{k=0}^{\infty}\nu_{\sss M}^{[i}P^{j]K}w_{\sss M}^K\nonumber\\
  &&+\sum_{k=0}^{\infty}\frac{1}{(k+1)!}\varepsilon_{ipj}C^{pK}w_{\sss M}^K+2\mathfrak{G}^{[i,j]}(\bm{w}_{\sss M})\;,
\er
The first two terms in the right side of  equation (\ref{5.18yt})
describe the Lense-Thirring (gravitomagnetic) and the de Sitter (geodetic) precessions caused by Earth. The third term is the Thomas precession due to the non-geodesic motion of the origin of the SRF. The forth and the fifth terms describe the de Sitter
precessions caused by the external mass-type multipoles of Sun and other planets. The sixth term is the Lense-Thirring precession caused by the external current-type multipoles of Sun and other planets. The last term is the relativistic precession caused by the time evolution of the external multipoles as the origin of the SRF moves along its world line. We should make a choice of the rotation of the SRF axes. Our preference is to make it dynamically non-rotating so that the Coriolis and centrifugal forces are eliminated from the equations of motion of artificial satellites of Moon. This preference is realized with equating $C^i_{\sss M}=0$. With this choice equation (\ref{5.18yt}) defines the matrix of the kinematic rotation $F^{ij}_{\sss M}$.

\subsubsection{The External Multipoles}

Matching determines the external selenocentric multipoles in terms of the derivatives of gravitational potentials of external bodies that are Earth, Sun and other planets. The external multipoles of the scalar field are
\be
 P_{\sss M}^L=U_{\sss E}^{,L}({\bm w}_{\sss M})+\sum_{k=0}\frac{1}{k!}P^{LK}w^K_{\sss M}+O(\epsilon^2)\,,                          \label{3.13yt}
\en
where all functions are taken at the origin of the SRF, $w^i_{M}=w^i_{M}(u)$, at the instant of time $u$.

The matching equation determines the external dipole moment of the SRF metric tensor as follows
\br							\label{5.8yt}
 Q_{\sss M}^i & = & \hat{U}_{\sss E}^{,i}(\bm{w}_{\sss M})-\alpha^i_{\sss M}+\sum_{k=0}^{\infty}\frac{1}{k!}Q^{iK}w_{\sss M}^K\nonumber\\
  &&+\epsilon^2\bigg\{+\hat{\Psi}_{\sss E}^{,i}(\bm{w}_{\sss M})-\frac{1}{2}\hat{\chi}_{\sss E}^{,iuu}(\bm{w}_{\sss M})+Q_{\sss M}^i[A_{\sss M}-\nu_{\sss M}^2-2\hat{U}_{\sss E}(\bm{w}_{\sss M})]+2(1+\gamma)\frac{d\hat{U}_{\sss E}^i(\bm{w}_{\sss M})}{du}\nonumber\\
  &&\phantom{+\epsilon^2\bigg\{}-2(1+\gamma)\nu^k_{\sss M}\hat{U}_{\sss E}^{k,i}(\bm{w}_{\sss M})-(1+2\gamma)\nu^i_{\sss M}\frac{d\hat{U}_{\sss E}(\bm{w}_{\sss M})}{du}+(2-\gamma)\hat{U}_{\sss E}(\bm{w}_{\sss M})\hat{U}_{\sss E}^{,i}(\bm{w}_{\sss M})\nonumber\\
  &&\phantom{+\epsilon^2\bigg\{}+(2+\gamma)\nu^2_{\sss M}\hat{U}_{\sss E}^{,i}(\bm{w}_{\sss M})-\frac{1}{2}\nu^i_{\sss M}\nu^k_{\sss M}\hat{U}_{\sss E}^{,k}(\bm{w}_{\sss M})-\frac{1}{2}\nu^i_{\sss M}\nu^k_{\sss M}\alpha^k_{\sss M}-2\nu^2_{\sss M}\alpha^i_{\sss M}\nonumber\\
  &&\phantom{+\epsilon^2\bigg\{}-(4+\gamma)\alpha^i_{\sss M}\hat{U}_{\sss E}(\bm{w}_{\sss M})+F_{\sss M}^{ik}\hat{U}_{\sss E}^{,k}(\bm{w}_{\sss M})-F_{\sss M}^{ik}\alpha^k_{\sss M}+\sum_{k=0}^{\infty}\frac{1}{k!}F_{\sss M}^{im}Q^{mK}w_{\sss M}^K\nonumber\\
  &&\phantom{+\epsilon^2\bigg\{} -w_{\sss M}^i\bigg(\frac{d^2A}{du^2}+\frac{1-\gamma}{3}\frac{d^2P}{du^2}\bigg)-\nu_{\sss M}^i\bigg(\frac{dA}{du}+\frac{2}{3}\frac{dQ}{du}-\frac{1-\gamma}{3}\frac{dP}{du}\bigg)-\nu_{\sss M}^i\bigg[\frac{1}{2}\sum_{k=0}^{\infty}\nu_{\sss M}^mQ^{mK}w_{\sss M}^K\nonumber\\
  &&\phantom{+\epsilon^2\bigg\{} +\sum_{k=1}^{\infty}\frac{6k^2+7k+5}{(2k+3)(k+1)!}\frac{dQ^K}{du}w_{\sss M}^K+2(\gamma-1)\sum_{k=1}\frac{2k^2+3k+2}{(2k+3)(k+1)!}\frac{dP^K}{du}w_{\sss M}^K\bigg]\nonumber\\
  &&\phantom{+\epsilon^2\bigg\{} -\alpha_{\sss M}^i\bigg[2A+2Q+4\sum_{k=1}^{\infty}\frac{1}{k}Q^Kw_{\sss M}^K+2(\gamma-1)\sum_{k=1}^{\infty}\frac{1}{k!}P^Kw_{\sss M}^K\bigg]\nonumber\\
  &&\phantom{+\epsilon^2\bigg\{} +2\hat{U}_{\sss E}^{,i}(\bm{w}_{\sss M})\sum_{k=0}^{\infty}\frac{1}{k!}\bigg[Q^K+(\beta-1)P^K\bigg]w_{\sss M}^K-AQ^i_{\sss M}-Q_{\sss M}^i\sum_{k=0}^{\infty}\frac{1}{k!}\bigg[Q^K+(\gamma-1)P^K\bigg]w_{\sss M}^K\nonumber\\
  &&\phantom{+\epsilon^2\bigg\{} +\bigg[3\nu_{\sss M}^2+(4-\gamma)\hat{U}_{E}(\bm{w}_{\sss M})\bigg]\sum_{k=0}^{\infty}\frac{1}{k!}Q^{iK}w_{\sss M}^K+\bigg[(\gamma-1)\nu_{\sss M}^2+2(\beta-1)\hat{U}_{E}(\bm{w}_{\sss M})\bigg]\sum_{k=0}^{\infty}\frac{1}{k!}P^{iK}w_{\sss M}^K\nonumber\\
  &&\phantom{+\epsilon^2\bigg\{} -\sum_{k=1}^{\infty}\frac{1}{k!}\varepsilon_{ipi_k}\frac{dC^{pK-1}}{du}w_{\sss M}^K+\sum_{k=1}^{\infty}\frac{1}{k!}\varepsilon_{ipi_k}\nu_{\sss M}^pC^{K}w_{\sss M}^{<K-1>}+\sum_{k=0}^{\infty}\frac{1}{(k+2)k!}\varepsilon_{mpq}\nu_{\sss M}^mC^{ipK}w_{\sss M}^{<qK>}\nonumber\\
  &&\phantom{+\epsilon^2\bigg\{} +\sum_{k=0}^{\infty}\frac{1}{k!}\nu_{\sss M}^mZ^{imK}w_{\sss M}^K-\sum_{k=1}^{\infty}\frac{1}{k!}\frac{dZ^{iK}}{du}w_{\sss M}^K+\frac{1}{2}w_{\sss M}^2\sum_{k=0}^{\infty}\frac{1}{(2k+3)k!}\frac{d^2Q^{iK}}{du^2}w_{\sss M}^K\nonumber\\
  &&\phantom{+\epsilon^2\bigg\{} -\sum_{k=1}^{\infty}\frac{7k+3}{(2k+3)(k+1)!}\frac{d^2Q^K}{du^2}w_{\sss M}^{<iK>}-2(\gamma-1)\sum_{k=1}^{\infty}\frac{2k+1}{(2k+3)(k+1)!}\frac{d^2P^K}{du^2}w_{\sss M}^{<iK>}\nonumber\\
  &&\phantom{+\epsilon^2\bigg\{} +2\nu_{\sss M}^mw_{\sss M}^m\sum_{k=1}^{\infty}\frac{k(2k+1)}{(2k+3)(k+1)!}\bigg[2\frac{d}{du}Q^{iK-1}+(\gamma-1)\frac{d}{du}P^{iK-1}\bigg]w_{\sss M}^{<K-1>}\nonumber\\
  &&\phantom{+\epsilon^2\bigg\{} -4\sum_{k=1}^{\infty}\frac{k}{(2k+3)(k+1)!}\nu_{\sss M}^{i_k}\bigg[2\frac{d}{du}Q^{K}+(\gamma-1)\frac{d}{du}P^{K}\bigg]w_{\sss M}^{<iK-1>}\nonumber\\
  &&\phantom{+\epsilon^2\bigg\{} -2w_{\sss M}^2\sum_{k=2}^{\infty}\frac{(2k+1)k(k-1)}{(2k+3)(2k-1)(k+1)!}\nu_{\sss M}^m\bigg[2\frac{d}{du}Q^{imK-2}+(\gamma-1)\frac{d}{du}P^{imK-2}\bigg]w_{\sss M}^{<K-2>}\bigg\}+O(\epsilon^4)\;.
\er
The external dipole, $Q_{\sss M}^i$, is explicitly expressed in terms of the external gravitational potentials and acceleration $\alpha^i_{\sss M}$ of the origin of the SRF with respect to the EMB frame. The dipole $Q_{\sss M}^i$ can be chosen arbitrary and determines the magnitude and direction of the inertial force acting on a test particle being in a free fall with respect to the selenocentric frame. Thus, equation (\ref{5.8yt}) should be understood as the law of orbital motion of the origin of the SRF in the EMB coordinates, which is governed by a particular choice of the dipole moment $Q_{\sss M}^i$. As soon as $Q_{\sss M}^i$ is chosen, the coordinate acceleration $\alpha_{\sss M}^i$ of the origin of the SRF with respect to the EMB coordinates is fully defined. The choice $Q_{\sss M}^i=0$ means that the origin of the SRF moves along a geodesic world line. The Moon has intrinsic quadrupole moment $J_2$ interacting with tidal gravitational field of Earth, Sun, and other planets. Therefore, the selenocenter does not move along a geodesic world line. To keep the origin of the SRF at the selenocenter at any instant of time, $Q_{\sss M}^i$ must be determined from the solution of the internal problem of motion of the selenocenter with respect to the SRF. We shall discuss this issue somewhere else.

The mass-type external multipoles $Q^L_{\sss M}$ for the case $l\ge2$ are derived from the matching equations and are defined by the following equation
\br\label{5.9yt}
    Q_{\sss M}^L & = & \hat{U}_{\sss E}^{,<L>}(\bm{w}_{\sss M})+\sum_{k=0}^{\infty}\frac{1}{k!}Q^{LK}w_{\sss M}^K\nonumber\\
  & & +\epsilon^2\bigg\{+\hat{\Psi}_{\sss E}^{,<L>}(\bm{w}_{\sss M})-\frac{1}{2}\hat{\chi}_{\sss E}^{,uu<L>}(\bm{w}_{\sss M})+K_{\sss M}^L+\frac{dZ_{\sss M}^L}{du}+lA_{\sss M}\hat{U}_{\sss E}^{,<L>}(\bm{w}_{\sss M})\nonumber\\
  & & \phantom{+\epsilon^2\bigg\{} +2(1+\gamma)\frac{d}{du}\hat{U}_{\sss E}^{<i_l,L-1>}(\bm{w}_{\sss M})-2(1+\gamma)\nu_{\sss M}^k\hat{U}_{\sss E}^{k,<L>}(\bm{w}_{\sss M})\nonumber\\
  & & \phantom{+\epsilon^2\bigg\{} +(l-2\gamma-2)\nu_{\sss M}^{<i_l}\frac{d}{du}\hat{U}_{\sss E}^{,L-1>}(\bm{w}_{\sss M})+(1+\gamma)\nu_{\sss M}^2\hat{U}_{\sss E}^{,<L>}(\bm{w}_{\sss M})\nonumber\\
  & & \phantom{+\epsilon^2\bigg\{} -\frac{l}{2}\nu^k_{\sss M}\nu_{\sss M}^{<i_l}\hat{U}_{\sss E}^{,L-1>k}(\bm{w}_{\sss M})-l\gamma\hat{U}_{\sss E}(\bm{w}_{\sss M})\hat{U}_{\sss E}^{,<L>}(\bm{w}_{\sss M})\nonumber\\
  & & \phantom{+\epsilon^2\bigg\{} -(l^2-l+2\gamma+2)\alpha_{\sss M}^{<i_l}\hat{U}_{\sss E}^{,L-1>}(\bm{w}_{\sss M})-lF_{\sss M}^{k<i_l}\hat{U}_{\sss E}^{,L-1>k}(\bm{w}_{\sss M})\nonumber\\
  & & \phantom{+\epsilon^2\bigg\{} +\sum_{k=0}\frac{1}{k!}\nu_{\sss M}^m\varepsilon^{mp<i_l}C^{KL-1>p}w_{\sss M}^K-\sum_{k=0}^{\infty}\frac{1}{(k+l)k!}\varepsilon_{i_lpq}\frac{dC^{pKL-1}}{du}w_{\sss M}^{<qK>}\nonumber\\
  & & \phantom{+\epsilon^2\bigg\{} +\sum_{k=0}^{\infty}\frac{1}{k!}\nu_{\sss M}^mZ^{mLK}w_{\sss M}^K-\sum_{k=0}^{\infty}\frac{1}{k!}\frac{dZ^{LK}}{du}w_{\sss M}^K-lA\hat{U}_{\sss E}^{,<L>}(\bm{w}_{\sss M})\nonumber\\
  & & \phantom{+\epsilon^2\bigg\{} -lQ_{\sss M}^L\sum_{k=1}^{\infty}\frac{1}{k!}\bigg[Q^K+(\gamma-1)P^K\bigg]w_{\sss M}^K-l\sum_{k=0}^{\infty}\frac{1}{k!}F_{\sss M}^{m<i_l}Q^{L-1>mK}w_{\sss M}^K\nonumber\\
  & & \phantom{+\epsilon^2\bigg\{} +\bigg[l\bigg(A_{\sss M}-A-\gamma\hat{U}_{\sss E}(\bm{w}_{\sss M})\bigg)+2\nu_{\sss M}^2\bigg]\sum_{k=0}^{\infty}\frac{1}{k!}Q^{LK}w_{\sss M}^K\nonumber\\
  & & \phantom{+\epsilon^2\bigg\{} -\frac{l}{2}\sum_{k=0}^{\infty}\frac{1}{k!}\nu_{\sss M}^m\nu_{\sss M}^{<i_l}Q^{L-1>mK}w_{\sss M}^K+2\sum_{k=0}^l\sum_{r=0}^{\infty}\frac{l!}{(l-k)!k!r!}Q^{R<L-K}\hat{U}_{\sss E}^{,K>}(\bm{w}_{\sss M})w_{\sss M}^R\nonumber\\
  & & \phantom{+\epsilon^2\bigg\{} +2(\beta-1)\sum_{k=0}^l\sum_{r=0}^{\infty}\frac{l!}{(l-k)!k!r!}P^{R<L-K}\hat{U}_{\sss E}^{,K>}(\bm{w}_{\sss M})w_{\sss M}^R-2\frac{d}{du}\mathfrak{G}^{<i_l,L-1>}(\bm{w}_{\sss M})\nonumber\\
  & & \phantom{+\epsilon^2\bigg\{} -(l^2-l+4)\sum_{k=0}^{\infty}\alpha_{\sss M}^{<i_l}Q^{L-1>K}w_{\sss M}^K+(l-4)\sum_{k=0}^{\infty}\frac{1}{k!}\nu_{\sss M}^{<i_l}\frac{d}{du}Q^{L-1>K}w_{\sss M}^K\nonumber\\
  & & \phantom{+\epsilon^2\bigg\{} -2(\gamma-1)\sum_{k=0}^{\infty}\frac{1}{k!}\nu_{\sss M}^{<i_l}\frac{d}{du}P^{L-1>K}w_{\sss M}^K-2(\gamma-1)\sum_{k=0}^{\infty}\frac{1}{k!}\alpha_{\sss M}^{<i_l}P^{L-1>K}w_{\sss M}^K\nonumber\\
  & & \phantom{+\epsilon^2\bigg\{} +(\gamma-1)\nu_{\sss M}^2\sum_{k=0}^{\infty}\frac{1}{k!}P^{LK}w_{\sss M}^K+\frac{1}{2}w_{\sss M}^2\sum_{k=0}^{\infty}\frac{1}{(2k+2l+3)k!}\frac{d^2}{du^2}Q^{LK}w_{\sss M}^K\nonumber\\
  & & \phantom{+\epsilon^2\bigg\{} +\sum_{k=0}^{\infty}\frac{l}{(2k+2l+1)k!}w_{\sss M}^{<i_l}\frac{d^2}{du^2}Q^{L-1>K}w_{\sss M}^K\nonumber\\
  & & \phantom{+\epsilon^2\bigg\{} +2\sum_{k=0}^{\infty}\frac{2k+2l-1}{(2k+2l+1)k!}\bigg[2\nu_{\sss M}^{<i_l}\frac{d}{du}Q^{KL-1>}+(\gamma-1)\nu_{\sss M}^{<i_l}\frac{d}{du}P^{KL-1>}\bigg]w_{\sss M}^K\bigg\}+O\left(\epsilon^4\right)\;,
\er
where we have used notations
\br
    K_{\sss M}^{ij}&\equiv&3\alpha_{\sss M}^{<i}\alpha_{\sss M}^{j>}\,,                      \label{5.10yt}
\\\nonumber\\
    K_{\sss M}^{L}&\equiv&0\,,\qquad\qquad\qquad (l\ge3).                          \label{5.11yt}
\er

The current-type external multipoles $C_{\sss M}^L$ are given by
\br\label{3.29yt}
\varepsilon^{ipi_l}C_{\sss M}^{pL-1} & = & \frac{4l(1+\gamma)}{l+1}\bigg[\nu_{\sss M}^{[i}\hat{U}_{\sss E}^{,i_l]L-1}(\bm{w}_{\sss M})-\hat{U}_{\sss E}^{[i,i_l]L-1}(\bm{w}_{\sss M})-\frac{l-1}{l}\delta^{i_{l-1}[i}\dot{\hat{U}}_{\sss E}^{,i_l]L-2}(\bm{w}_{\sss M})\bigg]\nonumber\\
  &&+\sum_{k=0}^{\infty}\frac{l}{(k+l)k!}\varepsilon^{ipi_l}C^{pKL-1}w_{\sss M}^K\nonumber\\
  &&+\frac{4l}{l+1}\bigg[\mathfrak{G}^{[i,i_l]L-1}(\bm{w}_{\sss M})+\frac{2}{2l+1}\delta_{i_{l-1}[i}\mathfrak{G}^k_{\phantom{K},i_l]kL-2}(\bm{w}_{\sss M})\bigg]\nonumber\\
  &&+\frac{4l}{l+1}\sum_{k=0}^{\infty}\frac{1}{k!}\bigg[2\nu_{\sss M}^{[i}Q^{i_l]KL-1}+(\gamma-1)\nu_{\sss M}^{[i}P^{i_l]KL-1}\bigg]w_{\sss M}^K\nonumber\\
  &&+\frac{8l}{(l+1)(2l+1)}\sum_{k=0}^{\infty}\frac{1}{k!}\nu_{\sss M}^m\bigg[2\delta^{i_{l-1}[i}Q^{i_l]mKL-2}+(\gamma-1)\delta^{i_{l-1}[i}P^{i_l]mKL-2}\bigg]w_{\sss M}^K+O\left(\epsilon^2\right)\,.
\er
This finishes our study of the post-Newtonian reference frames in the Earth-Moon system. The research material of this paper will be used for derivation of the post-Newtonian equations of motion of Earth and Moon in the publication which is forthcoming.
\begin*{acknowledgments}\noindent
We appreciate the help of L. Iorio who has provided references 
related to the historical development of the lunar ephemeris theories. We are grateful to V.A. Brumberg, B. Mashhoon, Q.G. Bailey and J. Williams for valuable comments and suggestions for improving the paper. Y. Xie is thankful to the Department of Physics \& Astronomy of the University of Missouri-Columbia for hospitality and accommodation.
The work of Y. Xie was supported by the China Scholarship Council Grant No. 2008102243.
The work of S. Kopeikin was supported by the Research Council Grant No. C1669103 of the University of Missouri-Columbia and by 2008-09 faculty incentive grant of the Arts and Science Alumni Organization of the University of Missouri-Columbia.
\end*{acknowledgments}
\bibliography{Lunar_Motion_update3}
\end{document}